\documentclass{IEEEtran}
\usepackage{latexsym}
\usepackage{cite}
\usepackage{amssymb}
\usepackage{bbm}
\usepackage{bm}
\usepackage[usenames,dvipsnames]{color}
\usepackage{dsfont}
\usepackage{amsmath,amssymb}
\DeclareMathOperator{\Tr}{tr}
\usepackage[dvips]{graphics}
\usepackage{graphicx}
\usepackage{comment}
\usepackage{amsmath}
\usepackage{psfrag}
\usepackage{caption}
\usepackage{lipsum}
\usepackage{ntheorem}
\usepackage{subfig}
\usepackage{setspace}
\usepackage{algorithm}
\usepackage{algpseudocode}
\usepackage{theorem}

\usepackage{float}
\usepackage{mathtools}
 \allowdisplaybreaks

\newtheorem{definition}{Definition}

\newtheorem{theorem}{Theorem}
\newtheorem{lemma}{Lemma}
\newtheorem{bound}{Bound}
\newtheorem{remark}{Remark}

\newtheorem{corollary}{Corollary}
\newtheorem{proposition}{Proposition}

\newcommand\numberthis{\addtocounter{equation}{1}\tag{\theequation}}
\catcode`~=11 
\newcommand{\urltilde}{\kern -.15em\lower .7ex\hbox{~}\kern .04em}
\catcode`~=13 

\begin{document}

\makeatletter
\newcommand{\vasti}{\bBigg@{3}}
\newcommand{\vast}{\bBigg@{4}}
\newcommand{\Vast}{\bBigg@{5}}
\makeatother
\newcommand{\be}{\begin{equation}}
\newcommand{\ee}{\end{equation}}
\newcommand{\ba}{\begin{align}}
\newcommand{\ea}{\end{align}}
\newcommand{\baa}{\begin{align*}}
\newcommand{\eaa}{\end{align*}}
\newcommand{\bea}{\begin{eqnarray}}
\newcommand{\eea}{\end{eqnarray}}
\newcommand{\beaa}{\begin{eqnarray*}}
\newcommand{\eeaa}{\end{eqnarray*}}
\newcommand{\p}[1]{\left(#1\right)}
\newcommand{\pp}[1]{\left[#1\right]}
\newcommand{\ppp}[1]{\left\{#1\right\}}
\newcommand{\ber}{\ \mathsf{Ber}}
\newcommand\aatop[2]{\genfrac{}{}{0pt}{}{#1}{#2}}

\title{MIMO Gaussian Broadcast Channels with Common, Private and Confidential Messages}

\author{Ziv Goldfeld, \emph{Member, IEEE}, and Haim H. Permuter, \emph{Senior Member, IEEE}

\thanks{
		The work of Z. Goldfeld and H. H. Permuter was supported by the Israel Science Foundation (grant no. 2012/14), the European Research Council under the European Union's Seventh Framework Programme (FP7/2007-2013) / ERC grant agreement n$^\circ$337752, and the Cyber Center and at Ben-Gurion University of the Negev. Z. Goldfeld was also supported by the Rothschild postdoc fellowship and by a grant from Skoltech--MIT Joint Next Generation Program (NGP).
		\newline Z. Goldfeld is with the Department of Electrical Engineering and Computer Science, Massachusetts Institute of Technology, Cambridge, MA 02139 USA (e-mail: zivg@mit.edu). H. H. Permuter is with the Department of Electrical Engineering, Ben-Gurion University, Beer-Sheva, Israel 8410501 (e-mail: haimp@bgu.ac.il).}}
\maketitle


\begin{abstract}
The two-user multiple-input multiple-output (MIMO) Gaussian broadcast channel (BC) with common, private and confidential messages is considered. The transmitter sends a common message to both users, a confidential message to User 1 and a private (non-confidential) message to User 2. The secrecy-capacity region is characterized by showing that certain inner and outer bounds coincide and that the boundary points are achieved by Gaussian inputs, which enables the development of a tight converse. The proof relies on factorization of upper concave envelopes and a variant of dirty-paper coding (DPC). It is shown that the entire region is exhausted by using DPC to cancel out the signal of the non-confidential message at Receiver 1, thus making DPC against the signal of the confidential message unnecessary. A numerical example illustrates the secrecy-capacity results.
\end{abstract}



\section{Introduction}\label{SEC:introduction}

\begin{figure}[t!]
    \begin{center}
        \begin{psfrags}
            \psfragscanon
            \psfrag{A}[][][1]{\ \ \ \ $\mathbf{X}$}
            \psfrag{B}[][][1]{$\mathrm{G}_1$}
            \psfrag{C}[][][1]{\ $\mathrm{G}_2$}
            \psfrag{D}[][][1]{\ $\mathbf{Z}_1\sim\mathcal{N}(\mathbf{0},\mathrm{I})$}
            \psfrag{E}[][][1]{\ $\mathbf{Z}_2\sim\mathcal{N}(\mathbf{0},\mathrm{I})$}
            \psfrag{F}[][][1]{$\mathbf{Y}_1$}
            \psfrag{G}[][][1]{$\mathbf{Y}_2$}
            \psfrag{L}[][][1]{\ $(M_0,M_1,M_2)$}
            \psfrag{M}[][][1]{\ \ \ \ \ \ \ \ \ $\big(\hat{M}_0^{(1)},\hat{M}_1\big)$}
            \psfrag{N}[][][1]{\ \ \ \ \ \ \ \ \ $\big(\hat{M}_0^{(2)},\hat{M}_2\big)$}
            \psfrag{O}[][][1]{$M_1$}
            \hspace{-4mm}\includegraphics[scale = .75]{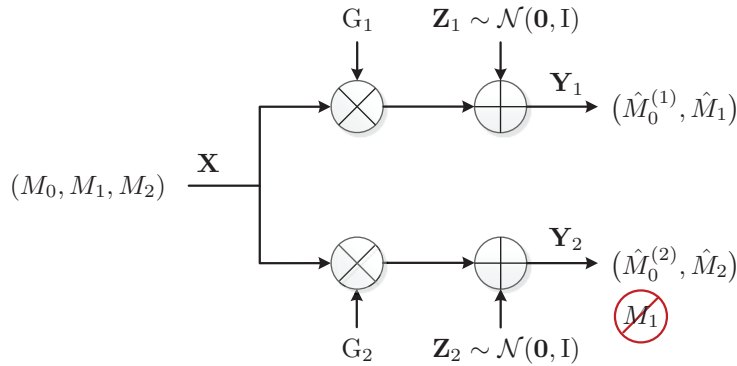}
            \caption{MIMO Gaussian BC.} \label{FIG:Gaussian_BC}
            \psfragscanoff
        \end{psfrags}
     \end{center}
 \end{figure}

Additive Gaussian channels are a common model for wireless communication, whose open nature makes it vulnerable to a variety of security threats, such as eavesdropping. However, eavesdroppers are not always a malicious entity from which \emph{all} transmissions are concealed. Rather, a legitimate recipient of one message may serve as an eavesdropper for other messages. We encapsulate this notion in a two-user multiple-input multiple-output (MIMO) Gaussian broadcast channel (BC) with common, private and confidential messages (Fig. \ref{FIG:Gaussian_BC}). The common message $M_0$ is intended to both users, while $M_1$ and $M_2$ are private messages that are sent to User 1 and User 2, respectively. Furthermore, $M_1$ is confidential and is kept secret from User 2. Many real-life scenarios fall within this framework. One such example is a banking site that simultaneously: (i) broadcasts an advertisement to all online users (modeled by $M_0$); (ii) offers public information (such as material on different banking programs, reports, forecasts, etc.) that is available only to users that are interested in it (modeled by the private message $M_2$); and (iii) provides an online banking service, by which users can access their account and perform transactions (this confidential information is modeled by $M_1$). Furthermore, 5th generation (5G) mobile technology \cite{gupta2015survey} puts significant emphasis on advanced MIMO capabilities and multiuser communication. In particular, schemes supporting multiple users exchanging various kinds of information over a MIMO communication system (and their fundamental limits) are of great interest. The studied MIMO Gaussian BC is an instance of a system where these aspects are jointly~incorporated.


\par In recent years, information-theoretic security over MIMO communication systems has been an active field of research (see \cite{Liang_MIMO_Secrecy2008} for a recent survey of progress in this area). Most noticeably, the secrecy-capacity of the Gaussian wiretap channel (WTC) was characterized in \cite{Yates_MISOWTC2007,Ulukus_221WTC2009,Khitsi_MISOWTC2010} for the multiple-input single-output scenario, and in \cite{Bustin_MMSE_WTC2009,Liu_Shamai_MIMOWTC2009,Khitsi_MIMOWTC2010,Hassibi_MINOWTC2011,Ulukus_Gaussian_Wiretap2011} for the MIMO case. The Gaussian MIMO WTC with a common message was studied in \cite{Liang_WTCCommon2009}. In \cite{Chong_MIMOGaussin2013}, the secrecy-capacity region for the setting with a degraded message set and an external eavesdropper (from which all messages are concealed) was derived. The MIMO Gaussian BC with confidential messages, in which the private message to each user is kept secret from the opposite user, without and with a common message, was solved in \cite{Poor_Shamai_Gaussian_MIMO_BC_Secrecy2010} and \cite{Ulukus_GaussianBCCommon2012}, respectively. As the capacity region of the MIMO Gaussian BC without secrecy requirements was derived in \cite{Weingarten_MIMOBC2006} with no common message present, and in \cite{Chandra_Gauss_BC2014} with a common message, this work settles the two remaining scenarios concerning secrecy. More specifically, focusing on the two-user MIMO Gaussian BC with or without a common message and where both, either or neither of the private messages are secret, we derive the secrecy-capacity regions of the only two instances that remained unsolved until now. A pointer to each past result and the contribution of this work are found in Table \ref{table_summary}.

Up until the more recent work of Geng and Nair \cite{Chandra_Gauss_BC2014}, all the aforementioned results established the optimality of Gaussian inputs based on channel enhancement arguments, originally used in \cite{Weingarten_MIMOBC2006} to characterize the private message capacity region of the MIMO Gaussian BC (without secrecy constraints). In a nutshell, the idea of \cite{Weingarten_MIMOBC2006} was to approximate the actual BC using enhanced BCs, for which the entropy power inequality applies and is invoked to establish the optimality of Gaussian inputs (similarly to the proof for the scalar case by Bergmans \cite{Bergmans_Gauss_BC1974}). Continuity arguments are then used to characterize the capacity region of the MIMO Gaussian BC of interest. The limitation of the channel enhancement technique seems to be the difficulty in generalizing it to account for both private and common messages. Attempted adaptations of this technique to scenarios comprising common and private messages include, e.g., \cite{ekrem2012outer} and \cite{ekrem2013multi}, where generally unmatching inner and outer bounds or constant gap-from-capacity results were derived for the MIMO Gaussian BC without and with security requirements, respectively.

Our goal is to fully characterize the secrecy-capacity region of the MIMO Gaussian BC with common, private and confidential messages and show that it is attained by Gaussian inputs. Since channel enhancement arguments are insufficient for this purpose, we adopt the approach of \cite{Chandra_Gauss_BC2014} for proving the optimality of Gaussians via factorization of upper concave envelopes (UCEs). We start by characterizing the secrecy-capacity region under an input covariance constraint for the setting with private and confidential messages only (i.e., when no common message is present). The derivation first describes the boundary points of a certain outer bound on the secrecy-capacity region as an UCE of a function of the input distribution. With this result at hand, we show that if this UCE satisfies a specific factorization property, then it is maximized by a Gaussian input distribution. Then, using an adaptation of dirty-paper coding (DPC) \cite{Cioffi_Precoding2004}, we establish the equivalence of the outer bound to a particular inner bound, thus characterizing the secrecy-capacity region. Interestingly, optimality is achieved by using DPC to cancel out the signal of the non-confidential message $M_2$ at Receiver 1 only. The other variant, i.e., DPC against the signal of the confidential message $M_1$, turns out to be unnecessary. This is in contrast to the case without secrecy requirements \cite{Chandra_Gauss_BC2014}, in which both variants of DPC are necessary to exhaust the entire region.

\begin{table*}[!t]
\begin{center}
\caption{MIMO Gaussian BCs with/without a common message and where none/some/all of the private messages are secret - Summary of Results}\label{table_summary}
\begin{tabular}{|c|c|c|c|}
\hline
$\bm{M_0}$ & $\bm{M_1}$ & $\bm{M_2}$ & \textbf{Solution} \\ \hline\hline
--- & Private & Private &  Weingarten-Steinberg-Shamai 2006 \cite{Weingarten_MIMOBC2006}\\ \hline
Public & Private & Private &  Geng-Nair 2014 \cite{Chandra_Gauss_BC2014}\\ \hline
Public & Secret from User 2& --- & Ly-Liu-Liang 2010 \cite{Liang_WTCCommon2009}\\\hline
--- & Secret from User 2 & Secret from User 1& Liu-Liu-Poor-Shamai 2010 \cite{Poor_Shamai_Gaussian_MIMO_BC_Secrecy2010}\\\hline
Public & Secret from User 2 & Secret from User 1& Ekrem-Ulukus 2012 \cite{Ulukus_GaussianBCCommon2012}\\\hline
--- & Secret from User 2& Private & This work\\\hline
Public & Secret from User 2& Private & This work\\\hline
\hline
\end{tabular}\vspace{-3mm}
\end{center}
\end{table*}

\par We then focus on the MIMO Gaussian BC with common, private and confidential messages (Fig. \ref{FIG:Gaussian_BC}) and derive our main result by characterizing its secrecy-capacity region. Although this is a generalization of the problem without a common message, the secrecy-capacity of the latter setting is solved first. In doing so, we use the result without a common message to show that Gaussian inputs are optimal for a certain portion of the region with a common message. The rest of the region is characterized by extending the tools from \cite{Chandra_Gauss_BC2014} and introducing the notion of a double-nested UCE. Gaussian inputs once again are shown to attain optimality. Finally, we visualize our results by a numerical example. Since the obtained regions are described as non-convex matrix optimization problems, we convert them into a computationally efficient form by relying on matrix decomposition properties from \cite{Khina_WTC_Decomposition2015}. 

\textbf{Organization:} The UCE factorization method introduced in \cite{Chandra_Gauss_BC2014} and further developed in this work relies on rather heavy machinery and many technical functional analysis results. The proofs of our main secrecy-capacity theorems (i.e., without and with a common message) frequently refer to these auxiliary results while also relying on additional information-theoretic arguments (e.g.,  DPC). In structuring this paper, it was important for us to distill the information-theoretic arguments from the functional analysis aspects of this work for two main reasons. First, this would ease the flow trough the information-theoretic proofs and highlight the usage of the UCE factorization method for showing that Gaussian inputs achieve capacity. Second, by aggregating the machinery behind this method in a separate section we hope to facilitate its application to additional research problems in future work.

In accordance to the above, the rest of the paper is organized as follows. Section \ref{SEC:preliminaries} gives definitions and describes the MIMO Gaussian BC with common, private and confidential messages. In Section \ref{SEC:capacity_results} we state our main results: the secrecy-capacity regions of the considered BC without and with a common message, given in Theorems \ref{TM:capacity_nocommon} and \ref{TM:capacity_common}, respectively. Discussions of these results and a numerical example to illustrate the obtained regions are also given in Section \ref{SEC:capacity_results}. Section \ref{SEC:optimality_of_Gaussians} presents the various definitions and properties related to UCEs used throughout this work (with proofs relegated to Section \ref{SEC:envelope_properties}). Section \ref{SEC:capacity_results_proofs} contains the proofs of our main secrecy-capacity results based on the technical background supplied in Section \ref{SEC:optimality_of_Gaussians}. Finally, Section \ref{SEC:summary} summarizes the main results and insights of this work.





\section{Notation and Preliminaries}\label{SEC:preliminaries}


\subsection{Notation}

We use the following notations. The set of natural numbers (which does not include 0) is denoted by $\mathbb{N}$, while $\mathbb{R}$ are the reals. We further define $\mathbb{R}_+\triangleq\{x\in\mathbb{R}|x\geq 0\}$. Given two real numbers $a,b$, we denote by $[a\mspace{-3mu}:\mspace{-3mu}b]$ the set of integers $\big\{n\in\mathbb{N}\big| \lceil a\rceil\leq n \leq\lfloor b \rfloor\big\}$; when $a=1$ we use the shorthand $[b]$. Calligraphic letters denote sets, e.g., $\mathcal{X}$, while $|\mathcal{X}|$ stands for the cardinality of $\mathcal{X}$. $\mathcal{X}^n$ denotes the $n$-fold Cartesian product of $\mathcal{X}$. An element of $\mathcal{X}^n$ is denoted by $x^n=(x_1,x_2,\ldots,x_n)$; whenever the dimension $n$ is clear from the context, vectors (or sequences) are denoted by boldface letters, e.g., $\mathbf{x}$. The transpose and the Euclidean norm of $\mathbf{x}$ are denoted by $\mathbf{x}^\top$ and $\|\mathbf{x}\|$, respectively. Random variables are denoted by uppercase letters, e.g., $X$, with similar conventions for random vectors. All the random variables considered in this work are real valued.

Probability density functions (PDFs) are denoted by the lowercase letters $p$ or $q$, with a subscript that identifies the random variable and its possible conditioning. For example, for two jointly continuous random vectors $\mathbf{X}$ and $\mathbf{Y}$, let $p_\mathbf{X}$, $p_{\mathbf{X},\mathbf{Y}}$ and $p_{\mathbf{X}|\mathbf{Y}}$ denote, respectively, the PDF of $\mathbf{X}$, the joint PDF of $(\mathbf{X},\mathbf{Y})$ and the conditional PDF of $\mathbf{X}$ given $\mathbf{Y}$. Expressions such as $p_{X,Y}=p_Xp_{Y|X}$ are to be understood pointwise, i.e., $p_{X,Y}(x,y)=p_X(x)p_{Y|X}(y|x)$, for all $(x,y)\in\mathcal{X}\times\mathcal{Y}$. Accordingly, when three random variables $X$, $Y$ and $Z$ satisfy $p_{X|Y,Z}=p_{X|Y}$, they form a Markov chain, which we denote by $X-Y-Z$. The subscripts of a PDF are omitted if its arguments are lowercase versions of the corresponding random variables. General (i.e., not necessarily continuous) probability distributions are denoted by the uppercase letters $P$ and $Q$, with conventions similar to those used for PDFs. The expectation of a random variable $X$ is $\mathbb{E}X$. When a random variable $X$ is normally distributed we write $X\sim\mathcal{N}(\mu,\sigma^2)$, where $\mu=\mathbb{E}X$ is the expectation of $X$ and $\sigma^2=\mathsf{var}(X)$ is its variance. Similarly, an $n$-dimensional Gaussian distribution of dimension is defined by the expectation $\bm{\mu}=\mathbb{E}\mathbf{X}\in\mathbb{R}^n$ and the covariance matrix $\mathrm{K}=\mathbb{E}\big[(\mathbf{X}-\bm{\mu})(\mathbf{X}-\bm{\mu})^\top\big]$, for which we write $\mathbf{X}\sim\mathcal{N}(\bm{\mu},\mathrm{K})$. Generally, non-italic capital letters, e.g., $\mathrm{A}$, denote matrices. We use $\mathrm{A}\succeq0$ to indicate that a matrix $\mathrm{A}$ is positive semi-definite, while $\mathrm{A}\preceq\mathrm{B}$ denotes ``less than or equal to'' in the positive semi-definite ordering, i.e., $ \mathrm{B}-\mathrm{A}\succeq0$. The determinant of a square matrix $\mathrm{A}$ is designated by $|\mathrm{A}|$.


\begin{definition}[Upper Concave Envelope]\label{DEF:upper_concave_envelope}
Let $f:\mathcal{D}\to\mathbb{R}$ be a function defined on a convex set $\mathcal{D}$. The UCE $\mathfrak{C}f:\mathcal{D}\to\mathbb{R}$ of $f$ is the pointwise smallest concave function such that $\big(\mathfrak{C}f\big)(x)\geq f(x),\ \forall x\in\mathcal{D}$.
\end{definition}

Another representation of the UCE $\mathfrak{C}f$ of $f$ relies on the supporting hyperplanes of $f$. Namely, for any $x\in\mathcal{D}$, we have $\big(\mathfrak{C}f\big)(x)=\sup\limits_{V:\ \mathbb{E}V=x} \mathbb{E}f(V)$.


\subsection{Problem Definition}\label{SUBSEC:Problem_Definition}

The outputs of a MIMO Gaussian BC at the $i$th channel use are:
\begin{equation}
\mathbf{Y}_j(i) =\mathrm{G}_j\mathbf{X}(i)+\mathbf{Z}_j(i),\quad j=1,2,\quad i\in[n],\label{EQ:channel_model}
\end{equation}
where $\mathrm{G}_1,\mathrm{G}_2\in\mathbb{R}^{t\times t}$ are channel gain matrices (assumed to be known to all parties), $\big\{\mathbf{Z}_j(i)\big\}_{i\in[n]}$, for $j=1,2$, is an independent and identically distributed (i.i.d.) sequence of Gaussian random vectors taking values in $\mathbb{R}^{t\times 1}$. For each $j=1,2$ and $i\in[n]$, the elements of $\mathbf{Z}_j(i)=\Big[Z_{j,1}(i)\ Z_{j,2}(i)\ \ldots\ Z_{j,t}(i)\Big]^\top$ are also i.i.d. Gaussian random variables, whose expected values and variance are specified by the parameters of the normal distribution of $\mathbf{Z}_j(i)$. The input sequence $\big\{\mathbf{X}(i)\big\}_{i\in[n]}$ is is subject to the covariance constraint
\begin{equation}
\frac{1}{n}\sum_{i=1}^n\mathbb{E}\left[\mathbf{X}(i)\mathbf{X}(i)^\top\right]\preceq \mathrm{K},\label{EQ:power_constraint}
\end{equation}
where $\mathrm{K}\succeq \mathrm{0}$.

\begin{remark}[Assumptions] We make the following assumptions on the channel gain matrices and the noise statistics:
\begin{enumerate}
\item $\mathrm{G}_1$ and $\mathrm{G}_2$ are square and invertible. The analysis in this work relies on showing that certain inner and outer bounds on the secrecy-capacity region coincide. These bounds are characterized in terms of mutual information terms between the channel input (or some auxiliary random variables) and the channel outputs. The mutual information terms, and hence the inner and outer bounds, are continuous functions of the channel gain matrices. For square matrices, recall that the set of invertible matrices is a dense open set in the set of all $t\times t$ matrices. Therefore, by continuity of the bounds, the inner and outer bounds coincide for all channel gain matrices. If $\mathrm{G}_1$ and $\mathrm{G}_2$ are not square, the singular value decomposition (SVD) allows rewriting the original MIMO BC as a MIMO BC with square gain matrices (of size corresponding to the number of transmitting antennas) using only reversible manipulations. These manipulations have no effect on the secrecy-capacity region of the channel (see, e.g., \cite[Section 5]{Weingarten_MIMOBC2006}).

\item For each $j=1,2$, the Gaussian noise vectors, $\big\{\mathbf{Z}_j(i)\big\}_{i\in[n]}$, are i.i.d. according to $\mathcal{N}(\mathbf{0},\mathrm{I})$, where $\mathrm{I}$ is the $t\times t$-identity matrix. This assumption is without loss of generality due to the following reasons: First, the mean of the Gaussian noise does not affect the capacity region. Second, when the covariance matrix is invertible, the noises can be whitened by multiplying \eqref{EQ:channel_model} by another invertible matrix. On the other hand, if the covariance matrix is non-invertible, the communication scenario degenerates since a suitable linear transformation converts the Gaussian channel into a noiseless channel with infinite capacity.
\end{enumerate}
\end{remark}

We study the MIMO Gaussian BC with common, private and confidential messages (Fig. \ref{FIG:Gaussian_BC}). The sender communicates three messages $(M_0,M_1,M_2)$ over the MIMO Gaussian BC from \eqref{EQ:channel_model}. $M_0$ is a common message that is intended to both users, while $M_j$, for $j=1,2$, is delivered to user $j$ only. The receivers are to recover their intended messages with arbitrarily small error probability. Moreover, $M_1$ is a confidential message that is to be kept secret from User 2, which is formally described by the weak-secrecy requirement
\begin{equation}
\frac{1}{n}I(M_1;\mathbf{Y}_2^n)\xrightarrow[n\to\infty]{} 0,\label{EQ:weak_secrecy}
\end{equation}
where $n$ is the number of channel uses. In \eqref{EQ:weak_secrecy}, the notation $\mathbf{Y}_2^n\triangleq\Big[\mathbf{Y}_2(1)\ \mathbf{Y}_2(2)\ \ldots\ \mathbf{Y}_2(n)\Big]^\top$ is used, where for each $i\in[n]$, $\mathbf{Y}_2(i)$ is the output vector (taking values in $\mathbb{R}^t$) observed by User 2 at the $i$th channel instance. For any covariance constraint $\mathrm{K}\succeq 0$, the secrecy-capacity region $\mathcal{C}_\mathrm{K}$ is the closure of all achievable rate triples $(R_0,R_1,R_2)\in\mathbb{R}^3_+$, where achievability is defined in a standard manner (see, e.g.,~\cite{Csiszar_Korner_Book2011}).

\begin{remark}[Weak versus Strong Secrecy] We set up the problem in term of the weak-secrecy metric \eqref{EQ:weak_secrecy}, merely because the general inner and outer bounds we use \cite{goldfeld2017broadcast} were originally proven under this paradigm\footnote{The work \cite{goldfeld2017broadcast} derived bounds on the admissible rate region of a BC with privacy leakage constraints. Zero leakage corresponds to secrecy, but since leakage can only be defined in terms of rate the resulting notion of security is weak-secrecy.}. Nonetheless, the results from \cite{goldfeld2017broadcast} are readily upgraded to strong secrecy using the approach of Maurer and Wolf from \cite{Maurer_Wolf_Strong_Secrecy2000}, while accounting for the channel being continuous in a manner similar to \cite{barros2008strong}. Since the focus of this paper is on the optimality of Gaussian inputs and computable secrecy-capacity expressions, we do not dwell on the employed notion of security.
\end{remark}


\section{Secrecy-Capacity Results}\label{SEC:capacity_results}


\subsection{MIMO Gaussian BCs with Private and Confidential Messages}\label{SUBSEC:result_nocommon_capacity}

The MIMO Gaussian BC with private and confidential messages but without a common message is defined as in Section \ref{SUBSEC:Problem_Definition}, while setting $R_0=0$. For any covariance constraint $\mathrm{K}\succeq 0$, let $\hat{\mathcal{C}}_\mathrm{K}$ be the corresponding secrecy-capacity region, and for any $0 \preceq\mathrm{K}^\star\preceq\mathrm{K}$ set the following shorthand notations:
\begin{subequations}
\begin{align}
\hat{r}_1(\mathrm{K}^\star)&\triangleq \frac{1}{2}\log\left|\frac{\mathrm{I}+\mathrm{G}_1\mathrm{K}^\star\mathrm{G}_1^\top}{\mathrm{I}+\mathrm{G}_2\mathrm{K}^\star\mathrm{G}_2^\top}\right|\label{EQ:region_nocommon_simpleRB1}\\
\hat{r}_2(\mathrm{K}^\star)&\triangleq \frac{1}{2}\log\left|\frac{\mathrm{I}+\mathrm{G}_2\mathrm{K}\mathrm{G}_2^\top}{\mathrm{I}+\mathrm{G}_2\mathrm{K}^\star\mathrm{G}_2^\top}\right|.\label{EQ:region_nocommon_simpleRB2}
\end{align}\label{EQ:region_nocommon_simpleRB}%
\end{subequations}
Define also
\begin{equation}
\hat{\mathcal{C}}_\mathrm{K}(\mathrm{K}^\star)\triangleq \left\{(R_1,R_2)\in\mathbb{R}^2_+\vast| \begin{array}{ll}R_1\leq \hat{r}_1(\mathrm{K}^\star)\\ R_2\leq \hat{r}_2(\mathrm{K}^\star)\end{array}\right\}.
\end{equation}
The following theorem characterizes $\hat{\mathcal{C}}_\mathrm{K}$.

\begin{theorem}[Secrecy-Capacity without Common Message]\label{TM:capacity_nocommon}
The secrecy-capacity region $\hat{\mathcal{C}}_\mathrm{K}$ of the MIMO Gaussian BC with private and confidential messages under the covariance constraint \eqref{EQ:power_constraint} is
\begin{equation}
\hat{\mathcal{C}}_\mathrm{K}=\bigcup_{\mathrm{0}\preceq\mathrm{K}^\star\preceq\mathrm{K}}\hat{\mathcal{C}}_\mathrm{K}
(\mathrm{K}^\star)\label{EQ:region_nocommon}.
\end{equation}
\end{theorem}

The proof of Theorem \ref{TM:capacity_nocommon} (given in Section \ref{SUBSEC:proof_nocommon_capacity}) shows that certain inner and outer bounds of the secrecy-capacity region coincide, and that Gaussian inputs are optimal. First, we show that the boundary points of the outer bound are an UCE of a function of the input distribution. Based on some properties of UCEs (see Sections \ref{SUBSEC:optimality_of_Gaussians_example1} and \ref{SUBSEC:optimality_of_Gaussians_example2}) we deduce that a Gaussian input distribution maximizes the considered UCE. The secrecy-capacity region is then characterized by evaluating the boundary points of the inner bound under a Gaussian input vector and showing that they coincide with those of the outer bound.

\begin{remark}[Interpretation of Optimal Secrecy Rates]\label{REM:interper_nocommon}
The right-hand side (RHS) of \eqref{EQ:region_nocommon_simpleRB1} is the secrecy-capacity of the Gaussian MIMO WTC with input covariance $\mathrm{K}^\star$, where User 1 serves as the legitimate party and User 2 as the eavesdropper. The RHS of \eqref{EQ:region_nocommon_simpleRB2} is the capacity of the MIMO Gaussian point-to-point channel with input covariance $\mathrm{K}-\mathrm{K}^\star$ and noise covariance $\mathrm{I}+\mathrm{K}^\star$. Thus, $M_1$ being confidential forces User 1 to treat the second user as an eavesdropper. Then, the transmission rate of $M_2$ (to User 2) is maximized by consuming the remaining power, while treating the signal of the first user as noise. The optimization over $\mathrm{K}^\star$ corresponds to different choices of user prioritization.
\end{remark}


\begin{remark}[Relation to Dirty-Paper Coding]
As evident from the proof of Theorem \ref{TM:capacity_nocommon} (see Proposition \ref{PROP:Partial_DPC} in Section \ref{SUBSEC:proof_nocommon_capacity}), the entire secrecy-capacity region $\hat{\mathcal{C}}_\mathrm{K}$ is achieved by using DPC to cancel out the signal of the non-confidential message $M_2$ at Receiver 1 only. The other variant, i.e., DPC against the signal of the confidential message $M_1$ at Receiver 2, is unnecessary. This is in contrast to the situation without a secrecy requirement on $M_1$ (namely, the private message BC), for which the capacity region is exhausted by taking the convex hull of both variants (DPC against $M_1$ and DPC against $M_2$).
\end{remark}

\begin{remark}[Relation to Common Message Case] Theorem \ref{TM:capacity_nocommon} is a special case of the secrecy-capacity region of the MIMO Gaussian BC with common, private and confidential messages $\mathcal{C}_\mathrm{K}$ (given in Theorem \ref{TM:capacity_common}). Nonetheless, we separately state and prove Theorem \ref{TM:capacity_nocommon} since it is used as an auxiliary result for the proof of Theorem \ref{TM:capacity_common}, as it implies the optimality of Gaussian inputs for a certain portion of $\mathcal{C}_\mathrm{K}$. More specifically, when the private message rate $R_2$ is larger than the common message rate $R_0$, the optimizing distribution of $\mathcal{C}_\mathrm{K}$ coincides with that of $\hat{\mathcal{C}}_\mathrm{K}$; Theorem~\ref{TM:capacity_nocommon} shows that this distribution is Gaussian.
\end{remark}

As a corollary of Theorem \ref{TM:capacity_nocommon}, we characterize the secrecy-capacity region under the average total power constraint. This is a simple consequence of \cite[Lemma 1]{Weingarten_MIMOBC2006}.

\begin{corollary}[Average Total Power Constraint]\label{COR:capacity_nocommon_total_power}
The secrecy-capacity region of the MIMO Gaussian BC with private and confidential messages under the average total power constraint
\begin{subequations}
\begin{equation}
\frac{1}{n}\sum_{i=1}^n\big\|\mathbf{X}(i)\big\|^2\leq P,\label{EQ:nocommon_power}
\end{equation}
is given by
\begin{equation}
\hat{\mathcal{C}}_P=\bigcup_{\mathrm{0}\preceq\mathrm{K}:\ \Tr(\mathrm{K})\leq P}\hat{\mathcal{C}}_\mathrm{K}\label{EQ:region_nocommon_power}.
\end{equation}
\end{subequations}
\end{corollary}

\begin{remark}[Computing Secrecy-Capacity Region]
In general, it is hard to compute \eqref{EQ:region_nocommon} and \eqref{EQ:region_nocommon_power} as they involve non-convex matrix optimization problems. Nonetheless, in Section \ref{SUBSEC:numerical_example} we show how to convert \eqref{EQ:region_nocommon} and \eqref{EQ:region_nocommon_power} into a computationally efficient form based on matrix decomposition properties from \cite{Khina_WTC_Decomposition2015}. The simplified optimization problem is then used to illustrate the secrecy-capacity region under an average total power constraint $\hat{\mathcal{C}}_P$ on a numerical example.
\end{remark}


\subsection{MIMO Gaussian BCs with Common, Private and Confidential Messages}\label{SUBSEC:result_common_capacity}

To state the secrecy-capacity region $\mathcal{C}_\mathrm{K}$ of the MIMO Gaussian BC with common, private and confidential messages as defined in Section \ref{SUBSEC:Problem_Definition}, we define
\begin{subequations}
\begin{align}
r_0^{(j)}(\mathrm{K}_1,\mathrm{K}_2)&\triangleq\frac{1}{2}\log\left|\frac{\mathrm{I}+\mathrm{G}_j\mathrm{K}\mathrm{G}_j^\top}{\mathrm{I}+\mathrm{G}_j(\mathrm{K}_1+\mathrm{K}_2)\mathrm{G}_j^\top}\right|,\quad j=1,2\label{EQ:region_common_simpleRB0}\\
r_1(\mathrm{K}_2)&\triangleq\frac{1}{2}\log\left|\frac{\mathrm{I}+\mathrm{G}_1\mathrm{K}_2\mathrm{G}_1^\top}{\mathrm{I}+\mathrm{G}_2\mathrm{K}_2\mathrm{G}_2^\top}\right|\label{EQ:region_common_simpleRB1}\\
r_2(\mathrm{K}_1,\mathrm{K}_2)&\triangleq\frac{1}{2}\log\left|\frac{\mathrm{I}+\mathrm{G}_2(\mathrm{K}_1+\mathrm{K}_2)\mathrm{G}_2^\top}{\mathrm{I}+\mathrm{G}_2\mathrm{K}_2\mathrm{G}_2^\top}\right|,\label{EQ:region_common_simpleRB2}
\end{align}\label{EQ:region_common_simpleRB}
\end{subequations}
and $r_0(\mathrm{K}_1,\mathrm{K}_2)\triangleq\min\big\{r_0^{(1)}(\mathrm{K}_1,\mathrm{K}_2),r_0^{(2)}(\mathrm{K}_1,\mathrm{K}_2)\big\}$, and set
\begin{equation}
\mathcal{C}_\mathrm{K}(\mathrm{K}_1,\mathrm{K}_2)\triangleq \left\{(R_0,R_1,R_2)\in\mathbb{R}^3_+\Vast| \begin{array}{ll}R_0\leq r_0(\mathrm{K}_1,\mathrm{K}_2)\\R_1\leq r_1(\mathrm{K}_2)\\ R_2\leq r_2(\mathrm{K}_1,\mathrm{K}_2)\end{array}\right\}.\numberthis
\end{equation}

\begin{theorem}[Secrecy-Capacity with Common Message]\label{TM:capacity_common}
The secrecy-capacity region $\mathcal{C}_\mathrm{K}$ of the MIMO Gaussian BC with common, private and confidential messages under the covariance constraint \eqref{EQ:power_constraint} is
\begin{equation}
\mathcal{C}_\mathrm{K}=\bigcup_{\substack{\mathrm{0}\preceq\mathrm{K}_1,\mathrm{K}_2:\vspace{1mm}\\ \mathrm{K}_1+\mathrm{K}_2\preceq\mathrm{K}}}\mathcal{C}_\mathrm{K}(\mathrm{K}_1,\mathrm{K}_2)\label{EQ:region_common}.
\end{equation}
\end{theorem}

Theorem \ref{TM:capacity_common} is proven in Section \ref{SUBSEC:proof_common_capacity}.

\begin{remark}[Interpretation of Optimal Secrecy Rates]
Our interpretation of the structure of $\mathcal{C}_\mathrm{K}$ is reminiscent of Remark \ref{REM:interper_nocommon}. First, \eqref{EQ:region_common_simpleRB1} indicates that User 1 achieves rates up to the secrecy-capacity of the MIMO Gaussian WTC with input covariance $\mathrm{K}_2$. The 2nd user treats this signal as an additive Gaussian noise when decoding its private message $M_2$, which is transmitted using another (independent) Gaussian signal with covariance $\mathrm{K}_1$ (see \eqref{EQ:region_common_simpleRB2}). According to \eqref{EQ:region_common_simpleRB0}, the remaining portion of the total covariance matrix, that is, $\mathrm{K}-(\mathrm{K}_1+\mathrm{K}_2)$, is employed to encode the common message $M_0$, which is decoded by each receiver while treating all other signals as noise. As in the case without a common message, a layered coding scheme, when optimized over the choices of $\mathrm{K}_1$ and $\mathrm{K}_2$, exhausts the entire secrecy-capacity region.
\end{remark}


As before, Theorem \ref{TM:capacity_common} produces a characterization of the secrecy-capacity region under the average total power constraint.

\begin{corollary}[Average Total Power Constraint]
The secrecy-capacity region of the MIMO Gaussian BC with common, private and confidential messages under the average total power constraint \eqref{EQ:nocommon_power} is given by
\begin{equation}
\mathcal{C}_P=\bigcup_{\mathrm{0}\preceq\mathrm{K}:\ \Tr(\mathrm{K})\leq P}\mathcal{C}_\mathrm{K}\label{EQ:region_common_power}.
\end{equation}
\end{corollary}

\subsection{Numerical Example}\label{SUBSEC:numerical_example}


\begin{figure}[t!]
    \begin{center}
        \begin{psfrags}
            \psfragscanon
            \psfrag{A}[][][0.8]{$R_1$ [bits/use]}
            \psfrag{B}[][][0.8]{$R_2$ [bits/use]}
            \psfrag{W}[][][0.6]{\ \ \ \ \ \ \ \ \ \ \ \ \ \ \ \ \ \ $M_1$ Confidential}
            \psfrag{X}[][][0.6]{\ \ \ \ \ \ \ \ \ \ \ \ \ \ \ \ \ \ \ \ Both Confidential}
            \includegraphics[scale = 0.5]{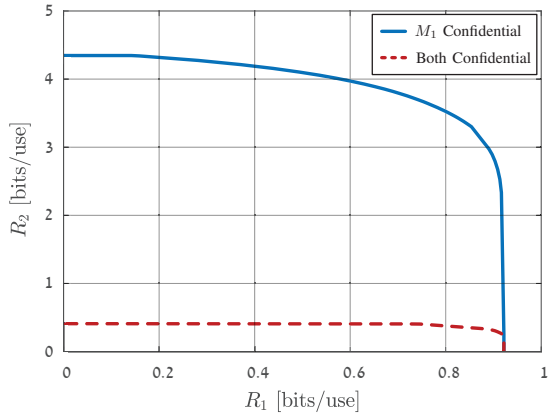}
            \caption{Secrecy-capacity region under an average total power constraint of the MIMO Gaussian BC without a common message, where: $M_1$ is confidential and $M_2$ is private (solid blue) vs. $M_1$ and $M_2$ are both confidential (dashed red).}\label{FIG:region}
            \psfragscanoff
        \end{psfrags}
     \end{center}
 \end{figure}


We illustrate the secrecy-capacity region $\hat{\mathcal{C}}_P$ of the MIMO Gaussian BC with private and confidential messages (without a common message) under an average total power constraint $P$ (Corollary \ref{COR:capacity_nocommon_total_power}). The region is described in \eqref{EQ:region_nocommon_power} as the union of all secrecy-capacity regions $\hat{\mathcal{C}}_\mathrm{K}$, each under a covariance constraint $\mathrm{K}\succeq 0$ with $\Tr(\mathrm{K})\leq P$. However, $\hat{\mathcal{C}}_\mathrm{K}$ itself is described as matrix optimization problems that is not convex in general, and is therefore, hard to compute.

We overcome the computational inefficiency of $\hat{\mathcal{C}}_\mathrm{K}$ by leveraging the decomposition proposed in \cite[Equation (10)]{Khina_WTC_Decomposition2015}: Every positive semi-definite matrix $\mathrm{K}^\star\in\mathbb{R}^{t\times t}$ with $\mathrm{K}^\star\preceq\mathrm{K}$ can be expressed as
\begin{equation}
\mathrm{K}^\star=\mathrm{K}^{\frac{1}{2}}\mathrm{V}\mathrm{D}\mathrm{V}^\top\mathrm{K}^{\frac{1}{2}^\top},\label{EQ:matrix_decomposition}
\end{equation}
where $\mathrm{V}\in\mathbb{R}^{t\times t}$ is a unitary matrix and $\mathrm{D}\in\mathbb{R}^{t\times t}$ is a diagonal matrix whose diagonal values are between 0 and 1. Since in the subsequent example the dimension is $t=2$, a unitary matrix $V$ is nothing but a rotation matrix, i.e., we set
\begin{equation}
\mathrm{V}=\begin{bmatrix} \cos(\theta) & -\sin(\theta) \\ \sin(\theta) & \cos(\theta)\end{bmatrix},\quad\theta\in[0,2\pi].\label{EQ:rotation_matrix}
\end{equation}
Running over all possible diagonal matrices $\mathrm{D}$ involves only two parameters, viz. the diagonal entries of $\mathrm{D}$. Finally, note that $\mathrm{K}^{\frac{1}{2}}$ is any matrix $\mathrm{B}$ satisfying $\mathrm{B}\mathrm{B}^\top= \mathrm{K}$. Obviously, there are many such matrices (in fact if $\mathrm{B}$ satisfies $\mathrm{B}\mathrm{B}^\top= \mathrm{K}$, then so does $\mathrm{BU}$, for any unitary $\mathrm{U}$). However, since the numerical calculation runs over all matrices $\mathrm{V}$ from \eqref{EQ:rotation_matrix} anyway, any choice of $\mathrm{B}$ would do. Our simulation uses the Cholesky decomposition of $\mathrm{K}$ to calculate~$\mathrm{B}$.

The region $\hat{\mathcal{C}}_P$ is computed according to \eqref{EQ:region_nocommon_power}, while noting that one may restrict the optimization domain to positive semi-definite matrices $\mathrm{K}$ with $\Tr(\mathrm{K})=P$. This observation follows because for every $\mathrm{K}'$ with $\Tr(\mathrm{K}')=\pi<P$, there is a $\mathrm{K}$ with $\Tr(\mathrm{K})=P$, such that
\begin{equation}
\hat{\mathcal{C}}_{\mathrm{K}'}\subseteq\hat{\mathcal{C}}_\mathrm{K}.\label{EQ:tr=P_optimal}
\end{equation}
The matrix $\mathrm{K}$ is constructed by increasing the $(1,1)$-th entry of $\mathrm{K}'$ by $P-\pi$, while all other entries of $\mathrm{K}'$ remain unchanged. The construction satisfies $\mathrm{K}'\preceq\mathrm{K}$ and the inclusion in \eqref{EQ:tr=P_optimal} follows because fixing $\mathrm{K}^\star\preceq\mathrm{K}'\preceq\mathrm{K}$ and replacing $\mathrm{K}'$ with $\mathrm{K}$ in \eqref{EQ:region_nocommon_simpleRB} does not alter \eqref{EQ:region_nocommon_simpleRB1} and strictly increases \eqref{EQ:region_nocommon_simpleRB2}.

In the numerical example we set
\begin{equation}
\mathrm{G}_1=\begin{bmatrix} 0.3 & 2.5 \\ 2.2 & 1.8\end{bmatrix}\ ,\ \ \mathrm{G}_2=\begin{bmatrix} 1.3 & 1.2 \\ 1.5 & 3.9\end{bmatrix}
\end{equation}
and $P=12$. The secrecy-capacity region $\hat{\mathcal{C}}_P$ is given by the solid blue curve in Fig. \ref{FIG:region}. For comparison, the secrecy-capacity region of the MIMO Gaussian BC with confidential messages \cite{Poor_Shamai_Gaussian_MIMO_BC_Secrecy2010} (i.e., when each user serves as the eavesdropped of the message to the other user) is depicted by the dashed red curve. As expected, Fig. \ref{FIG:region} shows that imposing a secrecy constraint on $M_2$ at the 1st receiver strictly shrinks the secrecy-capacity region. Although in \emph{both} regions the maximal value of $R_1$ is the secrecy-capacity of the corresponding MIMO Gaussian WTC (see \eqref{EQ:region_nocommon_simpleRB1} and \cite[Equation (4)]{Poor_Shamai_Gaussian_MIMO_BC_Secrecy2010}), the achievable values of $R_2$ drop if $M_2$ is also confidential.


\section{Optimality of Gaussian Inputs via Factorization of Concave Envelopes}\label{SEC:optimality_of_Gaussians}

This section provides the mathematical background for characterizing the secrecy-capacity regions of the considered MIMO Gaussian BC without and with a common message (Theorems \ref{TM:capacity_nocommon} and \ref{TM:capacity_common}). In the sequel we define some generic functions and show that they are maximized by Gaussian distributions. These functions are later used to describe the boundary points of certain outer bounds on the secrecy-capacity regions of interest. The properties established in this section are leveraged to show that optimality is achieved by Gaussian inputs, and that the resulting expressions are attainable by a corresponding inner bound.

Sections \ref{SUBSEC:optimality_of_Gaussians_example1} and \ref{SUBSEC:optimality_of_Gaussians_example2} focus on functions that are reminiscent of those studied in \cite[Sections II-B and II-C]{Chandra_Gauss_BC2014}. Therefore, to avoid verbatim repetition of arguments from \cite{Chandra_Gauss_BC2014}, we state some of the properties in Sections \ref{SUBSEC:optimality_of_Gaussians_example1} and \ref{SUBSEC:optimality_of_Gaussians_example2} without proofs. The focus of Section \ref{SUBSEC:optimality_of_Gaussians_example3} is on a new function that was not considered in \cite{Chandra_Gauss_BC2014}, the properties of which we prove in full detail. All the proofs for this section are relegated to Section \ref{SEC:envelope_properties}.

Establishing Gaussian inputs as maximizers relies on the notion of two-letter BCs \cite[Section I-A]{Chandra_Gauss_BC2014}, which is a special case of a product BC (PBC).\footnote{Henceforth, we omit the time index $i$.}

\begin{definition}[Product BC]\label{DEF:product_BC}
A PBC consists of a sender $(\mathbf{X}_1,\mathbf{X}_2)$ and two receivers $(\mathbf{Y}_{11},\mathbf{Y}_{12})$ and $(\mathbf{Y}_{21},\mathbf{Y}_{22})$, and is described by a conditional PDF of the form $q^{(1)}_{\mathbf{Y}_{11},\mathbf{Y}_{21}|\mathbf{X}_1}\times q^{(2)}_{\mathbf{Y}_{12},\mathbf{Y}_{22}|\mathbf{X}_2}$.
\end{definition}

A MIMO Gaussian PBC can be represented as
\begin{subequations}
\begin{align}
\begin{bmatrix}\mathbf{Y}_{11} \\ \mathbf{Y}_{12}\end{bmatrix}&=\begin{bmatrix}\mathrm{G}_{11} & 0 \\ 0 & \mathrm{G}_{12}\end{bmatrix}\begin{bmatrix}\mathbf{X}_1 \\ \mathbf{X}_2\end{bmatrix}+\begin{bmatrix}\mathbf{Z}_{11} \\ \mathbf{Z}_{12}\end{bmatrix}\\
\begin{bmatrix}\mathbf{Y}_{21} \\ \mathbf{Y}_{22}\end{bmatrix}&=\begin{bmatrix}\mathrm{G}_{21} & 0 \\ 0 & \mathrm{G}_{22}\end{bmatrix}\begin{bmatrix}\mathbf{X}_1 \\ \mathbf{X}_2\end{bmatrix}+\begin{bmatrix}\mathbf{Z}_{21} \\ \mathbf{Z}_{22}\end{bmatrix},
\end{align}
\end{subequations}
where $\mathbf{Z}_{11},\mathbf{Z}_{12},\mathbf{Z}_{21},\mathbf{Z}_{22}\sim\mathcal{N}(\mathbf{0},\mathrm{I})$ are i.i.d. and independent of $(\mathbf{X}_1,\mathbf{X}_2)$. A \emph{two-letter} version of a BC is a PBC in which the components are identical, i.e., $q^{(1)}_{\mathbf{Y}_{11},\mathbf{Y}_{21}|\mathbf{X}_1}=q^{(2)}_{\mathbf{Y}_{12},\mathbf{Y}_{22}|\mathbf{X}_2}$. In all subsequent definitions and results, the input covariance constraining matrix $\mathrm{K}\succeq 0 $ (see \eqref{EQ:power_constraint}) stays fixed.


\subsection{Difference of Mutual Information Terms}\label{SUBSEC:optimality_of_Gaussians_example1}

Consider a BC $q_{\mathbf{Y}_1,\mathbf{Y}_2|\mathbf{X}}$. For any $\eta>1$, let $\mathsf{s}_\eta^q$ be a functional of $\mathbf{X}\sim P_{\mathbf{X}}$ defined by
\begin{equation}
\mathsf{s}_\eta^q(\mathbf{X})\triangleq  I(\mathbf{X};\mathbf{Y}_2)-\eta I(\mathbf{X};\mathbf{Y}_1)\label{EQ:function1}.
\end{equation}

\begin{remark}\label{REM:adaptation}
The definition of $\mathsf{s}_\eta^q(\mathbf{X})$ in \eqref{EQ:function1} coincides with that of $\mathsf{s}_\lambda^q(\mathbf{X})\triangleq I(\mathbf{X};\mathbf{Y}_1)-\lambda I(\mathbf{X};\mathbf{Y}_2)$ from \cite[Section II-B]{Chandra_Gauss_BC2014} (only differing in the ordering of the mutual information terms and the labeling of the parameter). Accordingly, we restate and use some of the properties of $\mathsf{s}_\eta^q(\mathbf{X})$ established in \cite{Chandra_Gauss_BC2014} without providing proofs. Additional attributes of $\mathsf{s}_\eta^q(\mathbf{X})$ that were not proven in \cite{Chandra_Gauss_BC2014} are rigorously derived.
\end{remark}

For a pair of random variables $(V,\mathbf{X})$ such that $V-\mathbf{X}-(\mathbf{Y}_1,\mathbf{Y}_2)$ forms a Markov chain, set
\begin{equation}
\mathsf{s}_\eta^q(\mathbf{X}|V)\triangleq  I(\mathbf{X};\mathbf{Y}_2|V)-\eta I(\mathbf{X};\mathbf{Y}_1|V),
\end{equation}
and define the \emph{UCE} of $\mathsf{s}_\eta^q(\mathbf{X})$ as
\begin{equation}
\mathsf{S}_\eta^q(\mathbf{X})\triangleq  \Big(\mathfrak{C}\mathsf{s}_\eta^q\Big)(\mathbf{X})=\sup_{\substack{P_{V|\mathbf{X}}:\\V-\mathbf{X}-(\mathbf{Y}_1,\mathbf{Y}_2)}}\mathsf{s}_\eta^q(\mathbf{X}|V).\label{EQ:upper_concave_envelope1}
\end{equation}
The second equality in \eqref{EQ:upper_concave_envelope1} follows directly from Definition \ref{DEF:upper_concave_envelope}. For any discrete random variable $V$ we also set $\mathsf{S}_\eta^q(\mathbf{X}|V)\triangleq \sum_{v}P(v)\mathsf{S}_\eta^q(\mathbf{X}|V=v)$, and naturally extend this definition (merely an expectation) for an arbitrary~$V$.

\begin{proposition}[Concave Envelopes Properties]\label{PROP:Envelope_Properties}
The UCE $\mathsf{S}_\eta^q$ satisfies:
\begin{enumerate}
\item If $V-\mathbf{X}-(\mathbf{Y}_1,\mathbf{Y}_2)$ forms a Markov chain, then $\mathsf{S}_\eta^q(\mathbf{X}|V)\leq \mathsf{S}_\eta^q(\mathbf{X})$.
\item If $W-V-\mathbf{X}-(\mathbf{Y}_1,\mathbf{Y}_2)$ forms a Markov chain, then $\mathsf{S}_\eta^q(\mathbf{X}|V,W)= \mathsf{S}_\eta^q(\mathbf{X}|V)$.
\item $\mathsf{S}_\eta^q(\mathbf{X})$ is convex in $\eta$ inside $(0,2)$, for a fixed $P_\mathbf{X}$, and therefore it is continuous in $\eta$ at $\eta=1$.\footnote{The 3rd property can be established by considering any bounded, open interval containing 1, and not necessarily $(0,2)$.}
\end{enumerate}
\end{proposition}

The proof of Proposition \ref{PROP:Envelope_Properties} is given in Section \ref{SUBSEC:Envelope_Properties_proof}.

\begin{definition}[Maximized Concave Envelope]
For any MIMO Gaussian BC $q_{\mathbf{Y}_1,\mathbf{Y}_2|\mathbf{X}}$, define
\begin{equation}
V_\eta^q(\mathrm{K})\triangleq
\sup_{\mathbf{X}:\ \mathbb{E}[\mathbf{X}\mathbf{X}^\top]\preceq \mathrm{K}} \mathsf{S}_\eta^q(\mathbf{X})= \sup_{\substack{(V,\mathbf{X}):\ \mathbb{E}[\mathbf{X}\mathbf{X}^\top]\preceq \mathrm{K}, \\ V-\mathbf{X}-(\mathbf{Y}_1,\mathbf{Y}_2)}} \mathsf{s}_\eta^q(\mathbf{X}|V).\label{EQ:UCE1_optimum}
\end{equation}
\end{definition}

We subsequently show that \eqref{EQ:UCE1_optimum} is achieved by a Gaussian input distribution. Central to the proof is a certain factorization property of $\mathsf{S}_\eta^q$. To formulate this property, we first extend $\mathsf{S}_\eta^q$ to PBCs. For a PBC $q^{(1)}_{\mathbf{Y}_{11},\mathbf{Y}_{21}|\mathbf{X}_1}\times q^{(2)}_{\mathbf{Y}_{12},\mathbf{Y}_{22}|\mathbf{X}_2}$ we set,
\begin{align*}
\mathsf{s}_\eta&^{q_1\times q_2}(\mathbf{X}_1,\mathbf{X}_2)\\&\triangleq  I(\mathbf{X}_1,\mathbf{X}_2;\mathbf{Y}_{21},\mathbf{Y}_{22})-\eta I(\mathbf{X}_1,\mathbf{X}_2;\mathbf{Y}_{11},\mathbf{Y}_{12}),\numberthis
\end{align*}
and define the quantities $\mathsf{s}_\eta^{q_1\times q_2}(\mathbf{X}_1,\mathbf{X}_2|V)$, $\mathsf{S}_\eta^{q_1\times q_2}(\mathbf{X}_1,\mathbf{X}_2)$ and $\mathsf{S}_\eta^{q_1\times q_2}(\mathbf{X}_1,\mathbf{X}_2|V)$ analogously to the definitions of $\mathsf{s}_\eta^q(\mathbf{X}|V)$, $\mathsf{S}_\eta^q(\mathbf{X})$ and $\mathsf{S}_\eta^q(\mathbf{X}|V)$ from the above, respectively.

\begin{proposition}[Factorization Property]\label{PROP:factorization_property1}
For any PBC $q^{(1)}_{\mathbf{Y}_{11},\mathbf{Y}_{21}|\mathbf{X}_1}\times q^{(2)}_{\mathbf{Y}_{12},\mathbf{Y}_{22}|\mathbf{X}_2}$, the following chain of inequalities holds
\begin{align*}
\mathsf{S}_\eta^{q_1\times q_2}(\mathbf{X}_1,\mathbf{X}_2)&\leq \mathsf{S}_\eta^{q_1}(\mathbf{X}_1|\mathbf{Y}_{22})+\mathsf{S}_\eta^{q_2}(\mathbf{X}_2|\mathbf{Y}_{11})\\&\leq \mathsf{S}_\eta^{q_1}(\mathbf{X}_1)+\mathsf{S}_\eta^{q_2}(\mathbf{X}_2).\numberthis
\end{align*}
\end{proposition}
The proof of Proposition \ref{PROP:factorization_property1} follows by repeating the steps in the proof of \cite[Proposition 6]{Chandra_Gauss_BC2014}, while switching the roles of $\mathbf{Y}_1$ and $\mathbf{Y}_2$.

\begin{theorem}[Gaussian Maximizer]\label{TM:Gaussian_maximizer_Envelope1}
Let $\mathbf{X}\sim \mathcal{N}(\mathbf{0},\mathrm{K})$. There exists a decomposition $\mathbf{X}=\mathbf{X}^\star+\mathbf{X}'$, such that $\mathbf{X}^\star$ and $\mathbf{X}'$ are independent, $\mathbf{X}^\star\sim\mathcal{N}(\mathbf{0},\mathrm{K}^\star)$, $\mathbf{X}'\sim\mathcal{N}(\mathbf{0},\mathrm{K}-\mathrm{K}^\star)$, where $\mathrm{K}^\star\preceq \mathrm{K}$, and $\mathsf{S}_\eta^q(\mathbf{X})=\mathsf{s}_\eta^q(\mathbf{X}^\star)=V_\eta^q(\mathrm{K})$. Furthermore, this decomposition (i.e., the covariance matrix $\mathrm{K}^\star$) is unique.
\end{theorem}

The proof of Theorem \ref{TM:Gaussian_maximizer_Envelope1} is also omitted as it mimics the proofs of Theorem 1 and Corollary 1 in \cite{Chandra_Gauss_BC2014}.


\subsection{Nested Upper Concave Envelopes}\label{SUBSEC:optimality_of_Gaussians_example2}

The function considered in this subsection is used to derive the secrecy-capacity region of the considered MIMO Gaussian BC without a common message (see Section \ref{SUBSEC:proof_nocommon_capacity}).

For a BC $q_{\mathbf{Y}_1,\mathbf{Y}_2|\mathbf{X}}$, $\eta>1$, $\bm{\lambda}=(\lambda_1,\lambda_2)$, where $\lambda_j>0$, $j=1,2$, and any $\mathbf{X}\sim P_\mathbf{X}$ define
\begin{equation}
\mathsf{t}_{\bm{\lambda},\eta}^q(\mathbf{X})\triangleq  \lambda_1I(\mathbf{X};\mathbf{Y}_1)-(\lambda_1+\lambda_2)I(\mathbf{X};\mathbf{Y}_2)+\lambda_1\mathsf{S}_\eta^q(\mathbf{X}),\label{EQ:function2}
\end{equation}
where $\mathsf{S}_\eta^q(\mathbf{X})$ is given by \eqref{EQ:upper_concave_envelope1}. As before, for a pair of random variables $(V,\mathbf{X})$ for which $V-\mathbf{X}-(\mathbf{Y}_1,\mathbf{Y}_2)$ forms a Markov chain, let
\begin{align*}
\mathsf{t}&_{\bm{\lambda},\eta}^q(\mathbf{X}|V)\\&\triangleq \lambda_1I(\mathbf{X};\mathbf{Y}_1|V)-(\lambda_1+\lambda_2)I(\mathbf{X};\mathbf{Y}_2|V)+\lambda_1\mathsf{S}_\eta^q(\mathbf{X}|V),\numberthis\label{EQ:function2_conditioned}
\end{align*}
and set
\begin{equation}
\mathsf{T}_{\bm{\lambda},\eta}^q(\mathbf{X})\triangleq \mathfrak{C}\big(\mathsf{t}_{\bm{\lambda},\eta}^q(\mathbf{X})\big)=\sup_{\substack{P_{V|\mathbf{X}}:\\V-\mathbf{X}-(\mathbf{Y}_1,\mathbf{Y}_2)}}\mathsf{t}_{\bm{\lambda},\eta}^q(\mathbf{X}|V).\label{EQ:upper_concave_envelope2}
\end{equation}
Define $\mathsf{T}_{\bm{\lambda},\eta}^q(\mathbf{X}|V)\triangleq \sum_{v}P(v)\mathsf{T}_{\bm{\lambda},\eta}^q(\mathbf{X}|V=v)$, for a $V$ with a countable alphabet and consider its natural extension when $V$ is an arbitrary random variable.

\begin{remark}[Nested Concave Envelopes Properties]\label{REM:Nesterd_Envelope_Properties}
Similarly to the properties of $\mathsf{S}_\eta^q$ stated in Proposition \ref{PROP:Envelope_Properties}, since $\mathsf{T}_{\bm{\lambda},\eta}^q$ is concave in $P_\mathbf{X}$, Jensen's inequality implies that $\mathsf{T}_{\bm{\lambda},\eta}^q(\mathbf{X}|V)\leq \mathsf{T}_{\bm{\lambda},\eta}^q(\mathbf{X})$, for any $(V,\mathbf{X})$ satisfying $V-\mathbf{X}-(\mathbf{Y}_1,\mathbf{Y}_2)$. Moreover, if $W-V-\mathbf{X}$ forms a Markov chain, then $\mathsf{T}_{\bm{\lambda},\eta}^q(\mathbf{X}|W,V)=\mathsf{T}_{\bm{\lambda},\eta}^q(\mathbf{X}|V)$, because $P_{\mathbf{X}|W,V}=P_{\mathbf{X}|V}$. Finally, $\mathsf{T}_{\bm{\lambda},\eta}^q(\mathbf{X})$ is convex in $\eta$ inside $(0,2)$, for a fixed $P_\mathbf{X}$, and therefore it is continuous as a function of $\eta$ at $\eta=1$.
\end{remark}

\begin{definition}[Maximized Nested Concave Envelope]
For any MIMO Gaussian BC $q_{\mathbf{Y}_1,\mathbf{Y}_2|\mathbf{X}}$, define
\begin{equation}
\hat{V}_{\bm{\lambda},\eta}^q(\mathrm{K})\triangleq  \mspace{-10mu}\sup_{\mathbf{X}:\ \mathbb{E}[\mathbf{X}\mathbf{X}^\top]\preceq \mathrm{K}}\mspace{-10mu} \mathsf{T}_{\bm{\lambda},\eta}^q(\mathbf{X})= \mspace{-10mu}\sup_{\substack{(V,\mathbf{X}):\ \mathbb{E}[\mathbf{X}\mathbf{X}^\top]\preceq \mathrm{K}, \\ V-\mathbf{X}-(\mathbf{Y}_1,\mathbf{Y}_2)}} \mspace{-10mu}\mathsf{t}_{\bm{\lambda},\eta}^q(\mathbf{X}|V).\label{EQ:UCE2_optimum}
\end{equation}
\end{definition}

\begin{proposition}[Continuity of Maximal Value]\label{PROP:Maximum_Countinuos}
For any $\bm{\lambda}$ as before, $\hat{V}_{\bm{\lambda},\eta}(\mathrm{K})$ is continuous in $\eta$ at $\eta=1$.
\end{proposition}
The proof of Proposition \ref{PROP:Maximum_Countinuos} follows by arguments similar to those in the proof of Property 3 of Proposition \ref{PROP:Envelope_Properties}. Namely, the continuity of $\hat{V}_{\bm{\lambda},\eta}(\mathrm{K})$ at $\eta=1$ follows by verifying that $\hat{V}_{\bm{\lambda},\eta}(\mathrm{K})$ is convex in $\eta$ inside $(0,2)$ and using Proposition 17 from \cite[Chapter 5]{Royden_Real_Analysis1988}.

As before, to state the factorization property for nested UCEs, we extend some of the preceding definitions to PBCs. For a PBC $q^{(1)}_{\mathbf{Y}_{11},\mathbf{Y}_{21}|\mathbf{X}_1}\times q^{(2)}_{\mathbf{Y}_{12},\mathbf{Y}_{22}|\mathbf{X}_2}$, we set
\begin{align*}
&\mathsf{t}_{\bm{\lambda},\eta}^{q_1\times q_2}(\mathbf{X}_1,\mathbf{X}_2)\\
&\triangleq\lambda_1I(\mathbf{X}_1,\mathbf{X}_2;\mathbf{Y}_{11},\mathbf{Y}_{12})\mspace{-3mu}-\mspace{-3mu}(\lambda_1\mspace{-3mu}+\mspace{-3mu}\lambda_2)I(\mathbf{X}_1,\mathbf{X}_2;\mathbf{Y}_{21},\mathbf{Y}_{22})\\
&\mspace{230mu}+\lambda_1\mathsf{S}_\eta^{q_1\times q_2}(\mathbf{X}_1,\mathbf{X}_2).\numberthis
\end{align*}
Furthermore, define $\mathsf{t}_{\bm{\lambda},\eta}^{q_1\times q_2}(\mathbf{X}_1,\mathbf{X}_2|V)$, $\mathsf{T}_{\bm{\lambda},\eta}^{q_1\times q_2}(\mathbf{X}_1,\mathbf{X}_2)$ and $\mathsf{T}_{\bm{\lambda},\eta}^{q_1\times q_2}(\mathbf{X}_1,\mathbf{X}_2|V)$ in a similar manner to $\mathsf{t}_{\bm{\lambda},\eta}^q(\mathbf{X}|V)$, $\mathsf{T}_{\bm{\lambda},\eta}^q(\mathbf{X})$ and $\mathsf{T}_{\bm{\lambda},\eta}^q(\mathbf{X}|V)$, respectively. The following proposition states the $\mathsf{T}_{\bm{\lambda},\eta}^{q_1\times q_2}(\mathbf{X}_1,\mathbf{X}_2|V)$ factorization property of interest, which is used to prove the existence of a Gaussian maximizer for $\hat{V}_{\bm{\lambda},\eta}^q(\mathrm{K})$ from \eqref{EQ:UCE2_optimum}.

\begin{proposition}[Factorization Property]\label{PROP:factorization_property2}
For any PBC $q^{(1)}_{\mathbf{Y}_{11},\mathbf{Y}_{21}|\mathbf{X}_1}\times q^{(2)}_{\mathbf{Y}_{12},\mathbf{Y}_{22}|\mathbf{X}_2}$, the following chain of inequalities holds
\begin{align*}
\mathsf{T}_
{\bm{\lambda},\eta}^{q_1\times q_2}(\mathbf{X}_1,\mathbf{X}_2)&\leq \mathsf{T}_{\bm{\lambda},\eta}^{q_1}(\mathbf{X}_1|\mathbf{Y}_{22})+\mathsf{T}_{\bm{\lambda},\eta}^{q_2}(\mathbf{X}_2|\mathbf{Y}_{11})\\
&\leq \mathsf{T}_{\bm{\lambda},\eta}^{q_1}(\mathbf{X}_1)+\mathsf{T}_{\bm{\lambda},\eta}^{q_2}(\mathbf{X}_2)\numberthis
\end{align*}
Furthermore, if the PBC is Gaussian and a triple $(V^\star,\mathbf{X}_1^\star,\mathbf{X}_2^\star)$ satisfies
\begin{equation}
\mathsf{t}_{\bm{\lambda},\eta}^{q_1\mspace{-3mu}\times \mspace{-1mu} q_2}\mspace{-1.5mu}(\mathbf{X}_1^\star,\mathbf{X}_2^\star|V^\star)\mspace{-3mu}=\mspace{-3mu} \mathsf{T}_{\bm{\lambda},\eta}^{q_1\mspace{-3mu}\times \mspace{-1mu} q_2}\mspace{-1.5mu}(\mathbf{X}_1^\star,\mathbf{X}_2^\star)\mspace{-3mu}=\mspace{-3mu} \mathsf{T}_{\bm{\lambda},\eta}^{q_1}(\mathbf{X}_1^\star)+\mathsf{T}_{\bm{\lambda},\eta}^{q_2}(\mathbf{X}_2^\star)
\end{equation}
then $\mathbf{X}_1^\star-V^\star-\mathbf{X}_2^\star$ and $\mathsf{t}_{\bm{\lambda},\eta}^{q_j}(\mathbf{X}_j^\star|V^\star)=\mathsf{T}_{\bm{\lambda},\eta}^{q_j}(\mathbf{X}_j^\star)$, for $j=1,2$.
\end{proposition}

See Section \ref{SUBSEC:factorization_property2_proof} for the proof of Proposition \ref{PROP:factorization_property2}. The existence of a Gaussian maximizer for $\hat{V}_{\bm{\lambda},\eta}^q(\mathrm{K})$ follows by repeating the proofs of Theorem 2 and Corollary 2 in \cite{Chandra_Gauss_BC2014} with respect to our definition of $\mathsf{T}_{\bm{\lambda},\eta}^q$. The existence is stated in the following Theorem, which we give without proof.

\begin{theorem}[Gaussian Maximizer]\label{TM:Gaussian_maximizer_Envelope2}
Let $\mathbf{X}\sim \mathcal{N}(\mathbf{0},\mathrm{K})$. There exists a unique decomposition $\mathbf{X}=\mathbf{X}_1^\star+\mathbf{X}_2^\star+\mathbf{X}'$ into independent random variables $(\mathbf{X}_1^\star,\mathbf{X}_2^\star,\mathbf{X}')$, where $\mathbf{X}_j^\star\sim\mathcal{N}(\mathbf{0},\mathrm{K}_j)$, $j=1,2$, and $\mathbf{X}'\sim\mathcal{N}(\mathbf{0},\mathrm{K}-(\mathrm{K}_1+\mathrm{K}_2)\big)$, $\mathrm{K}_1+\mathrm{K}_2\preceq \mathrm{K}$, such that
\begin{subequations}
\begin{align}
&\mathsf{T}_{\bm{\lambda},\eta}^q(\mathbf{X})=\mathsf{t}_{\bm{\lambda},\eta}^q(\mathbf{X}_1^\star+\mathbf{X}_2^\star)=\hat{V}_{\bm{\lambda},\eta}^q(\mathrm{K})\\
&\mathsf{S}_\eta^q(\mathbf{X}_1^\star+\mathbf{X}_2^\star)=\mathsf{s}_\eta^q(\mathbf{X}_1^\star)=V_\eta^q(\mathrm{K}_1+\mathrm{K}_2).
\end{align}
\end{subequations}
\end{theorem}


\subsection{Double-Nested Upper Concave Envelopes}\label{SUBSEC:optimality_of_Gaussians_example3}

The definitions and properties in this section are used to derive the secrecy-capacity region of the cooperative BC with common, private and confidential messages (as defined in Section \ref{SUBSEC:Problem_Definition}). Let $q_{\mathbf{Y}_1,\mathbf{Y}_2|\mathbf{X}}$, $\eta>1$ be a BC, $\bm{\lambda_0}=(\lambda_0,\lambda_1,\lambda_2)$, where $\lambda_j>0$ for $j=0,1,2$ and $\lambda_0>\lambda_2$, $\alpha\in[0,1]$ and $\bar{\alpha}=1-\alpha$. For any $\mathbf{X}\sim P_\mathbf{X}$ define
\begin{equation}
\mathsf{f}_{\bm{\lambda_0},\alpha,\eta}^q(\mathbf{X})\triangleq (\lambda_2-\bar{\alpha}\lambda_0)I(\mathbf{X};\mathbf{Y}_2)-\alpha\lambda_0I(\mathbf{X};\mathbf{Y}_1)+\mathsf{T}_{\bm{\lambda},\eta}^q(\mathbf{X}),\label{EQ:function3}
\end{equation}
where $\mathsf{T}_{\bm{\lambda},\eta}^q(\mathbf{X})$ is given by \eqref{EQ:upper_concave_envelope2} and $\bm{\lambda}=(\lambda_1,\lambda_2)$.

For $(V,\mathbf{X})$ that satisfy the Markov chain $V-\mathbf{X}-(\mathbf{Y}_1,\mathbf{Y}_2)$, we set $\mathsf{f}_{\bm{\lambda_0},\alpha,\eta}^q(\mathbf{X}|V)$ in an analogous manner to \eqref{EQ:function2_conditioned}, while $\mathsf{F}_{\bm{\lambda_0},\alpha,\eta}^q\triangleq \mathfrak{C}\mathsf{f}_{\bm{\lambda_0},\alpha,\eta}^q$ denotes the UCE of $\mathsf{f}_{\bm{\lambda_0},\alpha,\eta}^q$. We also set $\mathsf{F}_{\bm{\lambda_0},\alpha,\eta}^q(\mathbf{X}|V)\triangleq \sum_{v}P(v)\mathsf{F}_{\bm{\lambda_0},\alpha,\eta}^q(\mathbf{X}|V=v)$ for a discrete $V$ and, as before, consider its natural extension in the case where $V$ is arbitrary.

\begin{remark}[Double-Nested Concave Envelopes Properties]\label{REM:Double_Nesterd_Envelope_Properties}
The concavity of $\mathsf{F}_{\bm{\lambda_0},\alpha,\eta}^q$ in $P_\mathbf{X}$ and Jensen's inequality imply that for any $(V,\mathbf{X})$ with $V-\mathbf{X}-(\mathbf{Y}_1,\mathbf{Y}_2)$, we have $\mathsf{F}_{\bm{\lambda_0},\alpha,\eta}^q(\mathbf{X}|V)\leq \mathsf{F}_{\bm{\lambda_0},\alpha,\eta}^q(\mathbf{X})$. If the chain $W-V-\mathbf{X}$ is Markov, then $\mathsf{F}_{\bm{\lambda_0},\alpha,\eta}^q(\mathbf{X}|W,V)=\mathsf{F}_{\bm{\lambda_0},\alpha,\eta}^q(\mathbf{X}|V)$, because $P_{\mathbf{X}|W,V}=P_{\mathbf{X}|V}$. As a function of $\eta$, $\mathsf{F}_{\bm{\lambda_0},\alpha,\eta}^q(\mathbf{X})$ is convex inside $(0,2)$, for any fixed $\mathbf{X}$, and is thus continuous at~$\eta=1$.
\end{remark}

\begin{definition}[Maximized Double-Nested Concave Envelope]
For any MIMO Gaussian BC $q_{\mathbf{Y}_1,\mathbf{Y}_2|\mathbf{X}}$, define
\begin{align*}
\tilde{V}_{\bm{\lambda_0},\alpha,\eta}^q(\mathrm{K})&\triangleq  \sup_{\mathbf{X}:\ \mathbb{E}[\mathbf{X}\mathbf{X}^\top]\preceq \mathrm{K}} \mathsf{F}_{\bm{\lambda_0},\alpha,\eta}^q(\mathbf{X})\\&= \sup_{\substack{(V,\mathbf{X}):\ \mathbb{E}[\mathbf{X}\mathbf{X}^\top]\preceq \mathrm{K}, \\ V-\mathbf{X}-(\mathbf{Y}_1,\mathbf{Y}_2)}} \mathsf{f}_{\bm{\lambda_0},\alpha,\eta}^q(\mathbf{X}|V).\numberthis
\end{align*}
\end{definition}

\begin{remark}[Continuity of Maximal Value]\label{PROP:Maximum_Countinuos_double}
As before, one readily verifies that as a function of $\eta$, $\tilde{V}_{\bm{\lambda_0},\alpha,\eta}(\mathrm{K})$ is convex inside $(0,2)$, and deduce its continuity at $\eta=1$.
\end{remark}

The above notions are once again extended to PBCs. Namely, for any PBC $q^{(1)}_{\mathbf{Y}_{11},\mathbf{Y}_{21}|\mathbf{X}_1}\times q^{(2)}_{\mathbf{Y}_{12},\mathbf{Y}_{22}|\mathbf{X}_2}$, we set
\begin{align*}
\mathsf{f}_{\bm{\lambda_0},\alpha,\eta}&^{q_1\times q_2}(\mathbf{X}_1,\mathbf{X}_2)=(\lambda_2-\bar{\alpha}\lambda_0)I(\mathbf{X}_1,\mathbf{X}_2;\mathbf{Y}_{21},\mathbf{Y}_{22})\\
&\mspace{-12mu}-\alpha\lambda_0I(\mathbf{X}_1,\mathbf{X}_2;\mathbf{Y}_{11},\mathbf{Y}_{12})+\mathsf{T}_{\bm{\lambda},\eta}^{q_1\times q_2}(\mathbf{X}_1,\mathbf{X}_2),\numberthis
\end{align*}
and define $\mathsf{f}_{\bm{\lambda_0},\alpha,\eta}^{q_1\times q_2}(\mathbf{X}_1,\mathbf{X}_2|V)$, $\mathsf{F}_{\bm{\lambda_0},\alpha,\eta}^{q_1\times q_2}(\mathbf{X}_1,\mathbf{X}_2)$ and $\mathsf{F}_{\bm{\lambda_0},\alpha,\eta}^{q_1\times q_2}(\mathbf{X}_1,\mathbf{X}_2|V)$ as the natural extensions to the PBC scenario of $\mathsf{f}_{\bm{\lambda_0},\alpha,\eta}^q(\mathbf{X}|V)$, $\mathsf{F}_{\bm{\lambda_0},\alpha,\eta}^q(\mathbf{X})$ and $\mathsf{F}_{\bm{\lambda_0},\alpha,\eta}^q(\mathbf{X}|V)$ given above, respectively.


Moving forward, the factorization property of $\mathsf{F}_{\bm{\lambda_0},\alpha,\eta}^{q_1\times q_2}$ is stated in Proposition \ref{PROP:factorization_property3}, while Proposition \ref{PROP:existence_of_maximizer} establishes the existence of its maximizer.

\begin{proposition}[Factorization Property]\label{PROP:factorization_property3}
For any PBC $q^{(1)}_{\mathbf{Y}_{11},\mathbf{Y}_{21}|\mathbf{X}_1}\times q^{(2)}_{\mathbf{Y}_{12},\mathbf{Y}_{22}|\mathbf{X}_2}$, the following chain of inequalities holds
\begin{align*}
\mathsf{F}_{\bm{\lambda_0},\alpha,\eta}^{q_1\times q_2}(\mathbf{X}_1,\mathbf{X}_2)&\leq \mathsf{F}_{\bm{\lambda_0},\alpha,\eta}^{q_1}(\mathbf{X}_1|\mathbf{Y}_{22})+\mathsf{F}_{\bm{\lambda_0},\alpha,\eta}^{q_2}(\mathbf{X}_2|\mathbf{Y}_{11})\\
&\leq \mathsf{F}_{\bm{\lambda_0},\alpha,\eta}^{q_1}(\mathbf{X}_1)+\mathsf{F}_{\bm{\lambda_0},\alpha,\eta}^{q_2}(\mathbf{X}_2)\numberthis
\end{align*}
Furthermore, if the PBC is Gaussian and a triple $(V^\star,\mathbf{X}_1^\star,\mathbf{X}_2^\star)$ satisfies
\begin{align*}
\mathsf{f}_{\bm{\lambda_0},\alpha,\eta}^{q_1\times q_2}(\mathbf{X}_1^\star,\mathbf{X}_2^\star|V^\star)&= \mathsf{F}_{\bm{\lambda_0},\alpha,\eta}^{q_1\times q_2}(\mathbf{X}_1^\star,\mathbf{X}_2^\star)\\
&=\mathsf{F}_{\bm{\lambda_0},\alpha,\eta}^{q_1}(\mathbf{X}_1^\star)+\mathsf{F}_{\bm{\lambda_0},\alpha,\eta}^{q_2}(\mathbf{X}_2^\star),\numberthis\label{EQ:factorization_property3_equality}
\end{align*}
then $\mathbf{X}_1^\star-V^\star-\mathbf{X}_2^\star$ and $\mathsf{f}_{\bm{\lambda_0},\alpha,\eta}^{q_j}(\mathbf{X}_j^\star|V^\star)=\mathsf{F}_{\bm{\lambda_0},\alpha,\eta}^{q_j}(\mathbf{X}_j^\star)$, for $j=1,2$.
\end{proposition}

See Section \ref{SUBSEC:factorization_property3_proof} for the proof of Proposition \ref{PROP:factorization_property3}.

\begin{proposition}[Existence of a Maximizer]\label{PROP:existence_of_maximizer}
There exists a pair $(V^\star,\mathbf{X}^\star)$ with $|\mathcal{V}^\star|\leq\frac{t(t+1)}{2}+1$ and $\mathbb{E}\big[\mathbf{X}\mathbf{X}^\top\big]\preceq\mathrm{K}$, such that
\begin{equation}
\tilde{V}_{\bm{\lambda_0},\alpha,\eta}^q(\mathrm{K})=\mathsf{f}_{\bm{\lambda_0},\alpha,\eta}^q(\mathbf{X}^\star|V^\star)\label{EQ:existence_maximizer}.
\end{equation}
Furthermore, one may assume that $\mathbb{E}[\mathbf{X}^\star|V^\star=v^\star]=0$, for every $v^\star\in\mathcal{V}^\star$.
\end{proposition}

The existence of a maximizer and the cardinality bound on $\mathcal{V}^\star$ are proven in Section \ref{SUBSEC:maximizer_exists}. A zero conditional expectation can be assumed because centering conditioned on each $V^\star = v^\star$ does not change the mutual information terms and hence $\mathsf{f}_{\bm{\lambda_0},\alpha,\eta}^q(\mathbf{X}^\star_2|V^\star)$ remains unchanged as well. In addition, the centered versions of the input continues to satisfy the covariance constraint.

To show that the distribution that achieves $\tilde{V}_{\bm{\lambda_0},\alpha,\eta}^q(\mathrm{K})$ is Gaussian, we use the invariance of $\mathsf{f}_{\bm{\lambda_0},\alpha,\eta}^q$ to rotation. This invariance property is stated in the context of the next Proposition, which is proven in Section \ref{SUBSEC:rotation_invariant_proof}.

\begin{proposition}[Invariance to Rotation]\label{PROP:rotation_invariant}
Let $(V,\mathbf{X})\sim P_{V,\mathbf{X}}^\star$ attain $\tilde{V}_{\bm{\lambda_0},\alpha,\eta}^q(\mathrm{K})$, with $\left|\mathcal{V}\right|=m\leq\frac{t(t+1)}{2}+1$, and let $\mathbf{X}_v$ be a centered
random variable (zero mean) distributed according to the conditional PMF $P^\star_{\mathbf{X}|V=v}$. Let $(V_1,\mathbf{X}_1,V_2,\mathbf{X}_2)\sim P_{V,\mathbf{X}}^\star\times P_{V,\mathbf{X}}^\star$ be two i.i.d. copies of $(V^\star,\mathbf{X}^\star)$. Define
\begin{align*}
\tilde{V}&=(V_1,V_2)\\
\mathbf{X}_{\theta_1}\big|\big\{\tilde{V}=(v_1,v_2)\big\}&\sim\frac{1}{\sqrt{2}}(\mathbf{X}_{v_1}+\mathbf{X}_{v_2})\\
\mathbf{X}_{\theta_2}\big|\big\{\tilde{V}=(v_1,v_2)\big\}&\sim\frac{1}{\sqrt{2}}(\mathbf{X}_{v_1}-\mathbf{X}_{v_2}),
\end{align*}
where $\mathbf{X}_{v_1}$ and $\mathbf{X}_{v_2}$ are taken to be independent random variables, i.e., $(\mathbf{X}_{v_1},\mathbf{X}_{v_2})\sim P^\star_{\mathbf{X}|V=v_1}\times P^\star_{\mathbf{X}|V=v_2}$. Then $\mathbf{X}_{\theta_1}-\tilde{V}-\mathbf{X}_{\theta_2}$ and $\tilde{V}_{\bm{\lambda_0},\alpha,\eta}^q(\mathrm{K})=\mathsf{f}_{\bm{\lambda_0},\alpha,\eta}^q(\mathbf{X}_{\theta_j}|\tilde{V})$, for $j=1,2$.
\end{proposition}

The existence of a Gaussian Maximizer for $\tilde{V}_{\bm{\lambda_0},\alpha,\eta}^q(\mathrm{K})$ is stated next.

\begin{theorem}[Existence of Gaussian Maximizer]\label{TM:Gaussian_maximizer_Envelope_pre}
There exists an $\mathbf{X}^\star\sim\mathcal{N}(\mathbf{0},\mathrm{K}^\star)$, where $\mathrm{K}^\star\preceq\mathrm{K}$, such that $\tilde{V}_{\bm{\lambda_0},\alpha,\eta}^q(\mathrm{K})=\mathsf{f}_{\bm{\lambda_0},\alpha,\eta}^q(\mathbf{X}^\star)$. Furthermore, the zero mean maximizer is unique.
\end{theorem}

See Section \ref{SUBSEC:Gaussian_maximizer_Envelope_pre_proof} for the proof.

\begin{corollary}[Gaussian Maximizer Properties]\label{COR:Gaussian_maximizer_Envelope3}
Let $\mathbf{X}\sim \mathcal{N}(\mathbf{0},\mathrm{K})$. There is a unique decomposition $\mathbf{X}=\mathbf{X}_1^\star+\mathbf{X}_2^\star+\mathbf{X}_3^\star+\mathbf{X}'$ into independent random variables $(\mathbf{X}_1^\star,\mathbf{X}_2^\star,\mathbf{X}_3^\star,\mathbf{X}')$, where $\mathbf{X}_j^\star\sim\mathcal{N}(\mathbf{0},\mathrm{K}_j)$, for $j=1,2,3$, and $\mathbf{X}'\sim\mathcal{N}\big(\mathbf{0},\mathrm{K}-(\mathrm{K}_1+\mathrm{K}_2+\mathrm{K}_3)\big)$, with $\mathrm{K}_1+\mathrm{K}_2+\mathrm{K}_3\preceq \mathrm{K}$, such that
\begin{subequations}
\begin{align}
&\mathsf{F}_{\bm{\lambda_0},\alpha,\eta}^q(\mathbf{X})=\mathsf{f}_{\bm{\lambda_0},\alpha,\eta}^q(\mathbf{X}_1^\star+\mathbf{X}_2^\star+\mathbf{X}_3^\star)=\tilde{V}_{\bm{\lambda_0},\alpha,\eta}^q(\mathrm{K})\\
&\mathsf{T}_{\bm{\lambda},\eta}^q(\mathbf{X}_1^\star\mspace{-4mu}+\mspace{-4mu}\mathbf{X}_2^\star\mspace{-4mu}+\mspace{-4mu}\mathbf{X}_3^\star)=\mathsf{t}_{\bm{\lambda},\eta}^q(\mathbf{X}_1^\star\mspace{-3mu}+\mspace{-3mu}\mathbf{X}_2^\star)=\hat{V}_{\bm{\lambda},\eta}^q(\mathrm{K}_1\mspace{-4mu}+\mspace{-4mu}\mathrm{K}_2\mspace{-4mu}+\mspace{-4mu}\mathrm{K}_3)\\
&\mathsf{S}_\eta^q(\mathbf{X}_1^\star+\mathbf{X}_2^\star)=\mathsf{s}_\eta^q(\mathbf{X}_1^\star)=V_\eta^q(\mathrm{K}_1+\mathrm{K}_2).
\end{align}\label{EQ:Gaussian_maximizer_Envelope3}
\end{subequations}
\end{corollary}

Corollary \ref{COR:Gaussian_maximizer_Envelope3}, which is a consequence of Theorem \ref{TM:Gaussian_maximizer_Envelope_pre}, is our main tool for characterizing the secrecy-capacity region of the MIMO Gaussian BC with common, private and confidential messages. The proof of the Corollary is provided in Section~\ref{SUBSEC:Gaussian_maximizer_Envelope3_proof}.


\section{Proofs of Secrecy-Capacity Results}\label{SEC:capacity_results_proofs}


\subsection{Proof of Theorem \ref{TM:capacity_nocommon}}\label{SUBSEC:proof_nocommon_capacity}

We establish the secrecy-capacity region of the MIMO Gaussian BC with private and confidential messages by showing that certain outer bound and inner bounds match. In particular, we consider special cases of the inner and outer bounds from Theorems 1 and 3 of \cite{goldfeld2017broadcast}, respectively. To state the bounds, let $\hat{\mathcal{C}}$ denote the secrecy-capacity region of the corresponding discrete-memoryless (DM) BC.

\begin{bound}[Outer Bound]\label{BND:outer_bound_nocommon}
Let $\hat{\mathcal{O}}$ be the closure of the union of rate pairs $(R_1,R_2)\in\mathbb{R}^2_+$ satisfying:
\begin{subequations}
\begin{align}
R_1&\leq I(U;Y_1|V)-I(U;Y_2|V)\label{EQ:region_outer_nocommon1}\\
R_2&\leq I(V;Y_2)\label{EQ:region_outer_nocommon2}
\end{align}\label{EQ:region_outer_nocommon}
\end{subequations}

\vspace{-6.5mm}
\noindent over all $(V,U)-X-(Y_1,Y_2)$. Then $\hat{\mathcal{C}}\subseteq\hat{\mathcal{O}}$.
\end{bound}


\begin{bound}[Inner Bound]\label{BND:inner_bound_nocommon}
Let $\hat{\mathcal{I}}$ be the closure of the union of rate pairs $(R_1,R_2)\in\mathbb{R}^2_+$ satisfying:
\begin{subequations}
\begin{align}
R_1&\leq I(U;Y_1)-I(U;V)-I(U;Y_2|V)\label{EQ:region_inner_nocommon11}\\
R_2&\leq I(V;Y_2)\label{EQ:region_inner_nocommon23}
\end{align}\label{EQ:region_inner_nocommon}
\end{subequations}

\vspace{-6.5mm}
\noindent over all $(V,U)-X-(Y_1,Y_2)$. Then $\hat{\mathcal{I}}\subseteq\hat{\mathcal{C}}$.
\end{bound}

The reader is referred to Appendix \ref{APPEN:inner_outer_proof} for the proofs of Bounds \ref{BND:outer_bound_nocommon} and \ref{BND:inner_bound_nocommon}. Let $\hat{\mathcal{C}}_\mathrm{K}$, $\hat{\mathcal{O}}_\mathrm{K}$ and $\hat{\mathcal{I}}_\mathrm{K}$ denote the secrecy-capacity region, the outer bound and the inner bound for a MIMO Gaussian BC computed under a covariance input constraint $\mathbb{E}\big[\mathbf{X}\mathbf{X}^\top\big] \preceq \mathrm{K}$. Accordingly, we have $\hat{\mathcal{I}}_\mathrm{K}\subseteq\hat{\mathcal{C}}_\mathrm{K}\subseteq\hat{\mathcal{O}}_\mathrm{K}$.

The opposite inclusion, i.e., $\hat{\mathcal{O}}_\mathrm{K}\subseteq\hat{\mathcal{I}}_\mathrm{K}$, is shown next. The regions $\hat{\mathcal{I}}_\mathrm{K}$ and $\hat{\mathcal{O}}_\mathrm{K}$ are closed, convex and bounded subsets of the first quadrant, and therefore, are characterized by the intersection of their supporting hyperplanes.

\begin{lemma}[Supporting Hyperplanes]\label{LEMMA:supporting_hyperplanes}
The following are supporting hyperplanes of $\hat{\mathcal{O}}_\mathrm{K}$ and $\hat{\mathcal{I}}_\mathrm{K}$:
\begin{equation}
R_1\geq 0\ \ ,\ \ R_1\leq \mathcal{H}_1^K\ \ ,\ \ R_2\geq 0\ \ ,\ \ R_2\leq \mathcal{H}_2^K,\label{EQ:nocommon_hyper}
\end{equation}
where
\begin{subequations}
\begin{align}
\mathcal{H}_1^K&\triangleq \max_{\substack{(V,U)-\mathbf{X}-(\mathbf{Y}_1,\mathbf{Y}_2): \\ \mathbb{E}[\mathbf{X}\mathbf{X}^\top]\preceq \mathrm{K}}} I(U;\mathbf{Y}_1|V)-I(U;\mathbf{Y}_2|V)\label{EQ:nocommon_hyper1}\\
\mathcal{H}_2^K&\triangleq \max_{\substack{\mathbf{X}:\ \mathbb{E}[\mathbf{X}\mathbf{X}^\top]\preceq \mathrm{K}}} I(\mathbf{X};\mathbf{Y}_2).\label{EQ:nocommon_hyper2}
\end{align}
\end{subequations}
Furthermore, $(\mathcal{H}_1^K,0)$ and $(0,\mathcal{H}_2^K)$ are boundary points of $\hat{\mathcal{O}}_\mathrm{K}$ and $\hat{\mathcal{I}}_\mathrm{K}$.
\end{lemma}
Lemma \ref{LEMMA:supporting_hyperplanes} is proven in Appendix \ref{APPEN:supporting_hyperplanes_proof}. It shows that if we set $R_1^\star\triangleq \max\big\{R_1\big|(R_1,0)\in\hat{\mathcal{O}}_{\mathrm{K}}\big\}$ and $R_2^\star\triangleq \max\big\{R_2\big|(0,R_2)\in\hat{\mathcal{O}}_{\mathrm{K}}\big\}$, then $(R_1^\star,0)$ and $(0,R_2^\star)$ are attainable in $\hat{\mathcal{I}}_{\mathrm{K}}$. Consequently, to show that the regions coincide, it suffices to establish
\begin{equation}
\max_{(R_1,R_2)\in\hat{\mathcal{O}}_\mathrm{K}}\lambda_1R_1+\lambda_2R_2\leq \max_{(R_1,R_2)\in\hat{\mathcal{I}}_\mathrm{K}}\lambda_1R_1+\lambda_2R_2,\label{EQ:nocommon_sufficeint}
\end{equation}
for $\lambda_1,\lambda_2>0$. Observe that
\begin{align*}
&\max_{(R_1,R_2)\in\hat{\mathcal{O}}_\mathrm{K}}\lambda_1R_1+\lambda_2R_2\\
&\stackrel{(a)}\leq\mspace{-10mu} \sup_{\substack{(V,U)-\mathbf{X}-(\mathbf{Y}_1,\mathbf{Y}_2):\\\mathbb{E}[\mathbf{X}\mathbf{X}^\top]\preceq \mathrm{K}}}\mspace{-20mu}\lambda_1\Big[I(U;\mspace{-1.5mu}\mathbf{Y}_1|V)\mspace{-3mu}-\mspace{-3mu}I(U;\mspace{-1.5mu}\mathbf{Y}_2|V)\Big]\mspace{-3mu}+\mspace{-3mu}\lambda_2I(V;\mspace{-1.5mu}\mathbf{Y}_2)\\
&\stackrel{(b)}= \sup_{\substack{(V,U)-\mathbf{X}-(\mathbf{Y}_1,\mathbf{Y}_2):\\\mathbb{E}[\mathbf{X}\mathbf{X}^\top]\preceq \mathrm{K}}}\lambda_1I(\mathbf{X};\mathbf{Y}_1|V)-(\lambda_1+\lambda_2)I(\mathbf{X};\mathbf{Y}_2|V)\\
&\mspace{40mu}+\lambda_1\Big[I(\mathbf{X};\mathbf{Y}_2|V,U)-I(\mathbf{X};\mathbf{Y}_1|V,U)\Big]+\lambda_2I(\mathbf{X};\mathbf{Y}_2)\\
&\stackrel{(c)}\leq \sup_{\substack{V-\mathbf{X}-(\mathbf{Y}_1,\mathbf{Y}_2):\\\mathbb{E}[\mathbf{X}\mathbf{X}^\top]\preceq \mathrm{K}}}\lambda_1I(\mathbf{X};\mathbf{Y}_1|V)-(\lambda_1+\lambda_2)I(\mathbf{X};\mathbf{Y}_2|V)\\
&\mspace{200mu}+\lim_{\eta\downarrow 1}\lambda_1\mathsf{S}_\eta^q(\mathbf{X}|V)+\lambda_2I(\mathbf{X};\mathbf{Y}_2)\\
&\leq \sup_{\mathbb{E}[\mathbf{X}\mathbf{X}^\top]\preceq \mathrm{K}}\lambda_2I(\mathbf{X};\mathbf{Y}_2)+\sup_{\substack{V-\mathbf{X}-(\mathbf{Y}_1,\mathbf{Y}_2):\\\mathbb{E}[\mathbf{X}\mathbf{X}^\top]\preceq \mathrm{K}}}\lim_{\eta\downarrow 1}\mathsf{t}_{\bm{\lambda},\eta}(\mathbf{X}|V)\\
&\stackrel{(d)}= \sup_{\mathbb{E}[\mathbf{X}\mathbf{X}^\top]\preceq \mathrm{K}}\lambda_2I(\mathbf{X};\mathbf{Y}_2)+\sup_{\mathbb{E}[\mathbf{X}\mathbf{X}^\top]\preceq\mathrm{K}}\lim_{\eta\downarrow 1}\mathsf{T}_{\bm{\lambda},\eta}(\mathbf{X})\\
&\stackrel{(e)}= \sup_{\mathbb{E}[\mathbf{X}\mathbf{X}^\top]\preceq \mathrm{K}}\lambda_2I(\mathbf{X};\mathbf{Y}_2)+\lim_{\eta\downarrow 1}\hat{V}_{\bm{\lambda},\eta}(\mathrm{K}),\numberthis
\end{align*}
where:\\
(a) uses \eqref{EQ:region_outer_nocommon};\\
(b) is because $(V,U)-\mathbf{X}-(\mathbf{Y}_1,\mathbf{Y}_2)$ forms a Markov chain;\\
(c) follows by the definition of $\mathsf{s}_\eta^q(\mathbf{X}|V)$ and since conditioned on $V$, $U-\mathbf{X}-(\mathbf{Y}_1,\mathbf{Y}_2)$ forms a Markov chain, i.e., it holds that $P_{\mathbf{Y}_1,\mathbf{Y}_2|V,U,\mathbf{X}}=P_{\mathbf{Y}_1,\mathbf{Y}_2|\mathbf{X}}$. Furthermore, (c) uses the continuity of $\mathsf{S}_\eta^q(\mathbf{X}|V)$ in $\eta$ at $\eta=1$ (see Property 3 of Proposition \ref{PROP:Envelope_Properties}), which implies that for any $(V,\mathbf{X})$
\begin{equation*}
\mathsf{S}_1^q(\mathbf{X}|V)\triangleq \mathsf{S}_{\lim_{\eta\downarrow 1}\eta}^q(\mathbf{X}|V)=\lim_{\eta\downarrow 1}\mathsf{S}_\eta^q(\mathbf{X}|V);
\end{equation*}
(d) is by the definition of $\mathsf{T}_{\bm{\lambda},\eta}^q(\mathbf{X})$, the Markov relation $V-\mathbf{X}-(\mathbf{Y}_1,\mathbf{Y}_2)$, and because $\mathsf{T}_{\bm{\lambda},\eta}^q(\mathbf{X})$ is continuous in $\eta$ at $\eta=1$ (see Remark \ref{REM:Nesterd_Envelope_Properties});\\
(e) follows by Proposition \ref{PROP:Maximum_Countinuos}.

\vspace{2mm}
By Theorem \ref{TM:Gaussian_maximizer_Envelope2}, for every $\eta>1$, there exist independent random variables $\mathbf{X}_1^\star\sim\mathcal{N}(\mathbf{0},\mathrm{K}_1)$, $\mathbf{X}_2^\star\sim\mathcal{N}(\mathbf{0},\mathrm{K}_2)$ and $\mathbf{X}'\sim\mathcal{N}\big(\mathbf{0},\mathrm{K}-(\mathrm{K}_1+\mathrm{K}_2)\big)$, $\mathrm{K}_1+\mathrm{K}_2\preceq\mathrm{K}$, such that $\hat{V}_\eta^q(\mathrm{K})=\mathsf{t}_{\bm{\lambda},\eta}^q(\mathbf{X}_1^\star+\mathbf{X}_2^\star)$ and $\mathsf{S}_\eta^q(\mathbf{X}_1^\star+\mathbf{X}_2^\star)=\mathsf{s}_\eta^q(\mathbf{X}_1^\star)$. Moreover, setting $\mathbf{X}=\mathbf{X}^\star_1+\mathbf{X}^\star_2+\mathbf{X}'$ maximizes $\lambda_2I(\mathbf{X};\mathbf{Y}_2)$ and attains $\hat{V}_\eta^q(\mathrm{K})$ simultaneously. In order to conform to the notation in the bounds, let $V^\star=\mathbf{X}'$. Taking the limit as $\eta\downarrow 1$, we have
\begin{align*}
&\max_{(R_1,R_2)\in\hat{\mathcal{O}}_\mathrm{K}}\lambda_1R_1+\lambda_2R_2\\
&\leq\lambda_1I(\mathbf{X};\mathbf{Y}_1|V^\star)-(\lambda_1+\lambda_2)I(\mathbf{X};\mathbf{Y}_2|V^\star)\\
&+\lambda_1\Big[I(\mathbf{X};\mathbf{Y}_2|V^\star,\mathbf{X}^\star_2)-I(\mathbf{X};\mathbf{Y}_2|V^\star,\mathbf{X}^\star_2)\Big]+\lambda_2I(\mathbf{X};\mathbf{Y}_2)\\
&\leq\lambda_1\Big[I(\mathbf{X}^\star_2;\mathbf{Y}_1|V^\star)-I(\mathbf{X}^\star_2;\mathbf{Y}_2|V^\star)\Big]+\lambda_2I(V^\star;\mathbf{Y}_2).\numberthis\label{EQ:nocommon_outer_UB}
\end{align*}

The following proposition is used to show that \eqref{EQ:nocommon_outer_UB} is achievable within $\hat{\mathcal{I}}_{\mathrm{K}}$.
\begin{proposition}[Partial Dirty-Paper Coding (P-DPC)]\label{PROP:Partial_DPC}
Fix a covariance matrix $\mathrm{K}$ and let $\mathbf{X}=\mathbf{X}^\star_1+\mathbf{X}^\star_2+V^\star$, where $\mathbf{X}^\star_1,\ \mathbf{X}^\star_2$ and $V$ are independent Gaussian random vectors with covariance matrices $\mathrm{K}_1$, $\mathrm{K}_2$ and $\mathrm{K}-(\mathrm{K}_1+\mathrm{K}_2)$, respectively, for some $\mathrm{0}\preceq\mathrm{K}_1,\mathrm{K}_2\preceq\mathrm{K}$ with $\mathrm{K}_1+\mathrm{K}_2\preceq\mathrm{K}$. Let $\mathbf{Y}_j=\mathrm{G}_j\mathbf{X}+\mathbf{Z}_j$, for $j=1,2$, where $\mathbf{Z}_j\sim\mathcal{N}(\mathbf{0},\mathrm{I})$ is independent of $(\mathbf{X}^\star_1,\mathbf{X}^\star_2,V^\star)$. Set $U=\mathbf{X}^\star_2+\mathrm{A}V^\star$, where $\mathrm{A}=\mathrm{K}_2\tilde{\mathrm{G}}_1^\top\Big[\mathrm{I}+\tilde{\mathrm{G}}_1\mathrm{K}_2\tilde{\mathrm{G}}_1^\top\Big]^{-1}$ and $\tilde{\mathrm{G}}_1=\Big[\mathrm{I}+\mathrm{G}_1\mathrm{K}_1\mathrm{G}_1^\top\Big]^{-\frac{1}{2}}\mathrm{G}_1$. Then
\begin{align*}
I(\mathbf{X}^\star_2;&\mathbf{Y}_1|V^\star)-I(\mathbf{X}^\star_2;\mathbf{Y}_2|V^\star)\\
&=I(U^\star;\mathbf{Y}_1)-I(U^\star;V^\star)-I(\mathbf{X}^\star_2;\mathbf{Y}_2|V^\star).\numberthis
\end{align*}
\end{proposition}

\begin{IEEEproof}
We first write
\begin{align*}
\mathbf{Y}_1&=\mathrm{G}_1\mathbf{X}+\mathbf{Z}_1\\
            &=\mathrm{G}_1(\mathbf{X}^\star_1+\mathbf{X}^\star_2+V^\star)+\mathbf{Z}_1\\
            &=\mathrm{G}_1(\mathbf{X}^\star_2+V^\star)+(\mathrm{G}_1\mathbf{X}^\star_1+\mathbf{Z}_1)\\
            &\stackrel{(a)}=\mathrm{G}_1\tilde{\mathbf{X}}+\mathbf{Z}'_1,\numberthis
\end{align*}
where (a) follows by setting $\tilde{\mathbf{X}}\triangleq \mathbf{X}^\star_2+V^\star$ and $\mathbf{Z}'_1\triangleq \mathrm{G}_1\mathbf{X}^\star_1+\mathbf{Z}_1$. By the independence of $\mathbf{X}^\star_1,\ \mathbf{X}^\star_2, \ V^\star$ and $\mathbf{Z}_1$, we have that $\tilde{\mathbf{X}}$ and $\mathbf{Z}'_1$ are also independent. Moreover, $\mathbf{Z}'_1\sim\mathcal{N}\big(\mathbf{0},\mathrm{I}+\mathrm{G}_1\mathrm{K}_1\mathrm{G}_1^\top\big)$, where the covariance matrix $\mathrm{I}+\mathrm{G}_1\mathrm{K}_1\mathrm{G}_1^\top$ is diagonalizable (due to its symmetry) and invertible (because it is positive definite). Denoting $\Sigma\triangleq  \mathrm{I}+\mathrm{G}_1\mathrm{K}_1\mathrm{G}_1^\top$, gives
\begin{equation}
\Sigma=\mathrm{Q}\Lambda\mathrm{Q}^\top,
\end{equation}
where $\mathrm{Q}$ is a unitary matrix and $\Lambda$ is diagonal, and furthermore $\Sigma^{-\frac{1}{2}}=\mathrm{Q}\Lambda^{-\frac{1}{2}}\mathrm{Q}^\top$. By defining $\tilde{\mathbf{Y}}_1=\Sigma^{-\frac{1}{2}}\mathbf{Y}_1$, we have
\begin{equation}
\tilde{\mathbf{Y}}_1=\tilde{\mathrm{G}}_1\tilde{\mathbf{X}}+\tilde{\mathbf{Z}}_1,
\end{equation}
where $\tilde{\mathrm{G}}_1=\Sigma^{-\frac{1}{2}}\mathrm{G}_1$, $\tilde{\mathbf{Z}}_1=\Sigma^{-\frac{1}{2}}\mathbf{Z}'_1$ and $\tilde{\mathbf{Z}}_1\sim\mathcal{N}(\mathbf{0},\mathrm{I})$. Setting $U^\star$ as above and invoking the classic Dirty-Paper Coding Theorem (here we use the formulation from \cite[Proposition 12]{Chandra_Gauss_BC2014}), we have
\begin{equation}
I(\tilde{\mathbf{X}};\tilde{\mathbf{Y}}_1|V^\star)=I(U^\star;\tilde{\mathbf{Y}}_1)-(U^\star;V^\star).\label{EQ:proof_Partial_DPC1}
\end{equation}
Furthermore,
\begin{equation}
I(\mathbf{X}^\star_2;\mathbf{Y}_2|V^\star)=I(\mathbf{X}^\star_2+\mathrm{A}V^\star;\mathbf{Y}_2|V^\star)=I(U^\star;\mathbf{Y}_2|V^\star).\label{EQ:proof_Partial_DPC2}
\end{equation}
Note that $\mathbf{Y}_1\mapsto \Sigma^{-\frac{1}{2}}\mathbf{Y}_1$ is an invertible mapping, and as such, preserves mutual information. We conclude the proof as follows:
\begin{align*}
I(\mathbf{X}^\star_2;&\mathbf{Y}_1|V^\star)-I(\mathbf{X}^\star_2;\mathbf{Y}_2|V^\star)\\
&\stackrel{(a)}=I(\tilde{\mathbf{X}};\mathbf{Y}_1|V^\star)-I(U^\star;\mathbf{Y}_2|V^\star)\\
&\stackrel{(b)}=I(\tilde{\mathbf{X}};\tilde{\mathbf{Y}}_1|V^\star)-I(U^\star;\mathbf{Y}_2|V^\star)\\
&\stackrel{(c)}=I(U^\star;\tilde{\mathbf{Y}}_1)-I(U^\star;V^\star)-I(U^\star;\mathbf{Y}_2|V^\star)\\
&\stackrel{(d)}=I(U^\star;\mathbf{Y}_1)-I(U^\star;V^\star)-I(U^\star;\mathbf{Y}_2|V^\star),\numberthis
\end{align*}
where (a) is because $\tilde{\mathbf{X}}=\mathbf{X}^\star_2+V^\star$ and by \eqref{EQ:proof_Partial_DPC2}, (b) and (d) follow since $\mathbf{Y}_1\mapsto \Sigma^{-\frac{1}{2}}\mathbf{Y}_1$ preserves mutual information, while (c) uses \eqref{EQ:proof_Partial_DPC1}.
\end{IEEEproof}

Inserting $U^\star$ as stated in Proposition \ref{PROP:Partial_DPC} into the RHS of \eqref{EQ:nocommon_outer_UB}, we obtain
\begin{align*}
&\max_{(R_1,R_2)\in\hat{\mathcal{O}}_\mathrm{K}}\lambda_1R_1+\lambda_2R_2\\
&\leq\lambda_1\Big[I(\mathbf{X}^\star_2;\mathbf{Y}_1|V^\star)-I(\mathbf{X}^\star_2;\mathbf{Y}_2|V^\star)\Big]+\lambda_2I(V^\star;\mathbf{Y}_2)\\
&=\lambda_1\Big[I(U^\star;\mspace{-3mu}\mathbf{Y}_1)\mspace{-3mu}-\mspace{-3mu}I(U^\star;\mspace{-1mu}V^\star)\mspace{-3mu}-\mspace{-3mu}I(U^\star;\mspace{-3mu}\mathbf{Y}_2|V^\star)\Big]\mspace{-3mu}+\mspace{-3mu}\lambda_2I(V^\star;\mspace{-3mu}\mathbf{Y}_2)\\
&\stackrel{(a)}\leq\max_{(R_1,R_2)\in\hat{\mathcal{I}}_\mathrm{K}}\lambda_1R_1+\lambda_2R_2,\numberthis\label{EQ:nocommon_outer_UB_inner}
\end{align*}
where (a) follows since $(U^\star,V^\star)-\mathbf{X}-(\mathbf{Y}_1,\mathbf{Y}_2)$ forms a Markov chain and $\mathbb{E}\big[\mathbf{X}\mathbf{X}^\top\big]\preceq \mathrm{K}$ is satisfied, which implies that the rate pair $R_1=I(U^\star;\mathbf{Y}_1)-I(U^\star;V^\star)-I(U^\star;\mathbf{Y}_2|V^\star)$ and $R_2=I(V^\star;\mathbf{Y}_2)$ belongs to $\hat{\mathcal{I}}_\mathrm{K}$. Concluding, we see that $\hat{\mathcal{I}}_\mathrm{K}=\hat{\mathcal{C}}_\mathrm{K}=\hat{\mathcal{O}}_\mathrm{K}$, which characterizes the secrecy-capacity region of the MIMO Gaussian BC with private and confidential messages.

Furthermore, equality (and hence the extreme points of $\hat{\mathcal{C}}_\mathrm{K}$) is achieved by Gaussian inputs as stated in Proposition \ref{PROP:Partial_DPC}, thus making the region computable. By evaluating $\hat{\mathcal{I}}_\mathrm{K}$ (or, equivalently $\hat{\mathcal{O}}_\mathrm{K}$) with respect to this input distribution, we describe the secrecy-capacity region $\hat{\mathcal{C}}_\mathrm{K}$ as the union of rate pairs $(R_1,R_2)\in\mathbb{R}^2_+$ satisfying:
\begin{subequations}
\begin{align}
R_1&\leq\frac{1}{2}\log\left|\frac{\mathrm{I}+\mathrm{G}_1(\mathrm{K}_1+\mathrm{K}_2)\mathrm{G}_1^\top}{\mathrm{I}+\mathrm{G}_1\mathrm{K}_1\mathrm{G}_1^\top}\right|\nonumber\\
&\mspace{120mu}-\frac{1}{2}\log\left|\frac{\mathrm{I}+\mathrm{G}_2(\mathrm{K}_1+\mathrm{K}_2)\mathrm{G}_2^\top}{\mathrm{I}+\mathrm{G}_2\mathrm{K}_1\mathrm{G}_2^\top}\right|\label{EQ:nocommon_ratebound1}\\
R_2&\leq \frac{1}{2}\log\left|\frac{\mathrm{I}+\mathrm{G}_2\mathrm{K}\mathrm{G}_2^\top}{\mathrm{I}+\mathrm{G}_2(\mathrm{K}_1+\mathrm{K}_2)\mathrm{G}_2^\top}\right|\label{EQ:nocommon_ratebound2},
\end{align}\label{EQ:nocommon_ratebound}%
\end{subequations}
where the union is over all positive semi-definite matrices $\mathrm{K}_1,\mathrm{K}_2$, such that $\mathrm{K}_1+\mathrm{K}_2\preceq\mathrm{K}$. We further simplify \eqref{EQ:nocommon_ratebound} by noting that the RHS of \eqref{EQ:nocommon_ratebound1} is the secrecy-capacity of the MIMO Gaussian WTC as derived in \cite[Appendix III]{Chandra_Gauss_BC2014}, which is maximized by setting $\mathrm{K}_1=\mathrm{0}$ (see \cite{Liu_Shamai_MIMOWTC2009,Khitsi_MIMOWTC2010,Hassibi_MINOWTC2011}). Further note that $\mathrm{K}_1=\mathrm{0}$ cannot decrease the RHS of \eqref{EQ:nocommon_ratebound2}. Thus, by relabeling $\mathrm{K}_2\triangleq \mathrm{K}^\star$ we establish \eqref{EQ:region_nocommon}.


\subsection{Proof of Theorem \ref{TM:capacity_common}}\label{SUBSEC:proof_common_capacity}

As in the case without a common message, the secrecy-capacity region $\mathcal{C}_\mathrm{K}$ is derived by showing that certain outer bound and inner bounds on $\mathcal{C}_\mathrm{K}$ coincide. Denoting by $\mathcal{C}$ the region of the DM-BC with common, private and confidential messages, we bound it as follows.


\begin{bound}[Outer Bound]\label{BND:outer_bound}
Let $\mathcal{O}$ be the closure of the union of rate triples $(R_0,R_1,R_2)\in\mathbb{R}^3_+$ satisfying:
\begin{subequations}
\begin{align}
R_0&\leq \min\big\{I(W;Y_1),I(W;Y_2)\big\}\label{EQ:region_outer0}\\
R_1&\leq I(U;Y_1|W,V)-I(U;Y_2|W,V)\label{EQ:region_outer11}\\
R_0+R_2&\leq I(V;Y_2|W)+\min\big\{I(W;Y_1),I(W;Y_2)\big\}\label{EQ:region_outer23}
\end{align}\label{EQ:region_outer}
\end{subequations}

\vspace{-6.5mm}
\noindent over all $(W,V,U)-X-(Y_1,Y_2)$. Then $\mathcal{C}\subseteq\mathcal{O}$.
\end{bound}

\begin{bound}[Inner Bound]\label{BND:inner_bound}
Let $\mathcal{I}$ be the closure of the union of rate triples $(R_0,R_1,R_2)\in\mathbb{R}^3_+$ satisfying:
\begin{subequations}
\begin{align}
R_0&\leq \min\big\{I(W;Y_1),I(W;Y_2)\big\}\label{EQ:region_inner0}\\
R_1&\leq I(U;Y_1|W)-I(U;V|W)-I(U;Y_2|W,V)\label{EQ:region_inner11}\\
R_2&\leq I(V;Y_2|W)\label{EQ:region_inner23}
\end{align}\label{EQ:region_inner}
\end{subequations}

\vspace{-6.5mm}
\noindent over all $(W,V,U)-X-(Y_1,Y_2)$. Then $\mathcal{I}\subseteq\mathcal{C}$.
\end{bound}

The proofs of Bounds \ref{BND:outer_bound} and \ref{BND:inner_bound} are relegated to Appendix \ref{APPEN:inner_outer_proof}. Denoting by $\mathcal{O}_\mathrm{K}$ and $\mathcal{I}_\mathrm{K}$ the adaptations of Bounds \ref{BND:outer_bound} and \ref{BND:inner_bound} to a MIMO Gaussian BC with a covariance input constraint $\mathbb{E}\big[\mathbf{X}\mathbf{X}^\top\big] \preceq \mathrm{K}$, we have $\mathcal{I}_\mathrm{K}\subseteq\mathcal{C}_\mathrm{K}\subseteq\mathcal{O}_\mathrm{K}$.

Next, we use the factorization of UCEs method to show that the opposite inclusion $\mathcal{O}_\mathrm{K}\subseteq\mathcal{I}_\mathrm{K}$ also holds. Given the supporting hyperplanes characterization of bounded and closed convex sets, using a similar reasoning as in Section \ref{SUBSEC:proof_nocommon_capacity} (see Lemma \ref{LEMMA:supporting_hyperplanes}), it suffices to study $\max_{(R_0,R_1,R_2)\in\mathcal{C}_\mathrm{K}} \lambda_0R_0+\lambda_1R_1+\lambda_2R_2$, for $\lambda_j>0$, $j=1,2,3$. Further note that it suffices to restrict attention to the case where $\lambda_0>\lambda_2$. This follows from the following observation: If a rate triple $(R_0,R_1,R_2)$ is in $\mathcal{C}_\mathrm{K}$ then so does the triple $(0,R_1,R_2+R_0)$, since one may always treat the common message as part of the (non-confidential) private message to Receiver 2. Assuming $\lambda_0\leq\lambda_2$, we have
\begin{equation}
\lambda_0R_0+\lambda_1R_1+\lambda_2R_2\leq 0\cdot R_0+\lambda_1R_1+\lambda_2(R_0+R_2),\label{EQ:common_reduces_nocommon1}
\end{equation}
and therefore,
\begin{align*}
\max_{(R_0,R_1,R_2)\in\mathcal{C}_\mathrm{K}} \mspace{-13mu} \lambda_0R_0\mspace{-2mu}+\mspace{-2mu}\lambda_1R_1\mspace{-2mu}+\mspace{-2mu}\lambda_2R_2&=\mspace{-10mu}\max_{(0,R_1,R_2)\in\mathcal{C}_\mathrm{K}} \mspace{-10mu} \lambda_1R_1+\lambda_2R_2\\
&=\mspace{-10mu}\max_{(R_1,R_2)\in\hat{\mathcal{C}}_\mathrm{K}} \lambda_1R_1+\lambda_2R_2,\numberthis\label{EQ:common_reduces_nocommon2}
\end{align*}
where $\hat{\mathcal{C}}_\mathrm{K}$ is the secrecy-capacity region without a common message that was characterized in Section \ref{SUBSEC:result_nocommon_capacity}.

Hence, it suffices to show that for all $\lambda_j>0$, $j=1,2,3$, with $\lambda_0>\lambda_2$, we have
\begin{align*}
\max_{(R_0,R_1,R_2)\in\mathcal{O}_\mathrm{K}}& \lambda_0R_0+\lambda_1R_1+\lambda_2R_2\\&\leq\max_{(R_0,R_1,R_2)\in\mathcal{I}_\mathrm{K}} \lambda_0R_0+\lambda_1R_1+\lambda_2R_2.\numberthis
\end{align*}

Now, for any $\alpha\in[0,1]$ set $\bar{\alpha}=1-\alpha$, and consider the following.
\begin{align*}
&\max_{(R_0,R_1,R_2)\in\mathcal{O}_\mathrm{K}}\lambda_0R_0+\lambda_1R_1+\lambda_2R_2\\
&\stackrel{(a)}\leq \sup_{\substack{(W,V,U)-\mathbf{X}-(\mathbf{Y}_1,\mathbf{Y}_2):\\\mathbb{E}[\mathbf{X}\mathbf{X}^\top]\preceq \mathrm{K}}}\lambda_0\Big[\alpha I(W;\mathbf{Y}_1)+\bar{\alpha} I(W;\mathbf{Y}_2)\Big]\\
&\mspace{10mu}+\lambda_1\Big[I(U;\mathbf{Y}_1|W,V)-I(U;\mathbf{Y}_2|W,V)\Big]+\lambda_2I(V;\mathbf{Y}_2|W)\\
&\stackrel{(b)}= \sup_{\substack{(W,V,U)-\mathbf{X}-(\mathbf{Y}_1,\mathbf{Y}_2):\\\mathbb{E}[\mathbf{X}\mathbf{X}^\top]\preceq \mathrm{K}}}
\alpha\lambda_0I(\mathbf{X};\mathbf{Y}_1)+\bar{\alpha}\lambda_0I(\mathbf{X};\mathbf{Y}_2)\\
&\mspace{33mu}+(\lambda_2-\bar{\alpha}\lambda_0)I(\mathbf{X};\mathbf{Y}_2|W)-\alpha\lambda_0I(\mathbf{X};\mathbf{Y}_1|W)\\
&\mspace{66mu}+\lambda_1I(\mathbf{X};\mathbf{Y}_1|W,V)-(\lambda_1+\lambda_2)I(\mathbf{X};\mathbf{Y}_2|W,V)\\
&\mspace{99mu}+\lambda_1I(\mathbf{X};\mathbf{Y}_2|W,V,U)-\lambda_1I(\mathbf{X};\mathbf{Y}_1|W,V,U)\\
&\stackrel{(c)}\leq \sup_{\substack{(W,V)-\mathbf{X}-(\mathbf{Y}_1,\mathbf{Y}_2):\\\mathbb{E}[\mathbf{X}\mathbf{X}^\top]\preceq \mathrm{K}}}\alpha\lambda_0I(\mathbf{X};\mathbf{Y}_1)+\bar{\alpha}\lambda_0I(\mathbf{X};\mathbf{Y}_2)\\
&\mspace{33mu}+(\lambda_2-\bar{\alpha}\lambda_0)I(\mathbf{X};\mathbf{Y}_2|W)-\alpha\lambda_0I(\mathbf{X};\mathbf{Y}_1|W)\\
&\mspace{66mu}+\lambda_1I(\mathbf{X};\mathbf{Y}_1|W,V)-(\lambda_1+\lambda_2)I(\mathbf{X};\mathbf{Y}_2|W,V)\\
&\mspace{290mu}+\lim_{\eta\downarrow 1}\lambda_1\mathsf{S}_\eta^q(\mathbf{X}|W,V)\\
&\stackrel{(d)}\leq \sup_{\substack{W-\mathbf{X}-(\mathbf{Y}_1,\mathbf{Y}_2):\\\mathbb{E}[\mathbf{X}\mathbf{X}^\top]\preceq \mathrm{K}}}\alpha\lambda_0I(\mathbf{X};\mathbf{Y}_1)+\bar{\alpha}\lambda_0I(\mathbf{X};\mathbf{Y}_2)\\
&\mspace{36mu}+(\lambda_2-\bar{\alpha}\lambda_0)I(\mathbf{X};\mathbf{Y}_2|W)-\alpha\lambda_0I(\mathbf{X};\mathbf{Y}_1|W)\\
&\mspace{310mu}+\lim_{\eta\downarrow 1}\mathsf{T}_{\bm{\lambda},\eta}^q(\mathbf{X}|W)\\
&\leq \sup_{\mathbb{E}[\mathbf{X}\mathbf{X}^\top]\preceq \mathrm{K}} \alpha\lambda_0I(\mathbf{X};\mathbf{Y}_1)+\bar{\alpha}\lambda_0I(\mathbf{X};\mathbf{Y}_2)\\
&\mspace{187mu}+\sup_{\substack{W-\mathbf{X}-(\mathbf{Y}_1,\mathbf{Y}_2):\\\mathbb{E}[\mathbf{X}\mathbf{X}^\top]\preceq \mathrm{K}}}\lim_{\eta\downarrow 1}\mathsf{f}_{\bm{\lambda_0},\alpha,\eta}^q(\mathbf{X}|W)\\
&\stackrel{(e)}\leq \sup_{\mathbb{E}[\mathbf{X}\mathbf{X}^\top]\preceq \mathrm{K}} \alpha\lambda_0I(\mathbf{X};\mathbf{Y}_1)+\bar{\alpha}\lambda_0I(\mathbf{X};\mathbf{Y}_2)\\
&\mspace{241mu}+\sup_{\mathbb{E}[\mathbf{X}\mathbf{X}^\top]\preceq \mathrm{K}}\lim_{\eta\downarrow 1}\mathsf{F}_{\bm{\lambda_0},\alpha,\eta}^q(\mathbf{X})\\
&\stackrel{(f)}\leq \sup_{\mathbb{E}[\mathbf{X}\mathbf{X}^\top]\preceq \mathrm{K}} \alpha\lambda_0I(\mathbf{X};\mspace{-2mu}\mathbf{Y}_1)\mspace{-2mu}+\mspace{-2mu}\bar{\alpha}\lambda_0I(\mathbf{X};\mspace{-2mu}\mathbf{Y}_2)\mspace{-2mu}+\mspace{-2mu}\lim_{\eta\downarrow 1}\tilde{V}_{\bm{\lambda_0},\alpha,\eta}^q\mspace{-2mu}(\mathrm{K}),\numberthis
\end{align*}
where:\\
(a) is by \eqref{EQ:region_outer};\\
(b) follows because $(W,V,U)-\mathbf{X}-(\mathbf{Y}_1,\mathbf{Y}_2)$ forms a Markov chain;\\
(c) is by the definition of $\mathsf{S}_\eta^q(\mathbf{X}|W,V)$ since conditioned on $(W,V)$, $U-\mathbf{X}-(\mathbf{Y}_1,\mathbf{Y}_2)$ forms a Markov chain, and because $\mathsf{S}_\eta^q(\mathbf{X}|V,W)$ is continuous in $\eta$ at $\eta=1$ (Property 3 of Proposition \ref{PROP:Envelope_Properties});\\
(d) follows by the definition of $\mathsf{T}_{\bm{\lambda},\eta}^q(\mathbf{X}|W)$ since conditioned on $V$, $W-\mathbf{X}-(\mathbf{Y}_1,\mathbf{Y}_2)$ forms a Markov chain. Furthermore, the continuity of $\mathsf{T}_{\bm{\lambda},\eta}^q(\mathbf{X}|W)$ in $\eta$ at $\eta=1$ (see Remark \ref{REM:Nesterd_Envelope_Properties}) is also exploited;\\
(e) is by the definition of $\mathsf{F}_{\bm{\lambda_0},\alpha,\eta}^q(\mathbf{X})$ (while noting that $W-\mathbf{X}-(\mathbf{Y}_1,\mathbf{Y}_2)$ forms a Markov chain), and because $\mathsf{F}_{\bm{\lambda},\eta}^q(\mathbf{X})$ is continuous at $\eta=1$ (Remark \ref{REM:Double_Nesterd_Envelope_Properties});\\
(f) makes use of the continuity argument from Remark \ref{PROP:Maximum_Countinuos_double}.

\vspace{2mm}
Recall that for any $\eta>1$, $\lambda_j>0$, for $j=0,1,2$, and $\lambda_0>\lambda_2$, Corollary \ref{COR:Gaussian_maximizer_Envelope3} implies that there exist independent random variables $\mathbf{X}_j^\star\sim\mathcal{N}(\mathbf{0},\mathrm{K}_j)$, $j=1,2,3$, and $\mathbf{X}'\sim\mathcal{N}\big(\mathbf{0},\mathrm{K}-(\mathrm{K}_1+\mathrm{K}_2+\mathrm{K}_3)\big)$, such that \eqref{EQ:Gaussian_maximizer_Envelope3} is satisfied. Furthermore, setting $\mathbf{X}=\mathbf{X}^\star_1+\mathbf{X}^\star_2+\mathbf{X}^\star_3+\mathbf{X}'$ not only attains $\tilde{V}_\eta^q(\mathrm{K})$, but it also simultaneously maximizes $\alpha\lambda_0I(\mathbf{X};\mathbf{Y}_1)$ and $\bar{\alpha}\lambda_0I(\mathbf{X};\mathbf{Y}_2)$. Relabeling $W^\star=\mathbf{X}'$ and $V^\star=\mathbf{X}^\star_3$ while taking the limit as $\eta\downarrow 1$, we have
\begin{align*}
&\max_{(R_0,R_1,R_2)\in\mathcal{O}_\mathrm{K}}\lambda_0R_0+\lambda_1R_1+\lambda_2R_2\\
&\leq\alpha\lambda_0I(\mathbf{X};\mathbf{Y}_1)+\bar{\alpha}\lambda_0I(\mathbf{X};\mathbf{Y}_2)\\
&\mspace{7mu}+(\lambda_2-\bar{\alpha}\lambda_0)I(\mathbf{X};\mathbf{Y}_2|W^\star)-\alpha\lambda_0I(\mathbf{X};\mathbf{Y}_1|W^\star)\\
&\mspace{14mu}+\lambda_1I(\mathbf{X};\mathbf{Y}_1|W^\star,V^\star)-(\lambda_1+\lambda_2)I(\mathbf{X};\mathbf{Y}_2|W^\star,V^\star)\\
&\mspace{21mu}+\lambda_1I(\mathbf{X};\mathbf{Y}_2|W^\star,V^\star,\mathbf{X}^\star_2)-\lambda_1I(\mathbf{X};\mathbf{Y}_1|W^\star,V^\star,\mathbf{X}^\star_2)\\
&\leq\lambda_0\Big[\alpha I(W^\star;\mathbf{Y}_1)+\bar{\alpha} I(W^\star;\mathbf{Y}_2)\Big]\\
&\mspace{15mu}+\lambda_1\Big[I(\mathbf{X}_2^\star;\mathbf{Y}_1|W^\star,V^\star)-I(\mathbf{X}_2^\star;\mathbf{Y}_2|W^\star,V^\star)\Big]\\
&\mspace{240mu}+\lambda_2I(V^\star;\mathbf{Y}_2|W^\star)\numberthis.\label{EQ:common_outer_UB}
\end{align*}

Using Proposition \ref{PROP:Partial_DPC}, we set $U=\mathbf{X}^\star_2+\tilde{\mathrm{A}}V^\star$ as before and obtain
\begin{align*}
I(&\mathbf{X}^\star_2;\mathbf{Y}_1|W^\star,V^\star)-I(\mathbf{X}^\star_2;\mathbf{Y}_2|W^\star,V^\star)\\
&=I(U^\star;\mathbf{Y}_1|W^\star)-I(U^\star;V^\star|W^\star)-I(\mathbf{X}^\star_2;\mathbf{Y}_2|W^\star,V^\star).\numberthis\label{EQ:common_DPC}
\end{align*}
Inserting \eqref{EQ:common_DPC} into \eqref{EQ:common_outer_UB}, yields
\begin{align*}
&\max_{(R_0,R_1,R_2)\in\mathcal{O}_\mathrm{K}}\lambda_0R_0+\lambda_1R_1+\lambda_2R_2\\
&\leq\lambda_0\Big[\alpha I(W^\star;\mathbf{Y}_1)+\bar{\alpha} I(W^\star;\mathbf{Y}_2)\Big]\\
&+\lambda_1\Big[I(U^\star;\mspace{-1.5mu}\mathbf{Y}_1|W^\star)\mspace{-2mu}-\mspace{-2mu}I(U^\star;V^\star|W^\star)\mspace{-2mu}-\mspace{-2mu}I(U^\star;\mspace{-1.5mu}\mathbf{Y}_2|W^\star\mspace{-2.5mu},\mspace{-2.5mu}V^\star)\Big]\\
&\mspace{294mu}+\lambda_2I(V^\star;\mathbf{Y}_2|W^\star)\\
&\leq\sup_{\substack{(W,V,U)-\mathbf{X}-(\mathbf{Y}_1,\mathbf{Y}_2):\\\mathbb{E}[\mathbf{X}\mathbf{X}^\top]\preceq \mathrm{K}}}\lambda_0\Big[\alpha I(W;\mathbf{Y}_1)+\bar{\alpha} I(W;\mathbf{Y}_2)\Big]\\
&+\lambda_1\Big[I(U;\mathbf{Y}_1|W)-I(U;V|W)-I(U;\mathbf{Y}_2|W,V)\Big]\\
&\mspace{265mu}+\lambda_2I(V;\mathbf{Y}_2|W).\numberthis\label{EQ:common_outer_UB2}
\end{align*}

Since \eqref{EQ:common_outer_UB2} holds for all $\alpha\in[0,1]$, we have
\begin{align*}
&\max_{(R_0,R_1,R_2)\in\mathcal{O}_\mathrm{K}}\lambda_0R_0+\lambda_1R_1+\lambda_2R_2\\&\leq\min_{\alpha\in[0,1]}\sup_{\substack{(W,V,U)-\mathbf{X}-(\mathbf{Y}_1,\mathbf{Y}_2):\\\mathbb{E}[\mathbf{X}\mathbf{X}^\top]\preceq \mathrm{K}}}\mspace{-10mu}\lambda_0\Big[\alpha I(W;\mathbf{Y}_1)+\bar{\alpha}I(W;\mathbf{Y}_2)\Big]\\
&+\lambda_1\Big[I(U;\mathbf{Y}_1|W)-I(U;V|W)-I(U;\mathbf{Y}_2|W,V)\Big]\\
&\mspace{265mu}+\lambda_2I(V;\mathbf{Y}_2|W).\numberthis\label{EQ:common_outer_UB_minalpha}
\end{align*}
Having \eqref{EQ:common_outer_UB_minalpha}, the desired equality $\mathcal{O}_\mathrm{K}=\mathcal{I}_\mathrm{K}$ is a consequence of the following Proposition \footnote{Proposition \ref{PROP:minimax_interchange} bears strong resemblance to Proposition 13 from \cite{Chandra_Gauss_BC2014} that was originally established in \cite{Nair_Marton_Optimality2014}}.

\begin{proposition}[Max-Min Interchanging]\label{PROP:minimax_interchange}
The following max-min interchanging holds
\begin{align*}
&\min_{\alpha\in[0,1]}\sup_{\substack{(W,V,U)-\mathbf{X}-(\mathbf{Y}_1,\mathbf{Y}_2):\\\mathbb{E}[\mathbf{X}\mathbf{X}^\top]\preceq \mathrm{K}}}\lambda_0\Big[\alpha I(W;\mathbf{Y}_1)+\bar{\alpha} I(W;\mathbf{Y}_2)\Big]\\
&+\lambda_1\Big[I(U;\mathbf{Y}_1|W)-I(U;V|W)-I(U;\mathbf{Y}_2|W,V)\Big]\\
&\mspace{265mu}+\lambda_2I(V;\mathbf{Y}_2|W)\\
&=\mspace{-3mu}\sup_{\substack{(W,V,U)-\mathbf{X}-(\mathbf{Y}_1,\mathbf{Y}_2):\\\mathbb{E}[\mathbf{X}\mathbf{X}^\top]\preceq \mathrm{K}}}\min_{\alpha\in[0,1]}\lambda_0\Big[\alpha I(W;\mathbf{Y}_1)\mspace{-3mu}+\mspace{-3mu}\bar{\alpha} I(W;\mathbf{Y}_2)\Big]\\
&+\lambda_1\Big[I(U;\mathbf{Y}_1|W)-I(U;V|W)-I(U;\mathbf{Y}_2|W,V)\Big]\\
&\mspace{265mu}+\lambda_2I(V;\mathbf{Y}_2|W)\\
&=\sup_{\substack{(W,V,U)-\mathbf{X}-(\mathbf{Y}_1,\mathbf{Y}_2):\\\mathbb{E}[\mathbf{X}\mathbf{X}^\top]\preceq \mathrm{K}}}\lambda_0\cdot\min\big\{I(W;\mathbf{Y}_1),I(W;\mathbf{Y}_2)\big\}\\
&+\lambda_1\Big[I(U;\mathbf{Y}_1|W)-I(U;V|W)-I(U;\mathbf{Y}_2|W,V)\Big]\\
&\mspace{265mu}+\lambda_2I(V;\mathbf{Y}_2|W).\numberthis
\end{align*}
\end{proposition}

The proof of Proposition \ref{PROP:minimax_interchange} is given in Appendix \ref{APPEN:minimax_interchange_proof}. Now, noting that $(W,V,U)-\mathbf{X}-(\mathbf{Y}_1,\mathbf{Y}_2)$ forms a Markov chain and $\mathbb{E}\big[\mathbf{X}\mathbf{X}^\top\big]\preceq\mathrm{K}$, we see that the triple
 \begin{align*}
 R_0&=\min\big\{I(W;\mathbf{Y}_1),I(W;\mathbf{Y}_2)\big\}\\
 R_1&=I(U;\mathbf{Y}_1|W)-I(U;V|W)-I(U;\mathbf{Y}_2|W,V)\\
 R_2&=I(V;\mathbf{Y}_2|W)
 \end{align*}
is inside the inner bound $\mathcal{I}_\mathrm{K}$. Hence
\begin{align*}
&\max_{(R_0,R_1,R_2)\in\mathcal{O}_\mathrm{K}}\lambda_0R_0+\lambda_1R_1+\lambda_2R_2\\&\leq\sup_{\substack{(W,V,U)-\mathbf{X}-(\mathbf{Y}_1,\mathbf{Y}_2):\\\mathbb{E}[\mathbf{X}\mathbf{X}^\top]\preceq \mathrm{K}}}\lambda_0\Big[\alpha I(W;\mathbf{Y}_1)+\bar{\alpha} I(W;\mathbf{Y}_2)\Big]\\
&+\lambda_1\Big[I(U;\mathbf{Y}_1|W)-I(U;V|W)-I(U;\mathbf{Y}_2|W,V)\Big]\\
&\mspace{285mu}+\lambda_2I(V;\mathbf{Y}_2|W)\\
&\leq\max_{(R_0,R_1,R_2)\in\mathcal{I}_\mathrm{K}}\lambda_0R_0+\lambda_1R_1+\lambda_2R_2,\numberthis\label{EQ:common_inner_outer_equality}
\end{align*}
which implies that $\hat{\mathcal{I}}_\mathrm{K}=\hat{\mathcal{C}}_\mathrm{K}=\hat{\mathcal{O}}_\mathrm{K}$ and characterizes the secrecy-capacity region of the MIMO Gaussian BC with common, private and confidential messages.

To obtain the description of $\mathcal{C}_\mathrm{K}$ stated in \eqref{EQ:region_common}, note that when $\lambda_0>\lambda_2$, equality (and hence the extreme points of $\mathcal{C}_\mathrm{K}$) is achieved by setting
\begin{equation}
\mathbf{X}=\mathbf{X}^\star_1+\mathbf{X}^\star_2+V^\star+W^\star,\label{EQ:optimal_input_common}
\end{equation}
where $\mathbf{X}_j^\star\sim\mathcal{N}(\mathbf{0},\mathrm{K}_j)$, $j=1,2$, $V^\star\sim\mathcal{N}(\mathbf{0},\mathrm{K}_3)$ and $W^\star\sim\mathcal{N}\big(\mathbf{0},\mathrm{K}-(\mathrm{K}_1+\mathrm{K}_2+\mathrm{K}_3)\big)$ are independent of each other, and $U=\mathbf{X}^\star_2+\tilde{\mathrm{A}}V^\star$, where $\tilde{\mathrm{A}}$ is the P-DPC matrix from Proposition \ref{PROP:Partial_DPC}. For the case when $\lambda_0\leq\lambda_2$, \eqref{EQ:common_reduces_nocommon1}-\eqref{EQ:common_reduces_nocommon2} imply that the boundary-achieving input distribution corresponds to the one that achieves the secrecy-capacity region when there is no common message (see Section \ref{SUBSEC:proof_nocommon_capacity}). Setting $\mathrm{K}_3=\mathrm{K}-(\mathrm{K}_1+\mathrm{K}_2)$ recovers the optimal input distribution for the case without common message.

By evaluating $\mathcal{I}_\mathrm{K}$ (or, equivalently $\mathcal{O}_\mathrm{K}$) with respect to \eqref{EQ:optimal_input_common}, we characterize the secrecy-capacity region $\mathcal{C}_\mathrm{K}$ as the union of rate triples $(R_0,R_1,R_2)\in\mathbb{R}^3_+$ satisfying:
\begin{subequations}
\begin{align}
R_0&\leq\min\Bigg\{\frac{1}{2}\log\left|\frac{\mathrm{I}+\mathrm{G}_1\mathrm{K}\mathrm{G}_1^\top}{\mathrm{I}+\mathrm{G}_1(\mathrm{K}_1+\mathrm{K}_2+\mathrm{K}_3)\mathrm{G}_1^\top}\right|\nonumber\\
&\mspace{80mu},\frac{1}{2}\log\left|\frac{\mathrm{I}+\mathrm{G}_1\mathrm{K}\mathrm{G}_1^\top}{\mathrm{I}+\mathrm{G}_1(\mathrm{K}_1+\mathrm{K}_2+\mathrm{K}_3)\mathrm{G}_1^\top}\right|\Bigg\}\label{EQ:region_common_complicatedRB0}\\
R_1&\leq\frac{1}{2}\log\left|\frac{\mathrm{I}+\mathrm{G}_1(\mathrm{K}_1+\mathrm{K}_2)\mathrm{G}_1^\top}{\mathrm{I}+\mathrm{G}_1\mathrm{K}_1\mathrm{G}_1^\top}\right|\nonumber\\
&\mspace{130mu}-\frac{1}{2}\log\left|\frac{\mathrm{I}+\mathrm{G}_2(\mathrm{K}_1+\mathrm{K}_2)\mathrm{G}_2^\top}{\mathrm{I}+\mathrm{G}_2\mathrm{K}_1\mathrm{G}_2^\top}\right|\label{EQ:region_common_complicatedRB1}\\
R_2&\leq\frac{1}{2}\log\left|\frac{\mathrm{I}+\mathrm{G}_2(\mathrm{K}_1+\mathrm{K}_2+\mathrm{K}_3)\mathrm{G}_2^\top}{\mathrm{I}+\mathrm{G}_2(\mathrm{K}_1+\mathrm{K}_2)\mathrm{G}_2^\top}\right|,\label{EQ:region_common_complicatedRB2}
\end{align}\label{EQ:region_common_comlicatedRB}%
\end{subequations}
where the union is over all positive semi-definite matrices $\mathrm{K}_1,\mathrm{K}_2,\mathrm{K}_3$, such that $\mathrm{K}_1+\mathrm{K}_2+\mathrm{K}_3\preceq\mathrm{K}$.

The region from \eqref{EQ:region_common_comlicatedRB} is further simplified using reasoning similar to this from Section \ref{SUBSEC:proof_nocommon_capacity}. First, \eqref{EQ:region_common_complicatedRB1} indicates that the signal to User 1 is a sum of two independent zero mean Gaussian random vector with covariance $\mathrm{K}_1$ and $\mathrm{K}_2$. The signal that corresponds to $\mathrm{K}_2$ carries the confidential message $M_1$, while the $\mathrm{K}_1$ signal is an artificial noise sent (on purpose) to confuse User 2 (which serves as an eavesdropper of $M_1$). The lack of structure in the artificial noise adds to the noise floor at both receivers (see also \cite[Remark 4]{Poor_Shamai_Gaussian_MIMO_BC_Secrecy2010}). However, since the RHS of \eqref{EQ:region_common_complicatedRB1} is the secrecy-capacity of the MIMO Gaussian WTC, it is maximized by setting $\mathrm{K}_1=\mathrm{0}$. Furthermore, since in both \eqref{EQ:region_common_complicatedRB0} and \eqref{EQ:region_common_complicatedRB2} $\mathrm{K}_1$ serves as noise (i.e., it is not used to encode any of the messages), setting $\mathrm{K}_1=\mathrm{0}$ achieves optimality. This is since $\mathrm{K}_1=\mathrm{0}$ corresponds to revealing the $\mathrm{K}_1$ signal to both receivers, which can only increase the transmission rates. Taking $\mathrm{K}_1=\mathrm{0}$ and recasting $\mathrm{K}_3$ as $\mathrm{K}_1$, recovers \eqref{EQ:region_common}.


\section{Proofs of Upper Concave Envelopes Properties}\label{SEC:envelope_properties}


\subsection{Proof of Proposition \ref{PROP:Envelope_Properties}}\label{SUBSEC:Envelope_Properties_proof}

Property 1 follows by Jensen's inequality since $\mathsf{S}_\eta^q$ is concave in $P_\mathbf{X}$, while for Property 2 we use the fact that $P_{\mathbf{X}|W,V}=P_{\mathbf{X}|V}$. To prove Property 3, fix $P_\mathbf{X}$ and let $\eta_1,\eta_2\in(0,2)$, $\alpha\in[0,1]$ and $\bar{\alpha}=1-\alpha$. Observe that
\begin{align*}
&\mathsf{S}_{\alpha\eta_1+\bar{\alpha}\eta_2}^q(\mathbf{X})\\
&=\sup_{\substack{P_{V|\mathbf{X}}:\\V-\mathbf{X}-(\mathbf{Y}_1,\mathbf{Y}_2)}}I(\mathbf{X};\mathbf{Y}_2|V)-(\alpha\eta_1+\bar{\alpha}\eta_2)I(\mathbf{X};\mathbf{Y}_1|V)\\
&\leq\mspace{10mu}\alpha\cdot\mspace{-35mu}\sup_{\substack{P_{V|\mathbf{X}}:\\V-\mathbf{X}-(\mathbf{Y}_1,\mathbf{Y}_2)}}\mspace{-25mu}I(\mathbf{X};\mathbf{Y}_2|V)-\eta_1I(\mathbf{X};\mathbf{Y}_1|V)\\
&\mspace{113mu}+\bar{\alpha}\cdot\mspace{-35mu}\sup_{\substack{P_{V|\mathbf{X}}:\\V-\mathbf{X}-(\mathbf{Y}_1,\mathbf{Y}_2)}}\mspace{-25mu}I(\mathbf{X};\mathbf{Y}_2|V)-\eta_2I(\mathbf{X};\mathbf{Y}_1|V)\\
&=\alpha \mathsf{S}_{\eta_1}^q(\mathbf{X})+\bar{\alpha}\mathsf{S}_{\eta_2}^q(\mathbf{X}).
\end{align*}
Clearly, $\mathsf{S}_\eta^q(\mathbf{X})$ is also bounded for every $\eta\in(0,2)$ and by invoking Proposition 17 from \cite[Chapter 5]{Royden_Real_Analysis1988}, we have that $\mathsf{S}_\eta^q(\mathbf{X})$ is continuous inside every closed subinterval of $(0,2)$, and in particular, at $\eta=1$.


\subsection{Proof of Proposition \ref{PROP:factorization_property2}}\label{SUBSEC:factorization_property2_proof}

Let $V-(\mathbf{X}_1,\mathbf{X}_2)-(\mathbf{Y}_{11},\mathbf{Y}_{12},\mathbf{Y}_{21},\mathbf{Y}_{22})$ form a Markov chain. We have,
\begin{align*}
&\mathsf{t}_{\bm{\lambda},\eta}^{q_1\times q_2}(\mathbf{X}_1,\mathbf{X}_2|V)\\
&=\lambda_1I(\mathbf{X}_1,\mathbf{X}_2;\mathbf{Y}_{11},\mathbf{Y}_{12}|V)\\
&\mspace{5mu}-(\lambda_1+\lambda_2)I(\mathbf{X}_1,\mathbf{X}_2;\mathbf{Y}_{21},\mathbf{Y}_{22}|V)+\lambda_1\mathsf{S}_\eta^{q_1\times q_2}(\mathbf{X}_1,\mathbf{X}_2|V)\\
&=\lambda_1\Big[I(\mathbf{X}_1,\mathbf{X}_2;\mathbf{Y}_{11}|V)+I(\mathbf{X}_1,\mathbf{X}_2;\mathbf{Y}_{12}|V,\mathbf{Y}_{11})\Big]\\
&-(\lambda_1+\lambda_2)\Big[I(\mathbf{X}_1,\mathbf{X}_2;\mathbf{Y}_{22}|V)+I(\mathbf{X}_1,\mathbf{X}_2;\mathbf{Y}_{21}|V,\mathbf{Y}_{22})\Big]\\
&\mspace{276mu}+\lambda_1\mathsf{S}_\eta^{q_1\times q_2}(\mathbf{X}_1,\mathbf{X}_2|V)\\
&\stackrel{(a)}=\lambda_1\Big[I(\mathbf{X}_1;\mathbf{Y}_{11}|V,\mathbf{Y}_{22})+I(\mathbf{X}_2;\mathbf{Y}_{12}|V,\mathbf{Y}_{11})\\
&+I(\mathbf{Y}_{11};\mathbf{Y}_{22}|\mathbf{Y}_{11})\Big]-(\lambda_1+\lambda_2)\Big[I(\mathbf{X}_2;\mathbf{Y}_{22}|V,\mathbf{Y}_{11})\\
&\mspace{120mu}+I(\mathbf{X}_1;\mathbf{Y}_{21}|V,\mathbf{Y}_{22})+I(\mathbf{Y}_{11};\mathbf{Y}_{22}|V)\Big]\\
&\mspace{276mu}+\lambda_1\mathsf{S}_\eta^{q_1\times q_2}(\mathbf{X}_1,\mathbf{X}_2|V)\\
&\stackrel{(b)}\leq\lambda_1I(\mathbf{X}_1;\mathbf{Y}_{11}|V,\mathbf{Y}_{22})-(\lambda_1+\lambda_2)I(\mathbf{X}_1;\mathbf{Y}_{21}|V,\mathbf{Y}_{22})\\
&\mspace{20mu}+\lambda_1\mathsf{S}_\eta^{q_1}(\mathbf{X}_1|V,\mathbf{Y}_{22})+\lambda_1I(\mathbf{X}_2;\mathbf{Y}_{12}|V,\mathbf{Y}_{11})\\
&\mspace{30mu}-(\lambda_1+\lambda_2)I(\mathbf{X}_2;\mathbf{Y}_{22}|V,\mathbf{Y}_{11})+\lambda_1\mathsf{S}_\eta^{q_2}(\mathbf{X}_2|V,\mathbf{Y}_{11})\\
&\mspace{300mu}-\lambda_2I(\mathbf{Y}_{11};\mathbf{Y}_{22}|V)\\
&\stackrel{(c)}\leq \mathsf{T}_{\bm{\lambda},\eta}^{q_1}(\mathbf{X}_1|\mathbf{Y}_{22})+\mathsf{T}_{\bm{\lambda},\eta}^{q_2}(\mathbf{X}_2|\mathbf{Y}_{11})-\lambda_2I(\mathbf{Y}_{11};\mathbf{Y}_{22}|V)\\
&\stackrel{(d)}\leq \mathsf{T}_{\bm{\lambda},\eta}^{q_1}(\mathbf{X}_1)+\mathsf{T}_{\bm{\lambda},\eta}^{q_2}(\mathbf{X}_2)-\lambda_2I(\mathbf{Y}_{11};\mathbf{Y}_{22}|V)\\
&\leq \mathsf{T}_{\bm{\lambda},\eta}^{q_1}(\mathbf{X}_1)+\mathsf{T}_{\bm{\lambda},\eta}^{q_2}(\mathbf{X}_2),\numberthis\label{EQ:fractorization_proof_ineq}
\end{align*}
where:\\
(a) is since given $V$ we have the Markov chain $(\mathbf{Y}_{11},\mathbf{Y}_{21})-\mathbf{X}_1-\mathbf{X}_2-(\mathbf{Y}_{12},\mathbf{Y}_{22})$;\\
(b) follows from Proposition \ref{PROP:factorization_property1} by the definition of $\mathsf{S}^q_\eta(\cdot|\cdot)$;\\
(c) is because $(V,\mathbf{Y}_{22})-\mathbf{X}_1-(\mathbf{Y}_{11},\mathbf{Y}_{21})$ and $(V,\mathbf{Y}_{11})-\mathbf{X}_2-(\mathbf{Y}_{12},\mathbf{Y}_{22})$ form Markov chains;\\
(d) follows by Remark \ref{REM:Nesterd_Envelope_Properties} due to the Markov chains $\mathbf{Y}_{22}-\mathbf{X}_1-(\mathbf{Y}_{11},\mathbf{Y}_{21})$ and $\mathbf{Y}_{11}-\mathbf{X}_2-(\mathbf{Y}_{12},\mathbf{Y}_{22})$.

Now for $(V^\star,\mathbf{X}_1^\star,\mathbf{X}_2^\star)$, an end-to-end equality holds in \eqref{EQ:fractorization_proof_ineq}. In particular, this implies that $I(\mathbf{Y}^\star_{11};\mathbf{Y}^\star_{22}|V^\star)=0$, i.e., that $\mathbf{Y}^\star_{11}-V^\star-\mathbf{Y}^\star_{22}$ forms a Markov chain. By Proposition 2 in \cite{Chandra_Gauss_BC2014}, we have that $\mathbf{X}^\star_{1}-V^\star-\mathbf{X}^\star_2$, which further implies the Markov chain
\begin{equation}
(\mathbf{Y}^\star_{11},\mathbf{Y}^\star_{21})-\mathbf{X}^\star_1-V^\star-\mathbf{X}^\star_2-(\mathbf{Y}^\star_{12},\mathbf{Y}^\star_{22}).\label{EQ:factorization_proof_Markov}
\end{equation}
The end-to-end equality in \eqref{EQ:fractorization_proof_ineq} also gives
\begin{align*}
&\mathsf{T}_{\bm{\lambda},\eta}^{q_1}(\mathbf{X}^\star_1)\\
&=\lambda_1I(\mathbf{X}^\star_1;\mathbf{Y}^\star_{11}|V^\star,\mathbf{Y}^\star_{22})-(\lambda_1+\lambda_2)I(\mathbf{X}^\star_1;\mathbf{Y}^\star_{21}|V^\star,\mathbf{Y}^\star_{22})\\
&\mspace{285mu}+\lambda_1\mathsf{S}_\eta^{q_1}(\mathbf{X}^\star_1|V^\star,\mathbf{Y}^\star_{22})\\
&\stackrel{(a)}=\lambda_1I(\mathbf{X}^\star_1;\mathbf{Y}^\star_{11}|V^\star)-(\lambda_1+\lambda_2)I(\mathbf{X}^\star_1;\mathbf{Y}^\star_{21}|V^\star)\\
&\mspace{265mu}+\lambda_1\mathsf{S}_\eta^{q_1}(\mathbf{X}^\star_1|V^\star),\numberthis\label{factorization_proof_equality}
\end{align*}
where (a) follows because \eqref{EQ:factorization_proof_Markov} implies that the chain $\mathbf{Y}^\star_{22}-V^\star-\mathbf{X}^\star_1-(\mathbf{Y}^\star_{11},\mathbf{Y}^\star_{21})$ is Markov. Similarly, it can be shown that $\mathsf{T}_{\bm{\lambda},\eta}^{q_2}(\mathbf{X}^\star_2)=\mathsf{t}_{\bm{\lambda},\eta}^{q_2}(\mathbf{X}^\star_2|V^\star)$.


\subsection{Proof of Proposition \ref{PROP:factorization_property3}}\label{SUBSEC:factorization_property3_proof}

Let $V-(\mathbf{X}_1,\mathbf{X}_2)-(\mathbf{Y}_{11},\mathbf{Y}_{12},\mathbf{Y}_{21},\mathbf{Y}_{22})$ be a Markov chain. We have,
\begin{align*}
&\mathsf{f}_{\bm{\lambda_0},\alpha,\eta}^{q_1\times q_2}(\mathbf{X}_1,\mathbf{X}_2|V)\\
&=(\lambda_2-\bar{\alpha}\lambda_0)I(\mathbf{X}_1,\mathbf{X}_2;\mathbf{Y}_{21},\mathbf{Y}_{22})\\
&\quad\quad\quad-\alpha\lambda_0I(\mathbf{X}_1,\mathbf{X}_2;\mathbf{Y}_{11},\mathbf{Y}_{12})+\mathsf{T}_{\bm{\lambda},\eta}^{q_1\times q_2}(\mathbf{X}_1,\mathbf{X}_2|V)\\
&\stackrel{(a)}=(\lambda_2-\bar{\alpha}\lambda_0)I(\mathbf{X}_1;\mathbf{Y}_{21}|V,\mathbf{Y}_{22})-\alpha\lambda_0I(\mathbf{X}_1;\mathbf{Y}_{11}|V,\mathbf{Y}_{22})\\
&\mspace{15mu}-\alpha\lambda_0I(\mathbf{Y}_{11};\mathbf{Y}_{22}|V)+(\lambda_2-\bar{\alpha}\lambda_0)I(\mathbf{X}_2;\mathbf{Y}_{22}|V,\mathbf{Y}_{11})\\
&\mspace{30mu}-\alpha\lambda_0I(\mathbf{X}_2;\mathbf{Y}_{21}|V,\mathbf{Y}_{11})+(\lambda_2-\bar{\alpha}\lambda_0)I(\mathbf{Y}_{11};\mathbf{Y}_{22}|V)\\
&\quad\quad\quad\quad\quad\quad\quad\quad\quad\quad\quad\quad\quad\quad\quad\quad\mspace{-2mu}+\mathsf{T}_{\bm{\lambda},\eta}^{q_1\times q_2}(\mathbf{X}_1,\mathbf{X}_2|V)\\
&\stackrel{(b)}\leq(\lambda_2-\bar{\alpha}\lambda_0)I(\mathbf{X}_1;\mathbf{Y}_{21}|V,\mathbf{Y}_{22})-\alpha\lambda_0I(\mathbf{X}_1;\mathbf{Y}_{11}|V,\mathbf{Y}_{22})\\
&\quad\quad+\mathsf{T}_{\bm{\lambda},\eta}^{q_1}(\mathbf{X}_1|V,\mathbf{Y}_{22})+(\lambda_2-\bar{\alpha}\lambda_0)I(\mathbf{X}_2;\mathbf{Y}_{22}|V,\mathbf{Y}_{11})\\
&\quad\quad\quad\quad\quad\mspace{10mu}-\alpha\lambda_0I(\mathbf{X}_2;\mathbf{Y}_{21}|V,\mathbf{Y}_{11})+\mathsf{T}_{\bm{\lambda},\eta}^{q_2}(\mathbf{X}_2|V,\mathbf{Y}_{11})\\
&\quad\quad\quad\quad\quad\quad\quad\quad\quad\quad\quad\quad\quad\mspace{7mu}-(\lambda_0-\lambda_2)I(\mathbf{Y}_{11};\mathbf{Y}_{22}|V)\\
&\stackrel{(c)}\leq \mathsf{F}_{\bm{\lambda_0},\alpha,\eta}^{q_1}(\mathbf{X}_1|\mathbf{Y}_{22})+\mathsf{F}_{\bm{\lambda_0},\alpha,\eta}^{q_2}(\mathbf{X}_2|\mathbf{Y}_{11})\\
&\quad\quad\quad\quad\quad\quad\quad\quad\quad\quad\quad\quad\quad\mspace{7mu}-(\lambda_0-\lambda_2)I(\mathbf{Y}_{11};\mathbf{Y}_{22}|V)\\
&\stackrel{(d)}\leq \mathsf{F}_{\bm{\lambda_0},\alpha,\eta}^{q_1}(\mathbf{X}_1)+\mathsf{F}_{\bm{\lambda_0},\alpha,\eta}^{q_2}(\mathbf{X}_2)-(\lambda_0-\lambda_2)I(\mathbf{Y}_{11};\mathbf{Y}_{22}|V)\\
&\leq \mathsf{F}_{\bm{\lambda_0},\alpha,\eta}^{q_1}(\mathbf{X}_1)+\mathsf{F}_{\bm{\lambda_0},\alpha,\eta}^{q_2}(\mathbf{X}_2),\numberthis\label{EQ:fractorization_proof_double_ineq}
\end{align*}
where:\\
(a) is similar to step (a) in the proof of Proposition \ref{PROP:factorization_property2};\\
(b) uses Proposition \ref{PROP:factorization_property2} and the definition of $\mathsf{T}_{\bm{\lambda},\eta}^q(\cdot|\cdot)$;\\
(c) is because $(V,\mathbf{Y}_{22})-\mathbf{X}_1-(\mathbf{Y}_{11},\mathbf{Y}_{21})$ and $(V,\mathbf{Y}_{11})-\mathbf{X}_2-(\mathbf{Y}_{12},\mathbf{Y}_{22})$ form Markov chains;\\
(d) follows by Remark \ref{REM:Double_Nesterd_Envelope_Properties} due to the Markov chains $\mathbf{Y}_{22}-\mathbf{X}_1-(\mathbf{Y}_{11},\mathbf{Y}_{21})$ and $\mathbf{Y}_{11}-\mathbf{X}_2-(\mathbf{Y}_{12},\mathbf{Y}_{22})$.

For $(V^\star,\mathbf{X}_1^\star,\mathbf{X}_2^\star)$ that satisfy \eqref{EQ:factorization_property3_equality}, an end-to-end equality holds in \eqref{EQ:fractorization_proof_double_ineq}, implying that $I(\mathbf{Y}^\star_{11};\mathbf{Y}^\star_{22}|V^\star)=0$. Invoking Proposition 2 from \cite{Chandra_Gauss_BC2014}, we deduce that $\mathbf{X}^\star_{1}-V^\star-\mathbf{X}^\star_2$ forms a Markov chain, which further implies the Markov chain
\begin{equation}
(\mathbf{Y}^\star_{11},\mathbf{Y}^\star_{21})-\mathbf{X}^\star_1-V^\star-\mathbf{X}^\star_2-(\mathbf{Y}^\star_{12},\mathbf{Y}^\star_{22}).\label{EQ:factorization_proof_double_Markov}
\end{equation}
By the end-to-end equality in \eqref{EQ:fractorization_proof_ineq}, we also have
\begin{align*}
&\mathsf{F}_{\bm{\lambda_0},\alpha,\eta}^{q_1}(\mathbf{X}^\star_1)\\
&=(\lambda_2\mspace{-2mu}-\mspace{-2mu}\bar{\alpha}\lambda_0)I(\mathbf{X}^\star_1;\mspace{-2.5mu}\mathbf{Y}^\star_{21}|V^\star\mspace{-4mu},\mspace{-1.5mu}\mathbf{Y}^\star_{22})\mspace{-2mu}-\mspace{-2mu}\alpha\lambda_0I(\mathbf{X}^\star_1;\mspace{-2.5mu}\mathbf{Y}^\star_{11}|V^\star\mspace{-4mu},\mspace{-1.5mu}\mathbf{Y}^\star_{22})\\
&\quad\quad\quad\quad\quad\quad\quad\quad\quad\quad\quad\quad\quad\quad\quad\mspace{-3mu}+\mathsf{T}_{\bm{\lambda},\alpha,\eta}^{q_1}(\mathbf{X}^\star_1|V^\star,\mathbf{Y}^\star_{22})\\
&\stackrel{(a)}=(\lambda_2-\bar{\alpha}\lambda_0)I(\mathbf{X}^\star_1;\mathbf{Y}^\star_{21}|V^\star)-\alpha\lambda_0I(\mathbf{X}^\star_1;\mathbf{Y}^\star_{11}|V^\star)\\
&\quad\quad\quad\quad\quad\quad\quad\quad\quad\quad\quad\quad\quad\quad\mspace{10mu}+\mathsf{T}_{\bm{\lambda},\alpha,\eta}^{q_1}(\mathbf{X}^\star_1|V^\star),\numberthis\label{factorization_proof_double_equality}
\end{align*}
where (a) uses \eqref{EQ:factorization_proof_double_Markov}, which implies that $\mathbf{Y}^\star_{22}-V^\star-\mathbf{X}^\star_1-(\mathbf{Y}^\star_{11},\mathbf{Y}^\star_{21})$ forms a Markov chain. Similarly, it can be shown that $\mathsf{F}_{\bm{\lambda_0},\alpha,\eta}^{q_2}(\mathbf{X}^\star_2)=\mathsf{f}_{\bm{\lambda_0},\alpha,\eta}^{q_2}(\mathbf{X}^\star_2|V^\star)$.


\subsection{Proof of Proposition \ref{PROP:existence_of_maximizer}}\label{SUBSEC:maximizer_exists}

A key arguments in the proof of Proposition \ref{PROP:existence_of_maximizer} is the continuity of the nested UCE $\mathsf{T}_{\bm{\lambda},\eta}^q(\mathbf{X})=\Big(\mathfrak{C}\mathsf{t}_{\bm{\lambda},\eta}^q\Big)(\mathbf{X})$ in $P_\mathbf{X}$. We shall establish this continuity using Proposition 21 from \cite{Chandra_Gauss_BC2014}, which we reproduced as follows.

\begin{proposition}[Boundedness and Continuity of UCE]\label{PROP:contiuity_nested_envelope}
Consider the space of all Borel probability distributions on $\mathbb{R}^t$ endowed with the topology induced by weak convergence.\footnote{A sequence $\big\{X_n\big\}_{n\in\mathbb{N}}$ of real-valued random variables is said to \emph{converge weakly} or, equivalently, converge in distribution to a random variable $X$ if $\lim_{n\to\infty}F_{X_n}(x)=F(x)$, for every $x\in\mathbb{R}$ for which $F_X$ is continuous, where $F_{X_n}$ and $F_X$ are the cumulative distribution functions of $X_n$ and $X$, respectively. This notion of convergence is denoted by $X_n\xrightarrow[n\to\infty]{\mathcal{D}}X$} Let $\big\{\mathbf{X}_n\big\}_{n\in\mathbb{N}}$ be a sequence of random variable that satisfies the following two properties: (i) $\exists\ p>1$, $B\in\mathbb{R}$ such that $\mathbb{E}\|\mathbf{X}_n\|^p\leq B$, $\forall n\in\mathbb{N}$ (i.e., the sequence has a uniformly bounded $p$th moment); (ii) $\mathbf{X}_n\xrightarrow[n\to\infty]{\mathcal{D}}\mathbf{X}^\star$. If $g:\mathbb{R}^t\to\mathbb{R}$ is a bounded real-valued function that satisfies $g(\mathbf{X}_n)\xrightarrow[n\to\infty]{}g(\mathbf{X}^\star)$, then its UCE $G=\mathfrak{C}g$ is bounded and satisfies $G(\mathbf{X}_n)\xrightarrow[n\to\infty]{} G(\mathbf{X}^\star)$.
\end{proposition}

Before proving Proposition \ref{PROP:existence_of_maximizer} at the end of this subsection, we first verify that Proposition \ref{PROP:contiuity_nested_envelope} applies to  $\mathsf{T}_{\bm{\lambda},\eta}^q(\mathbf{X})$. This is done by showing that $\mathsf{t}_{\bm{\lambda},\eta}^q(\mathbf{X})$ is bounded and continuous, as stated in the subsequent Lemma \ref{LEMMA:nested_envelope_bounded} and \ref{LEMMA:nested_envelope_continous}. The lemmas are proven, respectively, in Appendices \ref{APPEN:nested_envelope_bounded_proof} and \ref{APPEN:nested_envelope_continous_proof}.

\begin{lemma}[Boundedness of Nested Concave Envelopes]\label{LEMMA:nested_envelope_bounded}
For $\eta>1$ and $\lambda_1,\lambda_2>0$ there is a $B_{\bm{\lambda},\eta}\in\mathbb{R}$, such that $\mathsf{t}_{\bm{\lambda},\eta}^q(\mathbf{X})\leq B_{\bm{\lambda},\eta}$, for all $P_{\mathbf{X}}$.
\end{lemma}

\begin{lemma}[Continuity of Nested Concave Envelopes]\label{LEMMA:nested_envelope_continous}
Consider the space of all Borel probability distributions on $\mathbb{R}^t$ endowed with the topology induced by weak convergence and let $\big\{\mathbf{X}_n\big\}_{n\in\mathbb{N}}$ be a sequence of random variable that satisfies the following two properties: (i) $\exists\ p>1$, $B\in\mathbb{R}$ such that $\mathbb{E}\|\mathbf{X}_n\|^p\leq B$, $\forall n$; (ii) $\mathbf{X}_n\xrightarrow[n\to\infty]{\mathcal{D}}\mathbf{X}^\star$. Then $\mathsf{t}_{\bm{\lambda},\eta}^q(\mathbf{X}_n)\xrightarrow[n\to\infty]{} \mathsf{t}_{\bm{\lambda},\eta}^q(\mathbf{X}^\star)$.
\end{lemma}

Based on Lemmas \ref{LEMMA:nested_envelope_bounded} and \ref{LEMMA:nested_envelope_continous}, Proposition 21 from \cite{Chandra_Gauss_BC2014} states that $\mathsf{T}_{\bm{\lambda},\eta}^q(\mathbf{X}_n)$ is bounded and that it satisfies
$\mathsf{T}_{\bm{\lambda},\eta}^q(\mathbf{X}_n)\xrightarrow[n\to\infty]{} \mathsf{T}_{\bm{\lambda},\eta}^q(\mathbf{X}^\star)$. The existence of a unique maximizer of $\tilde{V}_{\bm{\lambda_0},\alpha,\eta}(\mathrm{K})$ is  established as follows. Let $\hat{\mathrm{K}}\succeq 0$ and define
\begin{align*}
&\tilde{\mathrm{v}}_{\bm{\lambda_0},\alpha,\eta}(\hat{\mathrm{K}})\\
&\triangleq \mspace{-5mu}\sup_{\mathbf{X}:\ \mathbb{E}[\mathbf{X}\mathbf{X}^\top]=\hat{\mathrm{K}}}\mspace{-10mu}\mathsf{f}_{\bm{\lambda_0},\alpha,\eta}(\mathbf{X})\\
&=\mspace{-5mu}\sup_{\mathbf{X}:\ \mathbb{E}[\mathbf{X}\mathbf{X}^\top]=\hat{\mathrm{K}}}\mspace{-10mu}(\lambda_2\mspace{-2.5mu}-\mspace{-2.5mu}\bar{\alpha}\lambda_0)I(\mathbf{X};\mspace{-2.5mu}\mathbf{Y}_2\mspace{-1.5mu})\mspace{-2.5mu}-\mspace{-2.5mu}\alpha\lambda_0I(\mathbf{X};\mspace{-2.5mu}\mathbf{Y}_1\mspace{-1.5mu})\mspace{-2.5mu}+\mspace{-2.5mu}\mathsf{T}_{\bm{\lambda},\eta}^q(\mathbf{X})\numberthis\label{EQ:exsists_maximizer_smallv}.
\end{align*}
Let $\big\{\mathbf{X}_n\big\}_{n\in\mathbb{N}}$ be a sequence of random variables with $\mathbb{E}\big[\mathbf{X}_n\mathbf{X}_n^\top\big]=\hat{\mathrm{K}}$, such that $\mathsf{f}_{\bm{\lambda_0},\alpha,\eta}(\mathbf{X}_n)\uparrow\tilde{\mathrm{v}}_{\bm{\lambda_0},\alpha,\eta}(\hat{\mathrm{K}})$, as $n\to\infty$. By \cite[Proposition 17]{Chandra_Gauss_BC2014} and since $\mathbb{E}\big[\mathbf{X}_n\mathbf{X}_n^\top\big]=\hat{\mathrm{K}}$ for every $n\in\mathbb{N}$, we have that $\big\{\mathbf{X}_n\big\}_{n\in\mathbb{N}}$ is a tight sequence\footnote{As defined in \cite{Chandra_Gauss_BC2014}, a sequence of random variables $\big\{\mathbf{X}_n\big\}_{n\in\mathbb{N}}$ taking values in $\mathbb{R}^t$ is \emph{tight} if for every $\epsilon>0$ there exists a compact set $\mathcal{C}_\epsilon\subset \mathbb{R}^t$, such that $\mathbb{P}\big(\mathbf{X}_n\notin\mathcal{C}_\epsilon\big)\leq \epsilon$, $\forall n\in\mathbb{N}$.}, and that there exist an $\mathbf{X}_{\hat{\mathrm{K}}}^\star$ and a convergent subsequence $\big\{\mathbf{X}_{n_m}\big\}_{m\in\mathbb{N}}$ such that $\mathbf{X}_{n_m}\xrightarrow[m\to\infty]{\mathcal{D}}\mathbf{X}_{\hat{\mathrm{K}}}^\star$. Invoking \cite[Proposition 18]{Chandra_Gauss_BC2014} once more we have that $h\left(\mathbf{Y}_{j,n_m}\right)\xrightarrow[m\to\infty]{} h\left(\mathbf{Y}_{j,\hat{\mathrm{K}}}^\star\right)$, for $j=1,2$, where $\mathbf{Y}_{1,n_m},\mathbf{Y}_{2,n_m},\mathbf{Y}_{1,\hat{\mathrm{K}}}^\star$ and $\mathbf{Y}_{2,\hat{\mathrm{K}}}^\star$ are the corresponding outputs. Thus,
\begin{equation}
\mathsf{f}_{\bm{\lambda_0},\alpha,\eta}(\mathbf{X}_{\hat{\mathrm{K}}}^\star)=\tilde{\mathrm{v}}_{\bm{\lambda_0},\alpha,\eta}(\hat{\mathrm{K}}).\label{EQ:exsists_maximizer_smallv_equality}
\end{equation}
By the definition of $\tilde{V}^q_{\bm{\lambda_0},\alpha,\eta}(\mathrm{K})$, we write
\begin{align*}
\tilde{V}_{\bm{\lambda_0},\alpha,\eta}^q(\mathrm{K})&= \sup_{\substack{(V,\mathbf{X}):\ \mathbb{E}[\mathbf{X}\mathbf{X}^\top]\preceq \mathrm{K}, \\ V-\mathbf{X}-(\mathbf{Y}_1,\mathbf{Y}_2)}} \mathsf{f}_{\bm{\lambda_0},\alpha,\eta}^q(\mathbf{X}|V)\\
&=\sup_{\substack{(V,\mathbf{X}):\ \mathbb{E}[\mathbf{X}\mathbf{X}^\top]\preceq \mathrm{K}, \\ V-\mathbf{X}-(\mathbf{Y}_1,\mathbf{Y}_2)}} \sum_vP(v)\mathsf{f}_{\bm{\lambda_0},\alpha,\eta}^q(\mathbf{X}|V=v).\numberthis
\end{align*}
Since $\tilde{V}_{\bm{\lambda_0},\alpha,\eta}^q(\mathrm{K})$ is a convex combination as above, to obtain the maximizer subject to the covariance constraint it suffices to restrict attention to the family of maximizers $\mathbf{X}_{\hat{\mathrm{K}}}^\star$, for $K\succeq\mathrm{0}$. Thus,
\begin{equation}
\tilde{V}_{\bm{\lambda_0},\alpha,\eta}^q(\mathrm{K})= \sup_{\substack{\{\alpha_i\},\{\hat{\mathrm{K}}_i\}:\ \alpha_i\geq 0\\ \sum_i\alpha_i=1,\ \sum_i\alpha_i\hat{\mathrm{K}}_i\preceq \mathrm{K}}}\sum_i\alpha_i\tilde{\mathrm{v}}_{\bm{\lambda_0},\alpha,\eta}^q(\hat{\mathrm{K}}).
\end{equation}
It takes $\frac{t(t+1)}{2}$ constraints to preserve the covariance matrix (due to its symmetry) and one other constraint to preserve $\sum_i\alpha_i\tilde{\mathrm{v}}_{\bm{\lambda_0},\alpha,\eta}^q(\hat{\mathrm{K}})$. Hence, by using the Bunt-Carathedory theorem \cite{Bunt_carathodory1934}, we can restrict ourselves to convex combinations of at most $m\triangleq \frac{t(t+1)}{2}+1$ points, i.e.,
\begin{equation}
\tilde{V}_{\bm{\lambda_0},\alpha,\eta}^q(\mathrm{K})= \sup_{\substack{\{\alpha_i\},\{\hat{\mathrm{K}}_i\}:\ \alpha_i\geq 0\\ \sum_{i=1}^m\alpha_i=1,\ \sum_{i=1}^m\alpha_i\hat{\mathrm{K}}_i\preceq \mathrm{K}}}\sum_{i=1}^m\alpha_i\tilde{\mathrm{v}}_{\bm{\lambda_0},\alpha,\eta}^q(\hat{\mathrm{K}}).
\end{equation}

Consider any sequence of convex combinations $\Big\{\big\{\alpha_i^{(n)}\big\}_{i\in[m]},\big\{\mathrm{K}_i^{(n)}\big\}_{i\in[m]}\Big\}_{n\in\mathbb{N}}$ that approaches the supremum as $n\to\infty$. The compactness of the $m$-dimensional simplex implies that $\alpha_i^{(n)}\xrightarrow[n\to\infty]{} \alpha_i^\star$, for all $i\in[m]$. Furthermore, we have the following property of the limiting points $\alpha_i^\star$; see Appendix \ref{APPEN:exsists_limits_notzero_proof} for the proof.

\begin{lemma}\label{LEMMA:exsists_limits_notzero}
For any $i\in[m]$, if $\alpha_i^\star=0$ then $\alpha_i^{(n)}\tilde{\mathrm{v}}_{\bm{\lambda_0},\alpha,\eta}^q\big(\mathrm{K}^{(n)}_i\big)\xrightarrow[n\to\infty]{}0$.
\end{lemma}

Based on Lemma \ref{LEMMA:exsists_limits_notzero}, we assume that $\alpha^\star\triangleq\min_{i\in[m]}\alpha_i^\star>0$, which implies that ${K}_i^{(n)}\preceq\frac{2}{\alpha^\star}\mathrm{K}$ uniformly in $i\in[m]$, for large enough values of $n$. Hence, for each $i\in[m]$ we can find a convergent subsequence $\big\{\mathrm{K}_i^{(n_k)}\big\}_{k\in\mathbb{N}}$, such that $\mathrm{K}_i^{(n_k)}\xrightarrow[k\to\infty]{}\mathrm{K}_i^\star$. Putting these
together, we have
\begin{equation}
\tilde{V}_{\bm{\lambda_0},\alpha,\eta}^q(\mathrm{K})=\sum_{i=1}^m\alpha_i^\star\tilde{\mathrm{v}}_{\bm{\lambda_0},\alpha,\eta}^q(\mathrm{K}_i^\star),
\end{equation}
i.e., one can always find a pair of random variables $(V^\star,\mathbf{X}^\star)$ with $|\mathcal{V}^\star|\leq\frac{t(t+1)}{2}+1$, such that $\tilde{V}_{\bm{\lambda_0},\alpha,\eta}^q(\mathrm{K})=\mathsf{f}_{\bm{\lambda_0},\alpha,\eta}^q(\mathbf{X}^\star|V^\star)$.


\subsection{Proof of Proposition \ref{PROP:rotation_invariant}}\label{SUBSEC:rotation_invariant_proof}

Denote $q_{\mathbf{Y}_1,\mathbf{Y}_2|\mathbf{X}}\triangleq q$ and consider the two-letter BC $q(\mathbf{y}_{11},\mathbf{y}_{21}|\mathbf{x}_1)\times q(\mathbf{y}_{12},\mathbf{y}_{22}|\mathbf{x}_2)$. We have
\begin{align*}
2\tilde{V}_{\bm{\lambda_0},\alpha,\eta}^q(\mathrm{K})&\stackrel{(a)}=\mathsf{f}_{\bm{\lambda_0},\alpha,\eta}^q(\mathbf{X}_1|V_1)+\mathsf{f}_{\bm{\lambda_0},\alpha,\eta}^q(\mathbf{X}_2|V_2)\\
                                             &\stackrel{(b)}=\mathsf{f}_{\bm{\lambda_0},\alpha,\eta}^{q\times q}(\mathbf{X}_1,\mathbf{X}_2|V_1,V_2)\\
                                             &\stackrel{(c)}=\mathsf{f}_{\bm{\lambda_0},\alpha,\eta}^{q\times q}(\mathbf{X}_{\theta_1},\mathbf{X}_{\theta_2}|\tilde{V})\\
                                             &\stackrel{(d)}\leq \mathsf{F}_{\bm{\lambda_0},\alpha,\eta}^{q\times q}(\mathbf{X}_{\theta_1},\mathbf{X}_{\theta_2})\\
                                             &\stackrel{(e)}\leq \mathsf{F}_{\bm{\lambda_0},\alpha,\eta}^q(\mathbf{X}_{\theta_1})+\mathsf{F}_{\bm{\lambda_0},\alpha,\eta}^q(\mathbf{X}_{\theta_2})\\
                                             &\stackrel{(f)}\leq \tilde{V}_{\bm{\lambda_0},\alpha,\eta}^q(\mathrm{K})+\tilde{V}_{\bm{\lambda_0},\alpha,\eta}^q(\mathrm{K})=2\tilde{V}_{\bm{\lambda_0},\alpha,\eta}^q(\mathrm{K}),\numberthis\label{EQ:rotation_invariant_endtoend}
\end{align*}
where:\\
(a) is because $P_{V,\mathbf{X}}^\star$ achieves $\tilde{V}_{\bm{\lambda_0},\alpha,\eta}^q(\mathrm{K})$;\\
(b) uses the independence of $(V_1,\mathbf{X}_1)$ and $(V_2,\mathbf{X}_2)$;\\
(c) is a consequence of \cite[Proposition 1]{Chandra_Gauss_BC2014} (namely, the invariance with respect to rotation of the mutual information between the input and output of an additive Gaussian channel);\\
(d) follows by the definition of the double-nested UCE;\\
(e) uses Proposition \ref{PROP:factorization_property3};\\
(f) is the definition of $\tilde{V}_{\bm{\lambda_0},\alpha,\eta}^q(\mathrm{K})$, while noting that the independence of $\mathbf{X}_{v_1}$ and $\mathbf{X}_{v_2}$, for every $v_1\neq v_2\in\mathcal{V}$, implies that
\begin{align*}
\mathbb{E}\big[\mathbf{X}_{\theta_1}\mathbf{X}_{\theta_1}^\top\big]&=\mathbb{E}\Bigg[\frac{1}{2}\mathbb{E}\Big[\mathbf{X}_{V_1}\mathbf{X}_{V_1}^\top\Big|\tilde{V}\Big]+\frac{1}{2}\mathbb{E}\Big[\mathbf{X}_{V_2}\mathbf{X}_{V_2}^\top\Big|\tilde{V}\Big]\Bigg]\\
&=\mathbb{E}\big[\mathbf{X}_{\theta_2}\mathbf{X}_{\theta_2}^\top\big]\numberthis
\end{align*}
and
\begin{align*}
\mathbb{E}\big[\mathbf{X}_{\theta_2}\mathbf{X}_{\theta_2}^\top\big]=\sum_v P_V^\star(v)\mathrm{K}_v=\mathbb{E}\Big[\mathbb{E}\big[\mathbf{X}\mathbf{X}^\top\big|V\big]\Big]\preceq\mathrm{K},\label{EQ:rotation_invariant_cov}
\end{align*}
where we have denoted $\mathrm{K}_v\triangleq \mathbb{E}\big[\mathbf{X}_v\mathbf{X}_v^\top\big]$.

Since the extremes of the chain of inequalities in \eqref{EQ:rotation_invariant_endtoend} match, all inequalities are, in fact, equalities. Equality in step (d) implies that $P_{\tilde{V}|\mathbf{X}_{\theta_1},\mathbf{X}_{\theta_2}}$ achieves $\mathsf{F}_{\bm{\lambda_0},\alpha,\eta}^{q\times q}(\mathbf{X}_{\theta_1},\mathbf{X}_{\theta_2})$. Furthermore, by Proposition \ref{PROP:factorization_property3}, since (d) and (e) are equalities we have that $\mathbf{X}_{\theta_1}-\tilde{V}-\mathbf{X}_{\theta_2}$, and that $P_{\tilde{V}|\mathbf{X}_{\theta_1}}$ and $P_{\tilde{V}|\mathbf{X}_{\theta_2}}$ achieve $\mathsf{F}_{\bm{\lambda_0},\alpha,\eta}^q(\mathbf{X}_{\theta_1})$ and $\mathsf{F}_{\bm{\lambda_0},\alpha,\eta}^q(\mathbf{X}_{\theta_1})$, respectively. Finally, equality in (f) means that $\tilde{V}_{\bm{\lambda_0},\alpha,\eta}^q(\mathrm{K})=\mathsf{F}_{\bm{\lambda_0},\alpha,\eta}^q(\mathbf{X}_{\theta_j}|\tilde{V})=\mathsf{f}_{\bm{\lambda_0},\alpha,\eta}^q(\mathbf{X}_{\theta_j}|\tilde{V})$, for $j=1,2$.


\subsection{Proof of Theorem \ref{TM:Gaussian_maximizer_Envelope_pre}}\label{SUBSEC:Gaussian_maximizer_Envelope_pre_proof}

As a consequence of Proposition \ref{PROP:rotation_invariant}, for any fixed $(v_1,v_2)\in\mathcal{V}^2$, $\mathbf{X}_{v_1}+\mathbf{X}_{v_2}$ and $\mathbf{X}_{v_1}-\mathbf{X}_{v_2}$ are independent. Combined with $\mathbf{X}_{v_1}$ and $\mathbf{X}_{v_2}$ being independent zero mean random variables, Corollary 3 in Appendix I-A of \cite{Chandra_Gauss_BC2014} implies that $\mathbf{X}_{v_1}$ and $\mathbf{X}_{v_2}$ are Gaussian random vectors with the same covariance matrix. Since the pair $(v_1,v_2)\in\mathcal{V}^2$ is arbitrary, we see that all Gaussian vectors $\{\mathbf{X}_v\}_{v\in\mathcal{V}}$ have the same covariance matrix. Furthermore, we may assume that $\{\mathbf{X}_v\}_{v\in\mathcal{V}}$ are all centered, and therefore, we get that this is an i.i.d. set of Gaussian random variables. Denoting this common covariance matrix by $\mathrm{K}^\star$, it clearly satisfies $\mathrm{K}^\star\preceq\mathrm{K}$. Letting $\mathbf{X}^\star\sim\mathcal{N}(\mathbf{0},\mathrm{K}^\star)$, we have
\begin{align*}
\tilde{V}_{\bm{\lambda_0},\alpha,\eta}^q(\mathrm{K})&\stackrel{(a)}=\mathsf{f}_{\bm{\lambda_0},\alpha,\eta}^q(\mathbf{X}|V)\\
&\stackrel{(b)}=\sum_{i=1}^mP^\star_V(v_i)\mathsf{f}_{\bm{\lambda_0},\alpha,\eta}^q(\mathbf{X}_{v_i})\\
&\stackrel{(c)}=\sum_{i=1}^mP^\star_V(v_i)\mathsf{f}_{\bm{\lambda_0},\alpha,\eta}^q(\mathbf{X}^\star)\\
&=\mathsf{f}_{\bm{\lambda_0},\alpha,\eta}^q(\mathbf{X}^\star),\numberthis
\end{align*}
where (a) follows since $(V,\mathbf{X})\sim P_{V,\mathbf{X}}^\star$ attains $\tilde{V}_{\bm{\lambda_0},\alpha,\eta}^q(\mathrm{K})$ (Proposition \ref{PROP:existence_of_maximizer}), (b) follows by the definition of $\mathbf{X}_v$ in the statement of Proposition \ref{PROP:rotation_invariant}, while (c) follows since $\mathbf{X}^\star$ and $\mathbf{X}_{v_i}$ are identically distributed, for every $i\in[m]$.

To account for the uniqueness of the zero-mean maximizer we first show that if a zero mean random vector $\mathbf{X}$ is a maximizer, i.e., $\tilde{V}_{\bm{\lambda_0},\alpha,\eta}^q(\mathrm{K})=\mathsf{f}_{\bm{\lambda_0},\alpha,\eta}^q(\mathbf{X})$, it must be Gaussian. Let $\mathbf{X}_1$ and $\mathbf{X}_1$ be two i.i.d. copies of $\mathbf{X}$. Applying Proposition \ref{PROP:rotation_invariant} while taking $V$ to be a constant, we obtain that $\mathbf{X}_1+\mathbf{X}_2$ and $\mathbf{X}_1-\mathbf{X}_2$ are also independent. Hence, by \cite[Corollary 3]{Chandra_Gauss_BC2014}, $\mathbf{X}$ is Gaussian.

Next, suppose that $\tilde{V}_{\bm{\lambda_0},\alpha,\eta}^q(\mathrm{K})$ has two independent Gaussian maximizers denoted by $\mathbf{A}_1\sim\mathcal{N}(\mathbf{0},\mathrm{K}_1)$ and $\mathbf{A}_2\sim\mathcal{N}(\mathbf{0},\mathrm{K}_2)$, such that $\mathrm{K}_1,\mathrm{K}_2\preceq$ and $\mathrm{K}_1\neq\mathrm{K}_2$. Let $(V,\mathbf{X})$ be a pair of random variables, such that $V\sim\ber\left(\frac{1}{2}\right)$ on $\mathcal{V}=\{1,2\}$, $\mathbf{X}\big|\big\{V=1\big\}\sim\mathcal{N}(\mathbf{0},\mathrm{K}_1)$ and $\mathbf{X}\big|\big\{V=2\big\}\sim\mathcal{N}(\mathbf{0},\mathrm{K}_2)$. Note that $(V,\mathbf{X})$ also attains $\tilde{V}_{\bm{\lambda_0},\alpha,\eta}^q(\mathrm{K})$. Taking $v_1=1$ and $v_2=2$, Proposition \ref{PROP:rotation_invariant} implies that $\mathbf{A}_1+\mathbf{A}_2$ and $\mathbf{A}_1-\mathbf{A}_2$ are independent, which contradict Corollary 3 from \cite{Chandra_Gauss_BC2014} as $\mathrm{K}_1\neq\mathrm{K}_2$.


\subsection{Proof of Corollary \ref{COR:Gaussian_maximizer_Envelope3}}\label{SUBSEC:Gaussian_maximizer_Envelope3_proof}

By Theorem \ref{TM:Gaussian_maximizer_Envelope_pre}, there is an $\mathbf{X}^\star\sim\mathcal{N}(\mathbf{0},\mathrm{K}^\star)$, such that $\mathrm{K}^\star\preceq\mathrm{K}$ and $\mathsf{f}_{\bm{\lambda_0},\alpha,\eta}^q(\mathbf{X}^\star)=\tilde{V}_{\bm{\lambda_0},\alpha,\eta}^q(\mathrm{K})$. Let $\mathbf{X}'\sim\mathcal{N}(\mathbf{0},\mathrm{K}-\mathrm{K}^\star)$ be independent of $\mathbf{X}^\star$ and set $\mathbf{X}=\mathbf{X}^\star+\mathbf{X}'$. Thus $\mathbf{X}\sim\mathcal{N}(\mathbf{0},\mathrm{K})$ and by definition we have $\mathsf{F}_{\bm{\lambda_0},\alpha,\eta}^q(\mathbf{X})\leq\tilde{V}_{\bm{\lambda_0},\alpha,\eta}^q(\mathrm{K})$.

On the other hand,
\begin{align*}
\mathsf{F}_{\bm{\lambda_0},\alpha,\eta}^q(\mathbf{X})&=\sup_{\substack{P_{V|\mathbf{X}}:\\ V-\mathbf{X}-(\mathbf{Y}_1,\mathbf{Y}_2)}} \mathsf{f}_{\bm{\lambda_0},\alpha,\eta}^q(\mathbf{X}|V)\\
&\stackrel{(a)}\geq\mathsf{f}_{\bm{\lambda_0},\alpha,\eta}^q(\mathbf{X}|\mathbf{X}')\\
&\stackrel{(b)}=\mathsf{f}_{\bm{\lambda_0},\alpha,\eta}^q(\mathbf{X}^\star)\\
&=\tilde{V}_{\bm{\lambda_0},\alpha,\eta}^q(\mathrm{K}),\numberthis\label{EQ:maximizer_prop_ineq1}
\end{align*}
where (a) follows since $\mathbf{X}'-\mathbf{X}-(\mathbf{Y}_1,\mathbf{Y}_2)$ forms a Markov chain, while (b) follows because $\mathbf{X}\big|\big\{\mathbf{X}'=\mathbf{x}'\big\}\sim\mathbf{X}^\star+\mathbf{x}'$. Thus,
\begin{align*}
\tilde{V}&_{\bm{\lambda_0},\alpha,\eta}^q(\mathrm{K})\\
&=\mathsf{F}_{\bm{\lambda_0},\alpha,\eta}^q(\mathbf{X})\\
&=\mathsf{f}_{\bm{\lambda_0},\alpha,\eta}^q(\mathbf{X}^\star)\\
&=(\lambda_2-\bar{\alpha}\lambda_0)I(\mathbf{X}^\star;\mathbf{Y}_2)-\alpha\lambda_0I(\mathbf{X}^\star;\mathbf{Y}_1)+\mathsf{T}_{\bm{\lambda},\eta}^q(\mathbf{X}^\star).\numberthis\label{EQ:maximizer_prop_eq1}
\end{align*}

By Theorem \ref{TM:Gaussian_maximizer_Envelope2}, one can decompose $\mathbf{X}^\star$ into independent $\mathbf{X}^\star_1\sim\mathcal{N}(\mathbf{0},\mathrm{K}_1)$ and $\mathbf{X}^\star_2\sim\mathcal{N}(\mathbf{0},\mathrm{K}_2)$, with $\mathrm{K}_1+\mathrm{K}_2\preceq\mathrm{K}^\star$, such that
\begin{equation}
\mathsf{T}_{\bm{\lambda},\eta}^q(\mathbf{X}^\star)=\mathsf{t}_{\bm{\lambda},\eta}^q(\mathbf{X}^\star_1+\mathbf{X}^\star_2)=\hat{V}_{\bm{\lambda},\eta}^q(\mathrm{K}^\star),\label{EQ:maximizer_prop_eq2}
\end{equation}
and
\begin{equation}
\mathsf{S}_\eta^q(\mathbf{X}^\star_1+\mathbf{X}^\star_2)=\mathsf{s}_\eta^q(\mathbf{X}^\star_1)=V_\eta^q(\mathrm{K}_1+\mathrm{K}_2).\label{EQ:maximizer_prop_eq3}
\end{equation}

The proof of existence is concluded by setting $\mathbf{X}_3^\star\sim\mathcal{N}(\mathbf{0},\mathrm{K}_3)$, where $\mathrm{K}_3=\mathrm{K}^\star-(\mathrm{K}_1+\mathrm{K}_2)$ and noting that $\mathbf{X}=\mathbf{X}^\star_1+\mathbf{X}^\star_2+\mathbf{X}^\star_3+\mathbf{X}'\sim\mathcal{N}(\mathbf{0},\mathrm{K})$ and $\mathbf{X}^\star=\mathbf{X}^\star_1+\mathbf{X}^\star_2+\mathbf{X}^\star_3\sim\mathcal{N}(\mathbf{0},\mathrm{K}^\star)$, which implies that \eqref{EQ:Gaussian_maximizer_Envelope3} holds. The uniqueness of the decomposition (i.e., of the covariance matrices $\mathrm{K}_1$, $\mathrm{K}_2$ and $\mathrm{K}_3$) is a direct consequence of Theorems \ref{TM:Gaussian_maximizer_Envelope1}, \ref{TM:Gaussian_maximizer_Envelope2} and \ref{TM:Gaussian_maximizer_Envelope_pre}.


\section{Summary and Concluding Remarks}\label{SEC:summary}

The two-user MIMO Gaussian BC with common, private and confidential messages was studied. The private message to Receiver 1 is confidential and kept secret from Receiver 2. The secrecy-capacity region without a common message was characterized first and Gaussian inputs were shown to achieve optimality. The proof relied on establishing an equivalence between certain inner and outer bounds using factorization of UCEs \cite{Chandra_Gauss_BC2014} and a variant of DPC \cite{Cioffi_Precoding2004}. Our results showed that using DPC to cancel out the signal of the non-confidential message at Receiver 1 exhausts the entire region, making DPC against the signal of the confidential message unnecessary.

This secrecy-capacity region without a common message was then used to characterize a portion of the region with a common message. The rest of the region was found using double-nested UCEs. The secrecy-capacity region without a common message was illustrated using a numerical example. To make the region (efficiently) computable, matrix decomposition properties from \cite{Khina_WTC_Decomposition2015} were leveraged. The region was shown to be strictly larger than the secrecy-capacity region of the MIMO Gaussian BC with confidential messages (in which each private message is kept secret from the opposite user).

\section*{Acknowledgements}
The authors would like to thank Anatoly Khina for insightful discussions about DPC and his advice regrading implementation of the numerical example. We also thank the Associate Editor and the anonymous reviewers for their suggestions and comments that helped us improve the presentation of this paper. 


\appendices


\section{Derivation of Inner and Outer Bounds}\label{APPEN:inner_outer_proof}


\subsection{Outer Bounds \ref{BND:outer_bound_nocommon} and \ref{BND:outer_bound}}\label{SUBAPPEN:outer_bound_proof}

We first establish Bound \ref{BND:outer_bound} as an outer bound on the secrecy-capacity region of the setting with a common message, and then use it to establish Bound \ref{BND:outer_bound_nocommon} as an outer bound on the region without a common message.

The result of \cite[Theorem 3]{goldfeld2017broadcast} characterizes an outer bound $\mathcal{R}_\mathsf{O}(L_1,L_2)$ on the $(L_1,L_2)$-leakage-capacity region of a DM-BC with common and private messages, for some leakage thresholds $(L_1,L_2)\in\mathbb{R}_+^2$. Setting $L_1=0$ and letting $L_2\to\infty$ in $\mathcal{R}_\mathsf{O}(L_1,L_2)$ (which corresponds to $M_1$ being confidential and $M_2$ not being subject to any secrecy requirements), we have that the closure of the union of rate triples $(R_0,R_1,R_2)\in\mathbb{R}^3_+$ satisfying:
\begin{subequations}
\begin{align}
R_0&\leq \min\big\{I(W;Y_1),I(W;Y_2)\big\}\label{EQ:region_full_outer0}\\
R_1&\leq I(U;Y_1|W,V)-I(U;Y_2|W,V)\label{EQ:region_full_outer11}\\
R_1&\leq I(U;Y_1|W)-I(U;Y_2|W)\label{EQ:region_full_outer12}\\
R_0+R_2&\leq I(V;Y_2|W)+\min\big\{I(W;Y_1),I(W;Y_2)\big\}\label{EQ:region_full_outer23}\\
R_0+R_1+R_2&\leq I(U;Y_1|W)+I(V;Y_2|W,U)\nonumber\\
&\quad\quad\quad\mspace{10mu}+\min\big\{I(W;Y_1),I(W;Y_2)\big\}\label{EQ:region_full_outer_sum2}
\end{align}\label{EQ:region_full_outer}%
\end{subequations}
over all $(W,U,V)-X-(Y_1,Y_2)$, is an outer bound on $\mathcal{C}$. By removing the rate bounds in \eqref{EQ:region_full_outer12} and \eqref{EQ:region_full_outer_sum2} from $\mathcal{R}_\mathsf{O}(0,\infty)$, one recovers the region $\mathcal{O}$ from \eqref{EQ:region_outer}. Clearly $\mathcal{R}_\mathsf{O}(0,\infty)\subseteq \mathcal{O}$, which shows that $\mathcal{C}\subseteq \mathcal{O}$ and establishes Bound \ref{BND:outer_bound}.

When there is no common message, one obtains Bound \ref{BND:outer_bound_nocommon}, i.e., that $\hat{\mathcal{C}}\subseteq\hat{\mathcal{O}}$, by setting $R_0=0$ into Bound \ref{BND:outer_bound}. This follows by noting that
\begin{equation}
I(V;Y_2|W)+\min\big\{I(W;Y_1),I(W;Y_2)\big\}\leq I(W,V;Y_2),
\end{equation}
and defining $\tilde{V}=(W,V)$.


\subsection{Inner Bounds \ref{BND:inner_bound_nocommon} and \ref{BND:inner_bound}}\label{SUBAPPEN:inner_bound_proof}

Referring to \cite[Theorem 1]{goldfeld2017broadcast}, we have $\mathcal{R}_\mathsf{I}(0,\infty)$ as an inner bound on $\mathcal{C}$, where $\mathcal{R}_\mathsf{I}(0,\infty)$ is the closure of the union of rate triples $(R_0,R_1,R_2)\in\mathbb{R}^3_+$ satisfying:
\begin{subequations}
\begin{align}
R_0&\leq \min\big\{I(W;Y_1),I(W;Y_2)\big\}\label{EQ:region_full_inner0}\\
R_1 &\leq I(U;Y_1|W)\mspace{-3mu}-\mspace{-3mu}I(U;V|W)\mspace{-3mu}-\mspace{-3mu}I(U;Y_2|W,V)\label{EQ:region_full_inner11}\\
R_0+R_1 &\leq I(U;Y_1|W)\mspace{-3mu}+\mspace{-3mu}\min\big\{I(W;Y_1),I(W;Y_2)\big\}\label{EQ:region_full_inner12}\\
R_0+R_2 &\leq I(V;Y_2|W)\mspace{-3mu}+\mspace{-3mu}\min\big\{I(W;Y_1),I(W;Y_2)\big\}\label{EQ:region_full_inner23}\\
R_0+R_1+R_2 &\leq I(U;Y_1|W)+I(V;Y_2|W)-I(U;V|W)\\
&\quad\quad\quad\mspace{10mu}+\min\big\{I(W;Y_1),I(W;Y_2)\big\}\label{EQ:region_full_inner_sum3}
\end{align}\label{EQ:region_full_inner}%
\end{subequations}
over all $(W,U,V)-X-(Y_1,Y_2)$. The inclusion $\mathcal{I}\subseteq\mathcal{R}_\mathsf{I}(0,\infty)$ immediately follows by noting that if $(R_1,R_2)$ satisfy  \eqref{EQ:region_inner} then they also satisfy \eqref{EQ:region_full_inner}. A simple consequence of the above is that Bound \ref{BND:inner_bound_nocommon} is an inner bound on the secrecy-capacity region without a common message, which follows by setting $R_0=0$ and $W=0$ into Bound~\ref{BND:inner_bound}.


\section{Proof of Lemma \ref{LEMMA:supporting_hyperplanes}}\label{APPEN:supporting_hyperplanes_proof}

First notice that \eqref{EQ:nocommon_hyper} are supporting hyperplanes of $\hat{\mathcal{O}}_\mathrm{K}$ and that the points $(\mathcal{H}_1^K,0)$ and $(0,\mathcal{H}_2^K)$ are on its boundary. Furthermore, $R_1\geq 0$, $R_2\geq 0$ and $R_2\leq \mathcal{H}_2^K$ are also supporting hyperplanes of $\hat{\mathcal{I}}_\mathrm{K}$, and since $(0,\mathcal{H}_2^K)\in\hat{\mathcal{I}}_\mathrm{K}$, it is a boundary point of $\hat{\mathcal{I}}_\mathrm{K}$. Note that $\mathcal{H}_1^K$ describes the secrecy-capacity of the MIMO Gaussian WTC, where User 1 serves as the legitimate receiver and User 2 as the eavesdropper. Therefore (see \cite{Liu_Shamai_MIMOWTC2009,Khitsi_MIMOWTC2010,Hassibi_MINOWTC2011}),
\begin{equation}
\mathcal{H}_1^K=\frac{1}{2}\max_{\mathrm{0}\preceq\mathrm{K}^\star\preceq\mathrm{K}}\log\left|\mathrm{I}+\mathrm{G}_1\mathrm{K}^\star\mathrm{G}_1^\top\right|-\log\left|\mathrm{I}+\mathrm{G}_2\mathrm{K}^\star\mathrm{G}_2^\top\right|.\label{EQ:nocommon_WTC_hyper}
\end{equation}
To see that $(\mathcal{H}_1^K,0)$ is also in $\hat{\mathcal{I}}_\mathrm{K}$ consider the following. For every $\mathrm{0}\preceq\mathrm{K}^\star\preceq\mathrm{K}$, let $\mathbf{X}_1$ and $\mathbf{X}_2$ be independent Gaussian random vectors with covariances $\mathrm{K}^\star$ and $\mathrm{K}-\mathrm{K}^\star$, respectively. Set
\begin{align*}
U&=\mathbf{X}_1+\mathrm{A}\mathbf{X}_2\\
V&=\mathbf{X}_2\\
\mathbf{X}&=\mathbf{X}_1+\mathbf{X}_2,
\end{align*}
where $\mathrm{A}=\mathrm{K}^\star\mathrm{G}_1^\top\big[I+\mathrm{G}_1\mathrm{K}^\star\mathrm{G}_1^\top\big]^{-1}$ is the precoding matrix for suppressing $V$ from $\mathbf{Y}_1$\cite[Theorem 1]{Cioffi_Precoding2004}. Evaluating the mutual information terms on the RHS of \eqref{EQ:region_inner_nocommon11}, we first have
\begin{equation}
I(U;\mathbf{Y}_1)-I(U;V)=I(\mathbf{X};\mathbf{Y}_1|V)=\frac{1}{2}\log\left|\mathrm{I}+\mathrm{G}_1\mathrm{K}^\star\mathrm{G}_1^\top\right|.\label{EQ:nocommon_WTC_IT1}
\end{equation}
Moreover,
\begin{align*}
I(U;\mathbf{Y}_2|V)&=I\big(\mathbf{X}_1+\mathrm{A}\mathbf{X}_2;\mathrm{G}_2(\mathbf{X}_1+\mathbf{X}_2)+\mathbf{Z}_2\big|\mathbf{X}_2\big)\\
                   &=I(\mathbf{X}_1;\mathrm{G}_2\mathbf{X}_1+\mathbf{Z}_2|\mathbf{X}_2)\\
                   &\stackrel{(a)}=I(\mathbf{X}_1;\mathrm{G}_2\mathbf{X}_1+\mathbf{Z}_2)\\
                   &=\frac{1}{2}\log\left|\mathrm{I}+\mathrm{G}_2\mathrm{K}^\star\mathrm{G}_2^\top\right|,\numberthis\label{EQ:nocommon_WTC_IT2}
\end{align*}
where (a) follows because $(\mathbf{X}_1,\mathbf{Z}_2)$ and $\mathbf{X}_2$ are independent. Combining \eqref{EQ:nocommon_WTC_IT1} with \eqref{EQ:nocommon_WTC_IT2} yields
\begin{align*}
I(U;&\mathbf{Y}_1)-I(U;V)-I(U;\mathbf{Y}_2|V)\\
&=\frac{1}{2}\log\left|\mathrm{I}+\mathrm{G}_1\mathrm{K}^\star\mathrm{G}_1^\top\right|-\frac{1}{2}\log\left|\mathrm{I}+\mathrm{G}_2\mathrm{K}^\star\mathrm{G}_2^\top\right|,\numberthis\label{EQ:nocommon_WTC_WTC_achievable}
\end{align*}
which implies that $(\mathcal{H}_1^K,0)\in\hat{\mathcal{I}}_\mathrm{K}$. Furthermore, since $(\mathcal{H}_1^K,0)$ is on the boundary of $\hat{\mathcal{O}}_\mathrm{K}$ and $\hat{\mathcal{I}}_\mathrm{K}\subseteq\hat{\mathcal{O}}_\mathrm{K}$, $(\mathcal{H}_1^K,0)$ must also be a boundary point of $\hat{\mathcal{I}}_\mathrm{K}$, and therefore, $R_1\leq\mathcal{H}_1^K$ is a supporting hyperplane of $\hat{\mathcal{I}}_\mathrm{K}$.


\section{Proof of Auxiliary Results for Proposition \ref{PROP:existence_of_maximizer} - Lemmas \ref{LEMMA:nested_envelope_bounded} and \ref{LEMMA:nested_envelope_continous}}\label{APPEN:maximizer_exists}


\subsection{Proof of Lemma \ref{LEMMA:nested_envelope_bounded}}\label{APPEN:nested_envelope_bounded_proof}

By Theorem \ref{TM:Gaussian_maximizer_Envelope2}, we have that if $\mathbb{E}\big[\mathbf{X}\mathbf{X}^\top\big]\preceq\mathrm{K}$, then
\begin{equation}
\hat{V}_{\bm{\lambda},\eta}^q(\mathrm{K})=\mathsf{t}_{\bm{\lambda},\eta}^q(\mathbf{X}^\star),\label{EQ:bounded_equality}
\end{equation}
where $\mathbf{X}^\star=\mathbf{X}^\star_1+\mathbf{X}^\star_2$, and $\mathbf{X}^\star_1$ and $\mathbf{X}^\star_2$ are independent random variables with $\mathbf{X}^\star_j\sim\mathcal{N}(\mathbf{0},\mathrm{K}_j)$, for $j=1,2$, such that $\mathrm{K}^\star\triangleq \mathrm{K}_1+\mathrm{K}_2\preceq\mathrm{K}$. Furthermore, by the definition of the UCE and the definition of $\hat{V}_{\bm{\lambda},\eta}^q(\mathrm{K})$, \eqref{EQ:bounded_equality} implies that for every $\mathbf{X}\sim P_{\mathbf{X}}$ with $\mathbb{E}\big[\mathbf{X}\mathbf{X}^\top\big]\preceq\mathrm{K}$, we have
\begin{equation}
\mathsf{t}_{\bm{\lambda},\eta}^q(\mathbf{X})\leq \mathsf{T}_{\bm{\lambda},\eta}^q(\mathbf{X})\leq\hat{V}_{\bm{\lambda},\eta}^q(\mathrm{K})=\mathsf{t}_{\bm{\lambda},\eta}^q(\mathbf{X}^\star).\label{EQ:bounded_inequality}
\end{equation}
Thus,
\begin{align*}
&\sup_{\mathbf{X}:\ \mathbb{E}[\mathbf{X}\mathbf{X}^\top]\preceq\mathrm{K}}\mathsf{t}_{\bm{\lambda},\eta}^q(\mathbf{X})\\
&\leq \lambda_1I(\mathbf{X}^\star;\mathbf{Y}_1)-(\lambda_1+\lambda_2)I(\mathbf{X}^\star;\mathbf{Y}_2)+\lambda_1\mathsf{S}_\eta^q(\mathbf{X}^\star)\\
&\stackrel{(a)}= \lambda_1I(\mathbf{X}^\star;\mathbf{Y}_1)-(\lambda_1+\lambda_2)I(\mathbf{X}^\star;\mathbf{Y}_2)+\lambda_1\mathsf{s}_\eta^q(\mathbf{X}_1^\star)\\
&\stackrel{(b)}\leq \lambda_1I(\mathbf{X}^\star;\mathbf{Y}_1)-(\lambda_1+\lambda_2)I(\mathbf{X}^\star;\mathbf{Y}_2)+\lambda_1C_\eta,\numberthis\label{EQ:bounded_inequality2}
\end{align*}
where (a) is by Theorem \ref{TM:Gaussian_maximizer_Envelope2}, while (b) follows from an adaptation of \cite[Proposition 19]{Chandra_Gauss_BC2014} to the function $\mathsf{s}_\eta^q(\mathbf{X})$ as defined in \eqref{EQ:function1}, which implies that for $\eta>1$ there exists a $C_\eta$, such that $\mathsf{s}_\eta^q(\mathbf{X})\leq C_\eta$, for all $P_{\mathbf{X}}$ (see Remark \ref{REM:adaptation}). By \eqref{EQ:bounded_inequality2}, we have
\begin{align*}
&\sup_{\mathbf{X}}\mathsf{t}_{\bm{\lambda},\eta}^q(\mathbf{X})\\
&\leq \sup_{\mathrm{0}\preceq\mathrm{K}:\ \mathbf{X}\sim\mathcal{N}(\mathbf{0},\mathrm{K})}\mspace{-15mu}\lambda_1I(\mathbf{X};\mathbf{Y}_1)-(\lambda_1+\lambda_2)I(\mathbf{X};\mathbf{Y}_2)+\lambda_1C_\eta.\numberthis\label{EQ:bounded_inequality_final}
\end{align*}
Let $\Sigma_j=(\rm{G}_j^\top\rm{G}_j)^{-1}$, $j=1,2$. For $\mathbf{X}\sim\mathcal{N}(\mathbf{0},\mathrm{K})$, we write
\begin{align*}
&2\lambda_1I(\mathbf{X};\mathbf{Y}_1)-2(\lambda_1+\lambda_2)I(\mathbf{X};\mathbf{Y}_2)\\
&=\lambda_1\log|\mathrm{I}+\mathrm{G}_1\mathrm{K}\mathrm{G}_1^\top|-(\lambda_1+\lambda_2)\log|\mathrm{I}+\mathrm{G}_2\mathrm{K}\mathrm{G}_2^\top|\\
&=\lambda_1\log|\mathrm{I}+\mathrm{K}\mathrm{G}_1\mathrm{G}_1^\top|-(\lambda_1+\lambda_2)\log|\mathrm{I}+\mathrm{K}\mathrm{G}_2\mathrm{G}_2^\top|\\
&\stackrel{(a)}=-\lambda_1\log|\Sigma_1|+(\lambda_1+\lambda_2)\log|\Sigma_2|\\
&\quad\quad\quad\quad\quad\quad+\lambda_1\big(\log|\Sigma_1+\mathrm{K}|-\lambda\log|\Sigma_2+K|\big),\numberthis\label{EQ:bounded_Gaussian_equality1}
\end{align*}
where (a) is by setting $\lambda\triangleq\frac{\lambda_1+\lambda_2}{\lambda_1}>1$. To bound the last two terms, we use the min-max theorem on eigenvalues: Let $\mu_i(\mathrm{A})$ be the $i$th smallest eigenvalue of the
symmetric matrix $\mathrm{A}\in\mathbb{R}^{t\times t}$, we have
\begin{equation}
\mu_i(\mathrm{A})=\min_{L_i}\max_{\mathbf{0}\neq\mathbf{u}\in L_i}\frac{\mathbf{u}^\top\mathrm{A}\mathbf{u}}{\mathbf{u}^\top\mathbf{u}}=\max_{L_{t+1-i}}\min_{\mathbf{0}\neq\mathbf{u}\in L_{t+1-i}}\frac{\mathbf{u}^\top\mathrm{A}\mathbf{u}}{\mathbf{u}^\top\mathbf{u}},
\end{equation}
where $L_i$ is an $i$-dimensional subspace of $\mathbb{R}^t$. Since the $t$-dimensional subspace of $\mathbb{R}^t$ is unique (that is, $L_t=\mathbb{R}_t$), we obtain
\begin{subequations}
\begin{align}
\mu_1(\mathrm{A})&=\max_{L_t}\min_{\mathbf{0}\neq\mathbf{u}\in L_t}\frac{\mathbf{u}^\top\mathrm{A}\mathbf{u}}{\mathbf{u}^\top\mathbf{u}}=\min_{\mathbf{0}\neq\mathbf{u}\in L_t}\frac{\mathbf{u}^\top\mathrm{A}\mathbf{u}}{\mathbf{u}^\top\mathbf{u}}\label{EQ:bounded_eigenvalue1}\\
\mu_t(\mathrm{A})&=\min_{L_t}\max_{\mathbf{0}\neq\mathbf{u}\in L_t}\frac{\mathbf{u}^\top\mathrm{A}\mathbf{u}}{\mathbf{u}^\top\mathbf{u}}=\max_{\mathbf{0}\neq\mathbf{u}\in L_t}\frac{\mathbf{u}^\top\mathrm{A}\mathbf{u}}{\mathbf{u}^\top\mathbf{u}}.\label{EQ:bounded_eigenvalue2}
\end{align}\label{EQ:bounded_eigenvalues}%
\end{subequations}
The RHSs of \eqref{EQ:bounded_eigenvalues} imply that for every non-zero $\mathbf{u}\in\mathbb{R}^t$ we have $\mu_1(\mathrm{A})\leq \frac{\mathbf{u}^\top\mathrm{A}\mathbf{u}}{\mathbf{u}^\top\mathbf{u}}\leq \mu_t(\mathrm{A})$. We upper and lower bound the $i$th eigenvalue of $\mathrm{K}+\Sigma_j$, for $j=1,2$, as follows
\begin{align*}
&\mu_i(\mathrm{K}+\Sigma_j)\\
&=\min_{L_i}\max_{\mathbf{0}\neq\mathbf{u}\in L_i}\left(\dfrac{\mathbf{u}^\top\mathrm{K}\mathbf{u}}{\mathbf{u}^\top\mathbf{u}}+\dfrac{\mathbf{u}^\top\Sigma_j\mathbf{u}}{\mathbf{u}^\top\mathbf{u}}\right)\\
&\mspace{-15mu}\begin{dcases}\geq \min_{L_i}\max\limits_{\mathbf{0}\neq\mathbf{u}\in L_i}\left(\dfrac{\mathbf{u}^\top\mathrm{K}\mathbf{u}}{\mathbf{u}^\top\mathbf{u}}+\mu_1(\Sigma_j)\right)=\mu_i(\mathrm{K})+\mu_1(\Sigma_j),\\
\leq \min_{L_i}\max_{\mathbf{0}\neq\mathbf{u}\in L_i}\left(\dfrac{\mathbf{u}^\top\mathrm{K}\mathbf{u}}{\mathbf{u}^\top\mathbf{u}}+\mu_t(\Sigma_j)\right)=\mu_i(\mathrm{K})+\mu_t(\Sigma_j).\end{dcases}\numberthis\label{EQ:bounded_eigenvalue_final}
\end{align*}
Hence the eigenvalues of $\mathrm{K}+\Sigma_j$, $j=1,2$, satisfy
\begin{equation}
\mu_i(\mathrm{K})+\mu_1(\Sigma_j)\leq \mu_i(\mathrm{K}+\Sigma_j)\leq \mu_i(\mathrm{K})+\mu_t(\Sigma_j),\label{EQ:bounded_eigenvalue_final_sum}
\end{equation}
where $i\in[t]$. We now bound the last two terms in \eqref{EQ:bounded_Gaussian_equality1} as
\begin{align*}
\log|\Sigma_1+\mathrm{K}|-\lambda&\log|\Sigma_2+K|\\
&=\sum_{i=1}^t\log\left(\frac{\mu_i(\mathrm{K}+\Sigma_1)}{\mu_i(\mathrm{K}+\Sigma_2)^\lambda}\right)\\
                                                 &\stackrel{(a)}\leq\sum_{i=1}^t\log\left(\frac{\mu_i(\mathrm{K})+\mu_t(\Sigma_1)}{\big(\mu_i(\mathrm{K})+\mu_1(\Sigma_2)\big)^\lambda}\right)\\
                                                 &\leq t\cdot \max_i\log\left(\frac{\mu_i(\mathrm{K})+\mu_t(\Sigma_1)}{\big(\mu_i(\mathrm{K})+\mu_1(\Sigma_2)\big)^\lambda}\right)\\
                                                 &\stackrel{(b)}\leq t\cdot\log\left(\frac{\mu^\star+\mu_t(\Sigma_1)}{\big(\mu^\star+\mu_1(\Sigma_2)\big)^\lambda}\right),\numberthis
\end{align*}
where (a) follows from by \eqref{EQ:bounded_eigenvalue_final}, and (b) is by setting $\mu^\star=\max\left\{0,\frac{1}{1-\lambda}\big(\mu_1(\Sigma_2)-\lambda\mu_t(\Sigma_1)\big)\right\}$ and noting that $\mu_i(\mathrm{K})\geq 0$ and that the derivative of the function $c(x)\triangleq \log\big(x+\mu_t(\Sigma_1)\big)-\lambda\log\big(x+\mu_1(\Sigma_2)\big)$ is zero at $x=\mu^\star$, negative when $x>\mu^\star$ and positive when $x<\mu^\star$. Since $\mu_t(\Sigma_1),\mu_1(\Sigma_2)> 0$ (which holds since the positive semi-definite matrices $\Sigma_1$ and $\Sigma_2$ are invertible), we conclude that for every $\mathbf{X}\sim P_\mathbf{X}$
\begin{align*}
&\mathsf{t}_{\bm{\lambda},\eta}^q(\mathbf{X})\\
&\leq -\lambda_1\log|\Sigma_1|+(\lambda_1+\lambda_2)\log|\Sigma_2|\\
&\quad\quad\quad\quad\quad\quad+\lambda_1\cdot t\cdot\log\left(\frac{\mu^\star+\mu_t(\Sigma_1)}{\big(\mu^\star+\mu_1(\Sigma_2)\big)^{\frac{\lambda_1+\lambda_2}{\lambda_1}}}\right)\\
&\triangleq  B_{\bm{\lambda},\eta}<\infty.\numberthis\label{eq:bounded_proved}
\end{align*}


\subsection{Proof of Lemma \ref{LEMMA:nested_envelope_continous}}\label{APPEN:nested_envelope_continous_proof}

The proof of Lemma \ref{LEMMA:nested_envelope_continous} follows immediately by an adaptation of \cite[Proposition 20]{Chandra_Gauss_BC2014} (see Remark \ref{REM:adaptation}) and by \cite[Theorem 5]{Chandra_Gauss_BC2014}, which, respectively, imply that $\mathsf{S}^q_\eta(\mathbf{X}_n)\xrightarrow[n\to\infty]{} \mathsf{S}^q_\eta(\mathbf{X}^\star)$, and that $h(\mathbf{Y}_{j,n})\xrightarrow[n\to\infty]{} h(\mathbf{Y}_j^\star)$, for $j=1,2$. Here $\mathbf{Y}_{1,n}$ and $\mathbf{Y}_{2,n}$ are the outputs of the MIMO Gaussian BC with input $\mathbf{X}_n$.


\subsection{Proof of Lemma \ref{LEMMA:exsists_limits_notzero}}\label{APPEN:exsists_limits_notzero_proof}

Let $i\in[m]$ be such that $\alpha_i^\star=0$ and note that for every $\hat{\mathrm{K}}\succeq\mathrm{0}$, we have
\begin{align*}
\tilde{\mathrm{v}}_{\bm{\lambda_0},\alpha,\eta}^q\big(\hat{\mathrm{K}}\big)&\stackrel{(a)}=\mathsf{f}_{\bm{\lambda_0},\alpha,\eta}^q\big(\mathbf{X}_{\hat{\mathrm{K}}}^\star\big)\\
                                                            &\stackrel{(b)}\leq\lambda_2I\big(\mathbf{X}_{\hat{\mathrm{K}}}^\star;\mathbf{Y}_2\big)+\hat{B}_{\bm{\lambda_0},\alpha,\eta}\\
                                                            &\leq \frac{\lambda_2}{2}\log\big|I+\mathrm{G}_2\hat{\mathrm{K}}\mathrm{G}_2^\top\big|+\hat{B}_{\bm{\lambda_0},\alpha,\eta},\numberthis\label{EQ:limnozero_ub1}
\end{align*}
where (a) follows from \eqref{EQ:exsists_maximizer_smallv_equality} and the non-negativity of mutual information, while (b) is since $\mathsf{T}_{\bm{\lambda},\eta}^q(\mathbf{X})$ is bounded. Let $\big\{\alpha^{(n_k)}_i\big\}_{k\in\mathbb{N}}$ be a subsequence of $\big\{\alpha^{(n)}_i\big\}_{n\in\mathbb{N}}$, such that $\alpha^{(n_k)}_i>0$ for every $k\in\mathbb{N}$ (if there is no such subsequence, the result of Lemma \ref{LEMMA:exsists_limits_notzero} is immediate). Since $\alpha_i^{(n_k)}\mathrm{K}_i^{(n_k)}\preceq \mathrm{K}$ and $\alpha_i^{(n_k)}>0$, we obtain
\begin{equation}
\mathrm{K}_i^{(n_k)}\preceq\frac{1}{\alpha_i^{(n_k)}}\mathrm{K},\quad\forall k\in\mathbb{N}.\label{EQ:limnozero_ub2}
\end{equation}
We thus conclude that
\begin{align*}
&\alpha_i^{(n_k)}\tilde{\mathrm{v}}_{\bm{\lambda_0},\alpha,\eta}^q\big(\mathrm{K}_i^{(n_k)}\big)\\
&\stackrel{(a)}\leq\alpha_i^{(n_k)}\left(\frac{\lambda_2}{2}\log\big|I+\mathrm{G}_2\mathrm{K}_i^{(n_k)}\mathrm{G}_2^\top\big|+\hat{B}_{\bm{\lambda_0},\alpha,\eta}\right)\\
&\stackrel{(b)}\leq\alpha_i^{(n_k)}\left(\frac{\lambda_2}{2}\log\left|I+\mathrm{G}_2\frac{\mathrm{K}}{\alpha_i^{(n_k)}}\mathrm{G}_2^\top\right|+\hat{B}_{\bm{\lambda_0},\alpha,\eta}\right)\\
&\stackrel{(c)}=\alpha_i^{(n_k)}\left(\frac{\lambda_2}{2}\sum_{j=1}^t\log\left(1+\frac{\mu_j}{\alpha_i^{(n_k)}}\right)+\hat{B}_{\bm{\lambda_0},\alpha,\eta}\right)\xrightarrow[n\to\infty]{}0,\numberthis
\end{align*}
where (a) and (b) follow from \eqref{EQ:limnozero_ub1} and \eqref{EQ:limnozero_ub2}, respectively, while (c) follows by denoting the eigenvalues of $\mathrm{G}_2\mathrm{K}\mathrm{G}_2^\top$ by $\{\mu_j\}_{j=1}^t$.


\section{Proof of Proposition \ref{PROP:minimax_interchange}}\label{APPEN:minimax_interchange_proof}

The proof relies on a result from \cite[Corollary 2]{Nair_Marton_Optimality2014}, which we reproduce in the following.

\begin{proposition}[Min-Max Interchange]\label{PROP:minmax_interchange}
Let $\Lambda_d\triangleq\left\{\bm{\lambda}\in\mathbb{R}_+^d\middle|\sum_{i=1}^d\lambda_i=1\right\}$ be the $d$-dimensional simplex. Let $\mathcal{P}$ be a set of distribution $P_U$ over a set $\mathcal{U}$. Let $\big\{g_i:\mathcal{P}\to\mathbb{R}\big\}_{i\in[d]}$ be a set of functionals such that
\begin{equation}
\mathcal{A}\triangleq\Big\{\mathbf{a}\in\mathbb{R}^d\Big|\forall i\in[d],\ \exists P_U\in\mathcal{P},\ a_i\leq g_i(P_U)\Big\}\label{EQ:convex_set_appen}
\end{equation}
is a convex set. Then
\begin{equation}
\sup_{P_U\in\mathcal{P}}\min_{\bm{\lambda}\in\Lambda_d}\sum_{i=1}^d\lambda_ig_i(P_U)=\min_{\bm{\lambda}\in\Lambda_d}\sup_{P_U\in\mathcal{P}}\sum_{i=1}^d\lambda_ig_i(P_U).
\end{equation}
\end{proposition}

Let $d=2$ and $\mathcal{P}$ be the set of PDFs $P_{W,V,U,\mathbf{X}}$ that satisfy $\mathbb{E}\big[\mathbf{X}\mathbf{X}^\top\big]\preceq\mathrm{K}$. Set
\begin{subequations}
\begin{align}
g_1&\left(P_{W,V,U,\mathbf{X}}\right)\nonumber\\
&=\lambda_0I(W;\mathbf{Y}_1)+\lambda_2I(V;\mathbf{Y}_2|W)\nonumber\\
&\quad+\lambda_1\Big[I(U;\mathbf{Y}_1|W)-I(U;V|W)-I(U;\mathbf{Y}_2|W,V)\Big]\\
g_2&\left(P_{W,V,U,\mathbf{X}}\right)\nonumber\\
&=\lambda_0I(W;\mathbf{Y}_2)+\lambda_2I(V;\mathbf{Y}_2|W)\nonumber\\
&\quad+\lambda_1\Big[I(U;\mathbf{Y}_1|W)-I(U;V|W)-I(U;\mathbf{Y}_2|W,V)\Big],
 \end{align}
\end{subequations}
and consider the corresponding set $\mathcal{A}$ from \eqref{EQ:convex_set_appen}. To show that $\mathcal{A}$ is convex, let $(a_1,a_2),(b_1,b_2)\in\mathcal{A}$ and $P_a,P_b\in\mathcal{P}$ be two distributions, such that $a_i\leq g_i(P_a)$ and $b_i\leq g_i(P_b)$, for $i=1,2$. Fix $\alpha\in[0,1]$ and consider a distribution $P$ given by
\begin{equation}
P(w,v,u,\mathbf{x})=\alpha P_a(w,v,u,\mathbf{x})+\bar{\alpha}P_b(w,v,u,\mathbf{x}),\label{EQ:new_PDF}
\end{equation}
for all $(w,v,u,\mathbf{x})\in\mathcal{W}\times\mathcal{V}\times\mathcal{U}\times\mathcal{X}^n$. Equivalently, $P$ can be represented by setting $\tilde{W}=(Q,W)$, where $Q\sim\ber(\alpha)$, and denoting $P\triangleq P_{\tilde{W},V,U,\mathbf{X}}=P_{(Q,W),V,U,\mathbf{X}}$, for which
\begin{subequations}
\begin{align}
P_{(Q,W),V,U,\mathbf{X}}\big((0,w),v,u,\mathbf{x}\big)&=\alpha P_a(w,v,u,\mathbf{x})\\
P_{(Q,W),V,U,\mathbf{X}}\big((1,w),v,u,\mathbf{x}\big)&=\bar{\alpha} P_b(w,v,u,\mathbf{x}),
\end{align}
\end{subequations}
for all $(w,v,u,\mathbf{x})\in\mathcal{W}\times\mathcal{V}\times\mathcal{U}\times\mathcal{X}^n$. First note that
\begin{equation}
\mathbb{E}_P\big[\mathbf{X}\mathbf{X}^T\big]=\alpha \mathbb{E}_{P_a}\big[\mathbf{X}\mathbf{X}^T\big]+\bar{\alpha}\mathbb{E}_{P_b}\big[\mathbf{X}\mathbf{X}^T\big]\preceq\mathrm{K},\label{EQ:new_PDF_expectation}
\end{equation}
where $\mathbb{E}_Q$ denotes that an expectation is taken with respect to $Q$. This implies that $P\in\mathcal{P}$.

Next, by evaluating $g_i$, $i=1,2$, with respect to $P$, we have
\begin{align*}
g_i(P)&=\lambda_0I_P(\tilde{W};\mathbf{Y}_i)+\lambda_1\Big[I_P(U;\mathbf{Y}_1|\tilde{W})-I_P(U;V|\tilde{W})\\
      &\quad\quad\quad\quad\quad\mspace{10mu}-I_P(U;\mathbf{Y}_2|\tilde{W},V)\Big]+\lambda_2I_P(V;\mathbf{Y}_2|\tilde{W})\\
      &=\lambda_0I_P(Q;\mathbf{Y}_i)+\alpha g_i(P_a)+\bar{\alpha} g_i(P_b)\\
      &\geq \alpha a_i+\bar{\alpha} b_i\numberthis,
\end{align*}
implying that $\alpha(a_1,a_2)+\bar{\alpha}(b_1,b_2)\in\mathcal{A}$, which establishes the convexity of $\mathcal{A}$. In the above, $I_P$ indicates that a mutual information term is taken with respect to an underlying distribution $P$. The proof of Proposition \ref{PROP:minimax_interchange} is completed by invoking Proposition \ref{PROP:minmax_interchange}, while noting that for every tuple of random variables $(W,V,U,\mathbf{X})$, with $(W,V,U)-\mathbf{X}-(\mathbf{Y}_1,\mathbf{Y}_2)$ and $\mathbb{E}\big[\mathbf{X}\mathbf{X}^\top\big]\preceq\mathrm{K}$, the minimum
\begin{align*}
\min_{\alpha\in[0,1]}&\Big\{\lambda_0\Big[\alpha I(W;\mathbf{Y}_1)+\bar{\alpha} I(W;\mathbf{Y}_2)\Big]\\
&\mspace{-30mu}+\lambda_1\Big[I(U;\mathbf{Y}_1|W)-I(U;V|W)-I(U;\mathbf{Y}_2|W,V)\Big]\\
&\quad\quad\quad\quad\quad\quad\quad\quad\quad\quad\quad\quad\quad\mspace{10mu}+\lambda_2I(V;\mathbf{Y}_2|W)\Big\}
\end{align*}
is attained by either $\alpha=0$ or $\alpha=1$.
\newpage

\bibliographystyle{unsrt}
\bibliographystyle{IEEEtran}
\bibliography{ref}

\begin{IEEEbiographynophoto}{Ziv Goldfeld}

Dr. Ziv Goldfeld is currently a postdoctoral fellow at the Laboratory for
Information and Decision Systems (LIDS) at MIT. He graduated with a B.Sc.
summa cum laude, an M.Sc. summa cum laude, and a Ph.D. in Electrical and
Computer Engineering from Ben-Gurion University, Israel, in 2012, 2015 and
2018, respectively. His research interest include theoretical machine
learning, information theory, complex dynamical systems, high-dimensional
and nonparametric statistics and applied probability. Honors include the

(S'13-M'17) received his B.Sc.\@ (summa cum laude), M.Sc.\@ (summa cum laude) and Ph.D. degrees in Electrical and Computer Engineering from the Ben-Gurion University, Israel, in 2012, 2014 and 2017,  respectively. He is currently a postdoctoral fellow at the Laboratory for Information and Decision Systems (LIDS) at MIT.

Ziv is a recipient of several awards, among them the are Rothschild postdoctoral fellowship, the Feder Award, a best student paper award in the IEEE 28-th Convention of Electrical and Electronics Engineers in Israel, the Basor fellowship for outstanding students in the direct Ph.D. program, the Lev-Zion fellowship and the Minerva Short-Term Research Grant (MRG).
\end{IEEEbiographynophoto}

\begin{IEEEbiographynophoto}{Haim H. Permuter}
(M'08-SM'13) received his B.Sc.\@ (summa cum laude) and M.Sc.\@ (summa cum laude) degrees in Electrical and Computer Engineering from the Ben-Gurion University, Israel, in 1997 and 2003, respectively, and a Ph.D. degree in Electrical Engineering from Stanford University, California in 2008.

Between 1997 and 2004, he was an officer at a research and development unit of the Israeli Defense Forces. Since 2009 he is with the department of Electrical and Computer Engineering at Ben-Gurion University where he is currently an associate professor.

Prof. Permuter is a recipient of several awards, among them are the Fullbright Fellowship, the Stanford Graduate Fellowship (SGF), Allon Fellowship, and the U.S.-Israel Binational Science Foundation Bergmann Memorial Award. Haim is currently serving on the editorial board of the IEEE Transactions on Information Theory.
\end{IEEEbiographynophoto}
\end{document}